\documentclass[12pt]{article}
\usepackage{amsfonts,amssymb,epsfig,amsmath}
\usepackage{color}

\renewcommand{\baselinestretch}{1.2}

\newcommand{\be}{\begin{eqnarray}}
\newcommand{\ee}{\end{eqnarray}}

\newcommand{\bn}{\begin{enumerate}}
\newcommand{\en}{\end{enumerate}}

\begin{document}

\makeatletter \@addtoreset{equation}{section} \makeatother
\renewcommand{\theequation}{\thesection.\arabic{equation}}
\renewcommand{\thefootnote}{\alph{footnote}}

\begin{titlepage}

\begin{center}
\hfill {\tt SNUTP17-001}\\

\vspace{3cm}

{\Large\bf Asymptotic M5-brane entropy from S-duality}

\vspace{2cm}

\renewcommand{\thefootnote}{\alph{footnote}}

{\large Seok Kim and June Nahmgoong}

\vspace{0.7cm}

\textit{Department of Physics and Astronomy \& Center for
Theoretical Physics\\
Seoul National University, 1 Gwanak-ro, Gwanak-gu, Seoul 08826, Korea.}\\

\vspace{0.7cm}

E-mails: {\tt skim@phya.snu.ac.kr, earendil25@snu.ac.kr }

\end{center}

\vspace{1cm}

\begin{abstract}

We study M5-branes compactified on $S^1$ from the D0-D4 Witten index in 
the Coulomb phase. We first show that the prepotential of this index is S-dual,
up to a simple anomalous part. This
is an extension of the well-known S-duality of the 4d 
$\mathcal{N}=4$ theory to the 6d $(2,0)$ theory on finite $T^2$. 
Using this anomalous S-duality,
we find that the asymptotic free energy scales like $N^3$ when various
temperature-like parameters are large. This shows that the number of 5d
Kaluza-Klein fields for light D0-brane bound states is proportional to $N^3$. 
We also compute some part of the asymptotic free energy from 6d chiral
anomalies, which precisely agrees with our D0-D4 calculus.

\end{abstract}

\end{titlepage}

\renewcommand{\thefootnote}{\arabic{footnote}}

\setcounter{footnote}{0}

\renewcommand{\baselinestretch}{1}

\tableofcontents

\renewcommand{\baselinestretch}{1.2}

\section{Introduction}

Strong/weak-coupling duality, or S-duality, exists in
a number of quantum systems. In 4d gauge theories,
it takes the form of electromagnetic duality,
which inverts the gauge coupling and exchanges the roles of elementary
charged particles and magnetic monopoles \cite{Montonen:1977sn}. It is  
realized in the simplest manner in maximally supersymmetric Yang-Mills (SYM)
theory \cite{Osborn:1979tq}. In this case, the spectrum of dyons in the Coulomb 
phase was shown to exhibit $SL(2,\mathbb{Z})$ duality \cite{Sen:1994yi},
providing a robust evidence of S-duality. S-duality in quantum field theories
has also been a cornerstone of developing string dualities \cite{Witten:1995ex}.
In both QFT and string theory, S-duality provides
valuable insights on the strongly coupled regions of the systems.

S-duality of maximal SYM has many implications. In particular, this duality 
is related to the existence of 6d superconformal field theories called $(2,0)$ 
theories \cite{Witten:1995zh}. 4d maximal SYM theories with gauge groups $U(N)$, 
$SO(2N)$, $E_N$ are realized by compactifying 6d $(2,0)$ theories on small $T^2$.
The $SL(2,\mathbb{Z})$ duality originates from the modular transformation on $T^2$.
On one side, this relation highlights the far-reaching implications of the 6d
CFTs to challenging lower dimensional systems. On the other hand, the 6d CFTs
lack microscopic definitions, so that S-duality can provide useful clues to
better understand the mysterious 6d CFTs. In this paper,
we study the S-duality of the 6d $(2,0)$ theories compactified on finite
$T^2$, and use it to explore some interesting properties of these systems.

Our key observable of the 6d $(2,0)$ theory is the partition function 
of the D0-D4 system. More precisely, we study the Witten index of the 
quantum mechanical $U(k)$ gauge theory for $k$ D0-branes bound to $N$ 
separated D4-branes and fundamental open strings, 
and also study their generating function for arbitrary $k$. In
M-theory, this system is made of $N$ M5-branes wrapping $S^1$. 
The D0-D4 systems describe the 6d $(2,0)$ CFT on M5-branes in
the sectors with nonzero Kaluza-Klein momenta. From the viewpoint of 
5d super-Yang-Mills theory on D4-branes, this partition function is also
known as Nekrasov's instanton partition function \cite{Nekrasov:2002qd}.
Although the 5d Yang-Mills description of its instanton solitons is UV incomplete,
the D0-D4 system provides a UV complete description for computing the instanton
partition function. The results in our paper rely only on the
UV complete D0-D4 quantum mechanics. 

The D0-D4 index was explored in \cite{Kim:2011mv}. From the 5d viewpoint,
this is a partition function on $\mathbb{R}^4\times S^1$, where $S^1$ is the
temporal circle for the Witten index. However, with D0-branes (Yang-Mills
instantons) providing the Kaluza-Klein (KK) modes of M-theory, there are evidences 
that this index (multiplied by a 5d perturbative part) is a 
partition function of the 6d $(2,0)$ SCFT
on $\mathbb{R}^4\times T^2$. See \cite{Kim:2011mv,Haghighat:2013gba} for the
$A_{N-1}$ theories, and \cite{Hwang:2016gfw} for the $D_N$ theories.

Regarding the D0-D4 index as a 6d partition function on $\mathbb{R}^4\times T^2$,
one can naturally ask if it transforms in a simple manner under 
the S-duality acting on $T^2$.
In this paper, we establish the S-duality of the prepotential of this index,
finding a simple anomaly of S-duality which we can naturally
interpret with 6d chiral anomalies. Note that the prepotential $F$
is the leading coefficient of the free energy
$-\log Z\sim\frac{F(a,m)}{\epsilon_1\epsilon_2}$ when the so-called
Omega deformation is taken to zero, $\epsilon_1,\epsilon_2\rightarrow 0$.
$a$ and $m$ are Coulomb VEV and 5d $\mathcal{N}=1^\ast$ mass parameter, 
explained in section 2. The anomaly of S-duality takes the following form. 
The prepotential $F$ can be divided into two parts,
$F(a,m)=F_{\textrm{S-dual}}(a,m)+F_{\rm anom}(m)$,
where $F_{\textrm{S-dual}}$ is related to its S-dual prepotential by a 6d 
generalization of the Legendre transformation. (See section 2 for details.) 
$F_{\rm anom}$ is a simple function which does not obey 
S-duality, thus named anomalous part. We find $F_{\rm anom}$ in a closed form 
in section 2, which in particular is independent of the Coulomb VEV $a$.

This finding has two major implications. Firstly, similar result was 
found for the prepotential $F^{\rm 4d}$ of the 4d $\mathcal{N}=2^\ast$ theory 
\cite{Billo':2015ria}, related to our prepotential by taking the small 
$T^2$ limit. $F^{\rm 4d}$ is also given by the sum 
$F^{\rm 4d}_{\textrm{S-dual}}(a,m)+F^{\rm 4d}_{\rm anom}(m)$, where
$F^{\rm 4d}_{\textrm{S-dual}}$ is self S-dual. Since
$F_{\rm anom}^{\rm 4d}$ is independent of $a$, $F^{\rm 4d}$ is S-dual 
in the Seiberg-Witten theory, which only sees $a$ derivatives of $F^{\rm 4d}$.
In our 6d uplift, $F$ appearing in the Seiberg-Witten theory also 
does not see $F_{\rm anom}(m)$ for the same reason.

Secondly, the partition function $Z$ itself is a Witten index of the 6d theory
on $\mathbb{R}^{4,1}\times S^1$.
So the full prepotential $F=F_{\textrm{S-dual}}+F_{\rm anom}$ including
the $a$ independent $F_{\rm anom}$ is physically meaningful, as the leading
part of the free energy $-\log Z$ when $\epsilon_1,\epsilon_2$
are small. At this stage, we note our key discovery that
$F_{\rm anom}$ contains a term proportional to $N^3$ in a suitabe
large $N$ limit, to be explained in section 2. In particular,
we further consider the
limit in which the KK modes on the circle become light. This amounts to
taking the chemical potential $\beta$ conjugate to the KK momentum
(D0-brane charge) to be much smaller than the inverse-radius of $S^1$. 
The small $\beta$ limit is
the strong coupling limit of the 5d Yang-Mills theory, or the limit in
which the sixth circle decompactifies. This is the regime in which 6d
CFT physics should be visible. The prepotential in this limit can
be computed from our anomalous S-duality, since it relates the small $\beta$
(strong coupling) regime to the well-understood large $\beta$ (weak coupling)
regime. $F_{\rm anom}$ determines the small $\beta$ limit 
of the free energy, and makes it scale like $N^3$.  
We also show that the term in the asymptotic free energy 
proportional to $N^3$ is related to the chiral anomaly of the 6d $(2,0)$ theory,
using the methods of \cite{DiPietro:2014bca}. These findings
show that the number of 5d KK fields for D0-brane bound states grows like
$N^3$, as we decompactify the system to 6d.

The rest of this paper is organized as follows. In section 2.1 and 2.2, 
we develop the anomalous S-duality of the prepotential and test it
either by expanding $F$ in the 5d $\mathcal{N}=1^\ast$ mass $m$, or 
by making the `M-string' expansion \cite{Haghighat:2013gba}. 
In section 2.3, we discuss the 6d $(2,0)$ theories of
$D_N$ and $E_N$ types. In section 3, we study the high temperature free energy
and show that it scales like $N^3$ in a suitable large $N$ limit. 
In section 3.1, we test our result for $U(1)$ 
theory. In section 3.2, we account for the imaginary part of the asymptotic free 
energy from 6d chiral anomalies. Section 4 concludes with
comments and future directions.

\section{S-duality of 6d $(2,0)$ theories on
$\mathbb{R}^4\times T^2$}

We shall study the Witten index of the D0-D4 quantum mechanics,
consisting of $k$ D0-branes and $N$ D4-branes. This system is a quantum 
mechanical $U(k)$ gauge theory with $8$ supersymmetry and $U(N)$ global symmetry.
See, for instance, \cite{Kim:2011mv} for the details of this system.
Here, we shall only explain some basic aspects.
The bosonic variables consist of four Hermitian $k\times k$ matrices
$a_m\sim a_{\alpha\dot\beta}$, two complex $k\times N$ matrices
$q_{\dot\alpha}$, five Hermitian $k\times k$ matrices $\varphi^I$,
and a quantum mechanical $U(k)$ gauge field $A_t$. Here, $m=1,2,3,4$ is
the vector index on $\mathbb{R}^4$ for the spatial worldvolume of
the D4-branes. $\alpha$ and $\dot\alpha$ are doublet indices of
$SU(2)_l$ and $SU(2)_r$ respectively, which form $SO(4)$ rotation of
$\mathbb{R}^4$. $I=1,\cdots,5$ is the vector index on $\mathbb{R}^5$
transverse to the D4-branes. When $\varphi^I$ are all diagonal
matrices, their eigenvalues are interpreted as D0-brane positions
transverse to D4-branes. Similarly, when
$a_m$ are all diagonal, their eigenvalues are interpreted as D0-brane
positions along D4-brane worldvolume. $q_{\dot\alpha}$ represent
internal degrees of freedom. The bosonic potential energy is given by
\begin{equation}
  V=\frac{1}{2}D^iD^i-\frac{1}{2}[\varphi^I,a_m]^2
  +\left|\varphi^Iq_{\dot\alpha}\right|^2-\frac{1}{4}[\varphi^I,\varphi^J]^2\ ,
\end{equation}
where traces are assumed if necessary, $i=1,2,3$ is the $SU(2)_r$ triplet 
index. $D^i$ are given by 
\begin{equation}\label{ADHM}
  D^i=(\tau^i)^{\dot\alpha}_{\ \dot\beta}\left(
  q_{\dot\alpha}q^{\dag\dot\beta}+\frac{1}{2}
  [a^{\dot\beta\alpha},a_{\alpha\dot\alpha}]
  \right)\ ,
\end{equation}
where $\tau^i$ are the Pauli matrices. 

This system flows in IR to two branches.
Classically, they are described by two branches of moduli space
satisfying $V=0$, or 
$D^i=0$, $[\varphi^I,a_m]=0$, $\varphi^Iq_{\dot\alpha}=0$ and
$[\varphi^I,\varphi^J]=0$. The first branch is obtained by taking
$q_{\dot\alpha}=0$, and $a_m$, $\varphi^I$ to be diagonal matrices.
The $k$ sets of eigenvalues of $(a_m,\varphi^I)$ represent the positions
of $k$ D0-branes on $\mathbb{R}^9$, unbound to the D4-branes.
The second branch is obtained by taking $\varphi^I=0$, and $q_{\dot\alpha}$,
$a_m$ to satisfy $D^i=0$. After modding out by the $U(k)$ gauge orbit,
one can show that this branch is described by $4Nk$ real parameters.
The two branches meet at $\varphi^I=0$, $q_{\dot\alpha}=0$. Far away from
this intersection, each branch is described by a nonlinear sigma model (NLSM)
on its moduli space. We are interested in the second branch,
describing 6d CFT on M5-branes in the sector with $k$ units of KK momentum.
The Witten index of the second branch can be computed
easily by deforming the system by a Fayet-Iliopoulos (FI) parameter,
shifting $D^i$ in (\ref{ADHM}) by three constant $\xi^i$. After this
deformation, the first branch becomes non-BPS, since $q_{\dot\alpha}=0$ cannot
solve $D^i=0$ with $\xi^i\neq 0$. So the Witten index acquires
contributions only from the second branch.

One can understand the second branch from the low energy field theory of 
D4-branes, the 5d maximal SYM theory. D0-branes
are realized in Yang-Mills theory as instanton solitons, classically
described by finite energy stationary solutions of the following BPS equation,
\begin{equation}\label{self-dual}
  F_{mn}=\pm\frac{1}{2}\epsilon_{mnpq}F_{pq}
  \ \ \ ,\ \ \ m,n,p,q=1,\cdots,4\ .
\end{equation}
The finite energy solutions are labeled by the instanton number $k$,
defined by
\begin{equation}
  k\equiv\frac{1}{16\pi^2}\int_{\mathbb{R}^4}{\rm tr}(F\wedge F)\in\mathbb{Z}\ .
\end{equation}
We shall consider D0-branes rather than anti-D0-branes, with $k>0$,
equivalently (\ref{self-dual}) with $+$ sign.
$k$ corresponds to the rank of the $U(k)$ gauge
group of the quantum mechanics. The solutions of
$D^i=0$, modded out by the $U(k)$ gauge orbit, provides the moduli
space of self-dual instantons. The NLSM on the second branch can be obtained 
by the moduli space approximation of the instanton solitons.
However, this moduli space is known to be singular, having the so-called
small instanton singularities. Due to these singularities, the NLSM description
is incomplete, and needs a UV completion. This is naturally interpreted as 
inheriting the UV incompleteness of the 5d Yang-Mills description. Although 
we do not know how to UV complete the full 5d SYM, the 
NLSM can be UV completed to the $U(k)$ quantum mechanics.

We shall study the D0-D4 system in the Coulomb phase, with scalar vacuum
expectation value (VEV). For $U(N)$ theories, the VEV is parametrized 
by constant $N\times N$ diagonal matrices 
$\Phi^I={\rm diag}(\Phi^I_1,\cdots,\Phi^I_{N})$.
In the D0-D4 system, $\Phi^I$ deforms the bosonic potential $V$ as 
\begin{equation}
  V=\frac{1}{2}D^iD^i-\frac{1}{2}[\varphi^I,a_m]^2
  -\frac{1}{4}[\varphi^I,\varphi^J]^2+\left|
  \varphi^Iq_{\dot\alpha}-q_{\dot\alpha}\Phi^I\right|^2\ .
\end{equation}
The $N$ eigenvalues of $\Phi^I$ correspond to the positions of
$N$ D4-branes on $\mathbb{R}^5$. We shall
separate the D4-branes along a line, giving nonzero VEV to
$\Phi^5$ only. In this setting,
we shall study the BPS bound states of the D0-branes and the
fundamental open strings stretched along the $\Phi^5$ direction,
suspended between a pair of D4-branes. The bound states preserve $4$ Hermitian
supercharges. In 6d $(2,0)$ theory, we compactify a spatial direction 
on a circle with radius $R^\prime$. The BPS states 
saturate the bound $E\geq \frac{P}{R^\prime}+v_iq_i$,
where $E$ is the energy, and
$P$ is the quantized momentum on $S^1$ which is $k$ in
the D0-D4 system. $v_i$ are the $N$ eigenvalues of the scalar $\Phi^5$, 
and $q_i$'s are the $U(1)^N$ electric charges in the Coulomb branch, 
satisfying $q_1+\cdots+q_N=0$. 
From the 6d viewpoint, they are the self-dual 
strings with charges $q_i$ coming from open M2-branes, with $P$ units of 
momenta on them. We also define $H\equiv R^\prime (E-v_iq_i)$, which is the 
(dimensionless) energy on the self-dual strings.

The 6d index is defined by
\begin{equation}
  Z(\tau,m,\epsilon_{1,2},v)={\rm Tr}\left[(-1)^F
  e^{2\pi i\tau\frac{H+P}{2}}e^{-2\pi i\bar\tau\frac{H-P}{2}}
  e^{\epsilon_1(J_1+J_R)+\epsilon_2(J_2+J_R)}e^{2mJ_L}e^{-v_iq_i}\right]\ .
\end{equation}
Here, $J_1,J_2$ are two Cartans rotating the two 2-planes of $\mathbb{R}^4$, 
$J_L,J_R$ are the Cartans of $SU(2)_L\times SU(2)_R=SO(4)\subset SO(5)$ unbroken 
by the VEV of $\Phi^5$. The measure is chosen so that it commutes with 
$2$ of the $4$ Hermitian supercharges preserved by the BPS states, 
or a complex supercharge $Q$ and its conjugate $Q^\dag$. See \cite{Kim:2011mv}
for the details. One also finds that $\frac{H-P}{2}\sim\{Q,Q^\dag\}$. 
Since only the states saturating the BPS bound $H\geq P$ contribute to the 
index, $Z$ is independent of $\bar\tau$. With $H=P$ understood, the factor
$e^{2\pi i\tau\frac{H+P}{2}}\rightarrow e^{2\pi i\tau P}$ 
weights the BPS states with the momentum $P$ along the circle. So 
$Z$ can be written as
\begin{equation}
  Z(\tau,m,\epsilon_{1,2},v)=Z_{\rm pert}(m,\epsilon_{1,2},v)
  \sum_{k=0}^\infty q^kZ_k(m,\epsilon_{1,2},v)
\end{equation}
where $q\equiv e^{2\pi i\tau}$, and $Z_0\equiv 1$ by definition. 
$Z$ can be computed in the weakly coupled
type IIA regime, in which D0-branes are much
heavier than the stretched fundamental strings.
$Z_k$ is computed as the nonperturbative Witten index of the D0-D4 system 
with fixed $k$.
$Z_{\rm pert}$ comes from the zero modes at $P=0$, the perturbative open 
string modes on the D4-branes. This factor can also be understoood as 
coming from the perturbative partition function of the 5d maximal SYM.
Since we are in the weakly coupled regime, $Z_{\rm pert}$ can be computed
unambiguously from the quadratic part of the Yang-Mills theory. 
Although we compute $Z_{\rm pert}$ and $Z_k$ in this special regime, 
we naturally expect the result to be valid at general type IIA coupling, 
since this is a Witten index independent of the continuous coupling. 

$Z_k$ and $Z_{\rm pert}$ are known for classical 
gauge groups. For $U(N)$, $Z_k$ is given by 
\cite{Flume:2002az,Bruzzo:2002xf,Kim:2011mv,Hwang:2014uwa}
\begin{equation}\label{Zk-Young}
  Z_k=\sum_{Y_i; \sum_{i=1}^N|Y_i|=k}
  \prod_{i,j=1}^N\prod_{s\in Y_i}\frac{\sinh\frac{E_{ij}(s)+m-\epsilon_+}{2}
  \sinh\frac{E_{ij}(s)-m-\epsilon_+}{2}}
  {\sinh\frac{E_{ij}(s)}{2}\sinh\frac{E_{ij}(s)-2\epsilon_+}{2}}
\end{equation}
where $\epsilon_\pm\equiv\frac{\epsilon_1\pm\epsilon_2}{2}$, and
\begin{equation}
  E_{ij}(s)=v_i-v_j-\epsilon_1h_i(s)+\epsilon_2(v_j(s)+1)\ .
\end{equation}
The summation is made over $N$ Young diagrams 
$Y_i$ with total number of boxes $k$, and $s$ runs over all boxes of 
the Young diagram $Y_i$. $h_i(s)$ is the distance from $s$ to the 
right end of the Young diagram $Y_i$, and $v_j(s)$ is the distance from 
$s$ to the bottom end of the Young diagram $Y_j$. See \cite{Kim:2011mv} 
for the details. One often calls 
$Z_{\rm inst}\equiv\sum_{k=0}^\infty q^kZ_k$ the instanton partition function.

$Z_{\rm pert}$ is given by \cite{Bullimore:2014awa,Bullimore:2014upa}
\begin{equation}
  Z_{\rm pert}=\prod_{\alpha\in{\bf adj}}
  \left[\frac{\tilde{\Gamma}_3
  (\frac{\alpha(v)+\epsilon_++m}{2\pi i}|
  \frac{\epsilon_1}{2\pi i},\frac{\epsilon_2}{2\pi i})
  \tilde{\Gamma}_3
  (\frac{\alpha(v)+\epsilon_+-m}{2\pi i}|
  \frac{\epsilon_1}{2\pi i},\frac{\epsilon_2}{2\pi i})}
  {\tilde{\Gamma}_3(\frac{\alpha(v)}{2\pi i}|\frac{\epsilon_1}{2\pi i},
  \frac{\epsilon_2}{2\pi i})
  \tilde{\Gamma}_3(\frac{\alpha(v)+2\epsilon_+}{2\pi i}|\frac{\epsilon_1}{2\pi i},
  \frac{\epsilon_2}{2\pi i})}\right]^{\frac{1}{2}}
\end{equation}
where 
$\tilde{\Gamma}_3(z|w_1,w_2)\equiv\Gamma_3(z|1,w_1,w_2)\Gamma_3(1-z|1,-w_1,-w_2)$,
and $\Gamma_N(z|w_1,\cdots,w_N)$ is the Barnes' Gamma 
function. As noted in \cite{Bullimore:2014upa}, $\alpha$ in the
adjoint representation includes Cartans, $\alpha=0$, for which 
`$\Gamma_3(0|\frac{\epsilon_1}{2\pi i},\frac{\epsilon_2}{2\pi i})$' in the 
denominator would diverge. For these $\alpha$, one replaces 
`$\tilde{\Gamma}_3(0|w_1,w_2)$' factors by 
$\tilde{\Gamma}_3^\prime(0|w_1,w_2)\equiv\lim[z\tilde{\Gamma}_3(z|w_1,w_2)]$.
See \cite{Bullimore:2014upa} for more details.
For $t_1\equiv e^{\epsilon_1}<1$, $t_2\equiv e^{\epsilon_2}<1$, 
$Z_{\rm pert}$ can be rewritten as
\begin{equation}\label{Zpert-barnes}
  Z_{\rm pert}=e^{-\mathcal{F}}\prod_{\alpha\in{\bf adj}}\prod_{n_1,n_2\geq 0}
  \left[\frac{(1-e^{\alpha(v)}t_1^{n_1}t_2^{n_2})^\prime
  (1-e^{\alpha(v)}t_1t_2t_1^{n_1}t_2^{n_2})}
  {(1-e^{\alpha(v)+\epsilon_+\pm m}t_1^{n_1}t_2^{n_2})}\right]^{\frac{1}{2}}\ ,
\end{equation}
where prime here again means excluding the zero modes 
at $n_1=n_2=0$ for the Cartans $\alpha=0$. The overall factor $\mathcal{F}$ 
is given for gauge group $G$ by \cite{Bullimore:2014awa}
\begin{eqnarray}
  \hspace*{-1.2cm}\mathcal{F}&=&-\frac{\pi i}{2}\sum_{\alpha\in{\bf adj}}
  \left[\zeta_3(0,\frac{\alpha(v)}{2\pi i}|1,
  \frac{\epsilon_1}{2\pi i},\frac{\epsilon_2}{2\pi i})
  +\zeta_3(0,\frac{\alpha(v)+2\epsilon_+}{2\pi i}|1,
  \frac{\epsilon_1}{2\pi i},\frac{\epsilon_2}{2\pi i})
  -\zeta_3(0,\frac{\alpha(v)+\epsilon_+\pm m}{2\pi i}|1,
  \frac{\epsilon_1}{2\pi i},\frac{\epsilon_2}{2\pi i})\right]\nonumber\\
  \hspace*{-1.2cm}&=&-\frac{\pi i}{2}\sum_{\alpha(v)\in{\bf adj}}
  \frac{\epsilon_+^2-m^2}{2\epsilon_1\epsilon_2}(1-2\alpha(v))
  =\frac{\pi i(m^2-\epsilon_+^2)}{4\epsilon_1\epsilon_2}|G|\ ,
\end{eqnarray}
where $\zeta_3$ is the Barnes' zeta function.
When $t_1,t_2<e^{\alpha(v)}$ for all $\alpha\in{\bf adj}$, 
$Z_{\rm pert}$ is rewritten as
\begin{equation}\label{Zpert}
  Z_{\rm pert}(v,\epsilon_{1,2},m)=e^{-\mathcal{F}}
  PE\left[\frac{1}{2}\frac{\sinh\frac{m+\epsilon_+}{2}
  \sinh\frac{m-\epsilon_+}{2}}{\sinh\frac{\epsilon_1}{2}\sinh\frac{\epsilon_2}{2}}
  \chi_{\bf adj}(e^{v})+\frac{r}{2}\right]\ ,
\end{equation}
where $PE[f(x,y,z,\cdots)]\equiv\exp\left[\sum_{n=1}^\infty\frac{1}{n}
f(nx,ny,nz,\cdots)\right]$, 
$\chi_{\bf adj}\equiv\sum_{\alpha\in{\bf adj}}e^{\alpha(v)}$, 
and $r$ is the rank of gauge group which is $r=N$ for $U(N)$.
The term $\frac{r}{2}$ in $PE$ comes from excluding $r$ fermionic zero modes 
for the Cartans.

One may multiply an alternative perturbative factor 
$\check{Z}_{\rm pert}\equiv e^{-\varepsilon_0}
[Z_{\rm pert}^{U(1)}]^N\hat{Z}_{\rm pert}$ to
$Z_{\rm inst}$, where $[Z_{\rm pert}^{U(1)}]^N$ is the 
perturbative partition function for the $N$ Cartans, 
$\hat{Z}_{\rm pert}$ is defined by
\begin{equation}\label{hat-Zpert}
  \hat{Z}_{\rm pert}=PE\left[\frac{\sinh\frac{m+\epsilon_+}{2}
  \sinh\frac{m-\epsilon_+}{2}}
  {\sinh\frac{\epsilon_1}{2}\sinh\frac{\epsilon_2}{2}}
  \sum_{\alpha>0}e^{-\alpha(v)}\right]\ ,
\end{equation}
and
\begin{equation}
  \varepsilon_0=\frac{m^2-\epsilon_+^2}{2\epsilon_1\epsilon_2}\sum_{\alpha>0}
  (\alpha(v)+\pi i)\ .
\end{equation}
Here all sums are over positive roots $\alpha>0$.
This expression is well defined when all $\alpha(v)$ are positive for positive 
roots and larger than $m,\epsilon_{1,2}$. This expression will be useful when 
studying S-duality from the M-string viewpoint, in section 2.2. 
$Z_{\rm pert}$ and $\check{Z}_{\rm pert}$ are different in subtle ways, 
which shall not affect the studies of prepotential in this paper but has 
implications on the S-duality of $Z$, which we comment on in section 2.2.
(\ref{hat-Zpert}) has a more natural 
interpretation as the Witten index of charged W-bosons in the Coulomb phase 
\cite{Kim:2011mv}. However, as an abstract partition function, $Z_{\rm pert}$ 
is more natural as it is manifestly Weyl-invariant.

It will also be useful to know the simple structures of the Abelian 
partition function, $Z^{U(1)}=Z_{\rm pert}^{U(1)}Z_{\rm inst}^{U(1)}$.
Firstly, the perturbative $U(1)$ partition function can be written as
\begin{equation}\label{Abelian-pert}
  Z_{\rm pert}^{U(1)}=e^{-\frac{\pi i(m^2-\epsilon_+^2)}{4\epsilon_1\epsilon_2}}
  PE\left[\frac{1}{2}\frac{\sinh\frac{m+\epsilon_+}{2}
  \sinh\frac{m-\epsilon_+}{2}}{\sinh\frac{\epsilon_1}{2}\sinh\frac{\epsilon_2}{2}}
  +\frac{1}{2}\right]=e^{-\frac{\pi i(m^2-\epsilon_+^2)}{4\epsilon_1\epsilon_2}}
  PE\left[\frac{1}{2}\frac{\sinh\frac{m+\epsilon_-}{2}
  \sinh\frac{m-\epsilon_-}{2}}{\sinh\frac{\epsilon_1}{2}\sinh\frac{\epsilon_2}{2}}
  \right]
\end{equation}
at $e^{\epsilon_1}<1$, $e^{\epsilon_2}<1$, by following the discussions 
till (\ref{Zpert}) for $N=1$. The instanton part can 
be written as \cite{Iqbal:2008ra}
\begin{equation}\label{Abelian-inst}
  Z_{\rm inst}^{U(1)}=\sum_{k=0}^\infty q^kZ_k=
  PE\left[\frac{\sinh\frac{m+\epsilon_-}{2}
  \sinh\frac{m-\epsilon_-}{2}}{\sinh\frac{\epsilon_1}{2}\sinh\frac{\epsilon_2}{2}}
  \frac{q}{1-q}\right]\ ,
\end{equation}
after summing over all Young diagrams in (\ref{Zk-Young}).

Given $Z=Z_{\rm pert}Z_{\rm inst}$, 
or $\check{Z}=\check{Z}_{\rm pert}Z_{\rm inst}$, one can write this 
partition function as
\begin{equation}\label{single-particle}
  Z=PE\left[\frac{f(\tau,m,\epsilon_{1,2},v)}
  {2\sinh\frac{\epsilon_1}{2}\cdot 2\sinh\frac{\epsilon_2}{2}}\right]
  \equiv\exp\left[\sum_{n=1}^\infty
  \frac{1}{n}\frac{f(n\tau,nm,n\epsilon_{1,2},nv)}
  {2\sinh\frac{n\epsilon_1}{2}\cdot 2\sinh\frac{n\epsilon_2}{2}}\right]\ .
\end{equation}
or a similar expression for $\check{Z}$ using $\check{f}$.
The expression appearing in $PE$ is called the single particle index, 
containing all the information on the BPS bound states. The coefficients 
of $f$ in fugacity expansion are also called Gopakumar-Vafa invariants 
\cite{Gopakumar:1998ii,Gopakumar:1998jq}.  
The factor $\frac{1}{2\sinh\frac{\epsilon_1}{2}\cdot 2\sinh\frac{\epsilon_2}{2}}$
comes from the center-of-mass zero modes of the particle on $\mathbb{R}^4$, 
which would have caused the path integral for $Z$ to diverge at $\epsilon_1=\epsilon_2=0$. So $\epsilon_{1,2}$ also plays
the role of IR regulators. $f(\tau,m,\epsilon_{1,2},v)$ takes into 
account the relative degrees of freedom of the bound state, in which 
$\epsilon_{1,2}$ are just chemical potentials. In particular, 
$\epsilon_{1,2}\rightarrow 0$ limit is smooth in $f$.

In this paper, we shall mostly discuss the limit 
$\epsilon_1,\epsilon_2\rightarrow 0$. In this limit, one finds
\begin{equation}\label{relation-Z-F-pert}
  Z_{\rm pert}\sim\exp\left[-\frac{F_{\rm pert}(v,m)}{\epsilon_1\epsilon_2}\right]
  \ \ ,\ \ Z_{\rm inst}\sim
  \exp\left[-\frac{F_{\rm inst}(\tau,v,m)}{\epsilon_1\epsilon_2}\right]
\end{equation}
from (\ref{single-particle}). $F=F_{\rm pert}+F_{\rm inst}$ is the prepotential.
$F_{\rm inst}$ can be obtained from (\ref{Zk-Young}) after a 
straightforward but tedious calculation. $F_{\rm pert}$ can be obtained 
from (\ref{Zpert}), which is given by
\begin{equation}\label{Fpert}
  F_{\rm pert}(v,m)=\frac{\pi im^2}{4}|G|+\sum_{\alpha\in{\bf adj}}
  \left({\rm Li}_3(e^{-\alpha(v)})-\frac{1}{2}{\rm Li}_3(e^{-(\alpha(v)+m)})
  -\frac{1}{2}{\rm Li}_3(e^{-(\alpha(v)-m)})\right)\ ,
\end{equation}
where ${\rm Li}_s(x)=\sum_{n=1}^\infty\frac{x^n}{n^s}$ for $|x|<1$, and can be 
continued to the complex $x$ plane with a branch cut.
The first term coming from $\mathcal{F}$ will play no role in this paper.
One way of obtaining (\ref{Fpert}) is to first take $v,m$ to be purely imaginary, 
to guarantee convergence of the sum in (\ref{Zpert}), and take the limit 
$\epsilon_{1,2}\rightarrow 0$ to obtain (\ref{Fpert}). Then, (\ref{Fpert}) can 
be analytically continued to complex $v,m$. One may alternatively start from 
$\check{Z}_{\rm pert}$ and obtain its prepotential,
\begin{equation}
  \check{F}_{\rm pert}=
  \frac{m^2}{2}(\pi i|\Delta_+|+\sum_{\alpha>0}\alpha(v))+\sum_{\alpha>0}
  \left(2{\rm Li}_3(e^{-\alpha(v)})-{\rm Li}_3(e^{-\alpha(v)\pm m})\right)
  +rF_{\rm pert}^{U(1)}\ .
\end{equation}
$\Delta_+$ is the set of positive roots. Here, from the identity
\begin{equation}
  {\rm Li}_n(e^{2\pi ix})+(-1)^n(e^{-2\pi ix})
  =-\frac{(2\pi i)^n}{n!}B_n(x)
\end{equation}
for $0<{\rm Re}(x)\leq 1$ and ${\rm Im}(x)<0$, where $B_n(x)$'s are 
Bernoulli polynomials, one finds
\begin{equation}
  {\rm Li}_3(e^x)-{\rm Li}_3(e^{-x})=-\frac{(2\pi i)^3}{6}
  B_3\left(\frac{x}{2\pi i}\right)
\end{equation}
for ${\rm Re}(x)>0$ and $0<{\rm Im}(x)\leq 2\pi$. So for simplicity, 
let us assume ${\rm Re}(\alpha(v))>\pm {\rm Re}(m)$ for all positive roots 
$\alpha$, and also ${\rm Im}(\alpha(v))$ is chosen such that all 
${\rm Im}(\alpha(v)\pm m)$ are within the range $(0,2\pi]$ for positive
roots. Then one finds 
\begin{equation}\label{pert-difference}
  F_{\rm pert}-\check{F}_{\rm pert}=
  -\frac{m^2}{2}\sum_{\alpha>0}\alpha(v)
  -\frac{(2\pi i)^3}{6}\sum_{\alpha>0}
  \left[B_3\left(\frac{\alpha(v)}{2\pi i}\right)
  -\frac{1}{2}B_3\left(\frac{\alpha(v)\pm m}{2\pi i}\right)\right]
  =-\frac{\pi im^2}{2}|\Delta_+|
\end{equation}
where we used $B_3(x)=x^3-\frac{3}{2}x^2+\frac{1}{2}x$. 
So at least in this setting, $F_{\rm pert}$ and $\check{F}_{\rm pert}$ 
differ only by a trivial constant independent of $v$. The last constant 
will play no role in this paper.

It will be helpful to consider the prepotential of the $U(1)$ theory separately.
From (\ref{Abelian-pert}) and (\ref{Abelian-inst}), the prepotential 
$f_{U(1)}=F_{\rm pert}^{U(1)}+F_{\rm inst}^{U(1)}$ for the 
$U(1)$ theory is given by
\begin{equation}\label{Abelian-f}
  f_{U(1)}=\sum_{n=1}^\infty\left(2{\rm Li}_3(q^n)-{\rm Li}_3(e^mq^n)-{\rm Li}_3(e^{-m}q^n)\right)
  +\frac{1}{2}\left(2{\rm Li}_3(1)-{\rm Li}_3(e^m)-{\rm Li}_3(e^{-m})
  \right)+\frac{\pi im^2}{4}\ .
\end{equation}
For studying the S-duality of this prepotential, 
it will be useful to make an expansion of $f_{U(1)}$ in $m$. One first finds 
that the instanton part is given by
\begin{eqnarray}
  \hspace*{-1cm}&&\sum_{n=1}^\infty\left(2{\rm Li}_3(q^n)\!-\!{\rm Li}_3(e^mq^n)\!-\!
  {\rm Li}_3(e^{-m}q^n)\right)=
  -m^2\sum_{n=1}^\infty{\rm Li}_1(q^n)-2\sum_{j=1}^\infty\sum_{n=1}^\infty
  \frac{m^{2j+2}}{(2j+2)!}{\rm Li}_{1-2j}(q^n)\nonumber\\
  &&=m^2\sum_{n=1}^\infty\log(1-q^n)
  -2\sum_{j,n,k=1}^\infty\frac{m^{2j+2}}{(2j+2)!}k^{2j-1}q^{nk}
  =m^2\log\phi(\tau)-2\sum_{j,k=1}^\infty\frac{m^{2j+2}}{(2j+2)!}
  \frac{k^{2j-1}q^k}{1-q^k}\nonumber\\
  &&=m^2\log\phi(\tau)+\sum_{j=1}^\infty\frac{m^{2j+2}}{2j(2j+2)!}
  (E_{2j}(\tau)-1)\ ,
\end{eqnarray}
where $\phi(\tau)=\prod_{n=1}^\infty(1-q^n)=q^{-\frac{1}{24}}\eta(\tau)$
is the Euler function, and we used the identity 
\begin{equation}
  \sum_{k=1}^\infty\frac{k^{2j-1}q^k}{1-q^k}=-\frac{B_{2j}}{4j}(E_{2j}(\tau)-1)
\end{equation}
for the Eistenstein series $E_{2n}(\tau)$. 
$B_{n}$ are the Bernoulli numbers: $B_{1}=\pm\frac{1}{2}$,
$B_{2n+1}=0$,
\begin{equation}
  B_0=1\ ,\ \ B_2=\frac{1}{6}\ ,\ \ B_4=-\frac{1}{30}\ ,\ \ B_6=\frac{1}{42}
  \ ,\ \ B_8=-\frac{1}{30}\ ,
\end{equation}
and so on. The perturbative 
prepotential can be expanded in $m$ by using 
\begin{equation}\label{Li-expand}
  {\rm Li}_n(e^z)=\frac{z^{n-1}}{(n-1)!}\left(H_{n-1}-\log(-z)\right)
  +\sum_{k=0;k\neq n-1}^\infty\frac{\zeta(n-k)}{k!}z^k\ ,
\end{equation}
at small $z$,
with $H_n=\sum_{p=1}^n\frac{1}{p}$. One finds
\begin{equation}
  \frac{1}{2}\left(2{\rm Li}_3(1)-{\rm Li}_3(e^m)-{\rm Li}_3(e^{-m})
  \right)=m^2\left(\frac{1}{2}\log m-\frac{3}{4}+\frac{1}{4}\log(-1)\right)
  +\sum_{j=1}^\infty\frac{B_{2j}m^{2j+2}}{2j(2j+2)!}
\end{equation}
Combining all, one obtains
\begin{equation}\label{anom-series}
  f_{U(1)}=m^2\left(\frac{1}{2}\log m-\frac{3}{4}+\frac{\pi i}{2}
  +\log\phi(\tau)\right)
  +\sum_{n=1}^\infty\frac{m^{2n+2}B_{2n}}{2n\cdot(2n+2)!}E_{2n}(\tau)\ .
\end{equation}
This will be useful later for understanding $Nf_{U(1)}$, 
as a part of the $U(N)$ prepotential.

One can understand the chemical potentials
from the viewpoint of the 4d effective action in the Coulomb branch. 
The dimensionless variables
$m,\epsilon_{1,2}$, $v$ take the form of
\begin{equation}
  m=RM\ ,\ \ \epsilon_{1,2}=R\varepsilon_{1,2}\ ,\ \ v=Ra\ ,
\end{equation}
where $R$ is the radius of the temporal circle of $\mathbb{R}^4\times S^1$.
$M$ is the mass deformation parameter of the 4d $\mathcal{N}=2^\ast$ Yang-Mills
theory, or the 5d $\mathcal{N}=1^\ast$ theory. 
(More precisely, $M$ is $2\pi$ times the mass.) 
$\varepsilon_{1,2}$ are the Omega 
deformation parameters which have dimensions of mass.
$a$ is the Coulomb VEV of the scalar field $\Phi^5$. $\tau$ is 
identified as
\begin{equation}
  \tau=i\frac{R}{R^\prime}\ ,
\end{equation}
where $R^\prime$ is the radius of the sixth circle. This is the 
inverse gauge coupling in 4d. $\tau$ can be complexified with a real part, 
given by the RR 1-form holonomy of type IIA theory on $S^1$. 

The 4 dimensional limit of the partition function is obtained by 
taking $R\rightarrow 0$ with fixed $\tau,M,\varepsilon_{1,2},a$.
From (\ref{Zk-Young}), one finds that all $\sinh$ functions of 
$v,\epsilon_{1,2},m$ are replaced by linear functions of $a,\varepsilon_{1,2},M$, 
and the $R$ dependences cancel between numerator and denominator. As a result, 
the 4d limit $Z_{k}^{\rm 4d}$ of the instanton partition function is given by a 
rational function of $M,\varepsilon_{1,2},a$ of degree $0$. This makes 
$Z^{\rm 4d}_{\rm inst}$
and $F_{\rm inst}^{\rm 4d}$ to enjoy a simple scaling property, 
\begin{equation}\label{4d-scaling}
  Z^{\rm 4d}_{\rm inst}(\tau,\lambda M,\lambda\varepsilon_{1,2},\lambda a)
  =Z_{\rm inst}^{\rm 4d}(\tau,M,\varepsilon_{1,2},a)\ \ ,\ \ \ 
  F^{\rm 4d}_{\rm inst}(\tau,\lambda M,\lambda a)
  =\lambda^{2}F_{\rm inst}^{\rm 4d}(\tau,M,a)\ .
\end{equation}
This will be used in section 2.1 to provide two interpretations 
of the 4d S-duality, and extend one version to 6d.
As for the perturbative part $F_{\rm pert}$, one can 
use (\ref{Li-expand}) to obtain
$F^{\rm 4d}_{\rm pert}\equiv\lim_{R\rightarrow 0}F_{\rm pert}$.
One finds
\begin{equation}\label{4d-Fpert}
  \hspace*{-.5cm}F^{\rm 4d}_{\rm pert}=\!\sum_{\alpha\in{\bf adj}}\left[
  M^2\left(\log R\!-\!\frac{3}{4}\right)\!-\!\frac{\alpha(a)^2}{2}\log \alpha(a)
  \!+\!\frac{(\alpha(a)\!+\!M)^2}{4}\log(\alpha(a)\!+\!M)
  \!+\!\frac{(\alpha(a)\!-\!M)^2}{4}\log(\alpha(a)\!-\!M)\right]
\end{equation}
where the first term independent of the Coulomb VEV is unphysical in the 
Seiberg-Witten theory. The perturbative prepotential satisfies the 
following pseudo-scaling property,
\begin{equation}
  F^{\rm 4d}_{\rm pert}(\lambda M,\lambda a)=\lambda^2
  \left(F^{\rm 4d}_{\rm pert}(M,a)+|G|\frac{M^2}{2}\log\lambda\right)\ ,
\end{equation}
which is homogenous and degree $2$ up to a Coulomb VEV independent shift.

$Z_{\rm inst}$ or $F_{\rm inst}$ are only known as $q$ expansion when $q\ll 1$,
or $\tau\rightarrow i\infty$. This is useful when the `temperature'
is much smaller than the Kaluz-Klein scale $\frac{1}{R^\prime}$,
when the KK modes are `heavy.' However,
to study 6d SCFT, it is more interesting to explore the regime
$q\rightarrow 1$, or $\tau\rightarrow i0^+$, in which case the
circle effectively decompactifies. The two regimes are weakly coupled and 
strongly coupled regimes, respectively.
So if there is S-duality for the partition function on
$\mathbb{R}^4\times T^2$, it will be helpful to study the interesting
decompactifying regime from the well-understood region
$\tau\rightarrow i\infty$. Developing the S-duality of the prepotential $F$
is the goal of this section. (In section 2.2, we also comment on
the S-duality of the full partition function.)

\subsection{S-duality and its anomaly}

Following \cite{Billo':2015ria}, we review the basic set up for studying 
the S-duality of 4 dimensional prepotential, and extend it to the 6d theory 
on $T^2$.

The prepotential $F$ of general 4d $\mathcal{N}=2$ gauge theory
determines the effective action in the Coulomb branch.
The magnetic dual description uses the dual Coulomb VEV $a_D(a)$ and the dual 
prepotential $F_D(a_D)$, defined by the following Legendre transformation,
\begin{equation}
  a_D=\frac{1}{2\pi i}\frac{\partial F}{\partial a}\ ,\ \
  F_D(a_D)=\mathcal{L}[F](a)\equiv F(a)-2\pi ia_Da=F-a\frac{\partial F}{\partial a}\ .
\end{equation}
For theories with higher rank $r>1$, $a$ has 
many components, $a_i$ with $i=1,\cdots,r$. Expressions like 
$a\frac{\partial}{\partial a}$ should be understood with contracted 
$i$ indices, i.e. $a\frac{\partial}{\partial a}\rightarrow 
\sum_{i=1}^ra_i\frac{\partial}{\partial a_i}$, whose sum structures will 
not be explicitly shown to make the notations simpler.
For generic $\mathcal{N}=2$ theories, $F,F_D$ depend on other parameters like
hypermultiplet masses and the coupling constant (or the dynamically generated
scale $\Lambda$ instead of the coupling).

For 4d $\mathcal{N}=2^\ast$ theory, the prepotential 
$F^{\rm 4d}$ (to be distinguished with the 6d prepotential $F$ which we shall 
consider later) depends on the microscopic coupling constant $\tau$ and the adjoint
hypermultiplet mass $M$. The prepotential can be divided into the classical,
perturbative, and instanton contributions,
\begin{equation}
  F^{\rm 4d}=F_{\rm cl}(\tau,a)+F_{\rm pert}^{\rm 4d}(a,M)+
  F_{\rm inst}^{\rm 4d}(\tau,a,M)\equiv F_{\rm cl}(\tau,a)+f^{\rm 4d}(\tau,a,M)
\end{equation}
where $F_{\rm cl}(\tau,a)=\pi i\tau a^2$, and $F^{\rm 4d}_{\rm pert}$.
$f^{\rm 4d}\equiv F^{\rm 4d}_{\rm pert}+F^{\rm 4d}_{\rm inst}$ is
the quantum  prepotential. To study self S-dual theories,
it is convenient to define $F_D^{\rm 4d}$ as a
function of the dual coupling $\tau_D=-\frac{1}{\tau}$. For the
4d $\mathcal{N}=2^\ast$ theory, $F_D^{\rm 4d}$ is defined by
\begin{equation}
  F_D^{\rm 4d}(\tau_D,a_D,M)=\mathcal{L}[F^{\rm 4d}](\tau,a,M)
  =F^{\rm 4d}(\tau,a,M)-a\frac{\partial F^{\rm 4d}}{\partial a}(\tau,a,M)\ .
\end{equation}
Then, self S-duality exists if $F_D^{\rm 4d}$ and $F^{\rm 4d}$ are 
same function, $F^{\rm 4d}_D(\tau,a,M)=F^{\rm 4d}(\tau,a,M)$.
This S-duality has been tested in detail in \cite{Billo':2015ria}.
More precisely, it was found that
\begin{equation}
  F^{\rm 4d}(\tau,a,M)=F^{\rm 4d}_{\textrm{S-dual}}(\tau,a,M)
  +F_{\rm anom}^{\rm 4d}(\tau,M)\ ,
\end{equation}
where $F_{\textrm{S-dual}}^{\rm 4d}$ satisfies
\begin{equation}\label{4d-S-dual}
  F^{\rm 4d}_{\textrm{S-dual}}(\tau_D,a_D,M)=F^{\rm 4d}_{\textrm{S-dual}}(\tau,a,M)
  -a\frac{\partial F^{\rm 4d}_{\textrm{S-dual}}}{\partial a}(\tau,a,M)\ ,
\end{equation}
and $F_{\rm anom}^{\rm 4d}$ is an anomalous part of S-duality, 
depending on $\tau,M$ but is independent of the Coulomb VEV $a$ 
\cite{Billo':2015ria}. Since the Coulomb branch effective action is
obtained by taking $a$ derivatives of $F^{\rm 4d}$, $F^{\rm 4d}$ and 
$F^{\rm 4d}_{\textrm{S-dual}}$ are identical in the Seiberg-Witten theory. 
This establishes the
S-duality of the 4d $\mathcal{N}=2^\ast$ theory in the Coulomb branch
effective action.

Let us rephrase the 4d S-duality in
a way that is suitable for 6d extension. $F^{\rm 4d}_{\rm inst}$ satisfies 
the scaling property (\ref{4d-scaling}).
Combining the perturbative part, one finds
\begin{equation}\label{homogeneous}
  F^{\rm 4d}(\tau,\lambda a,\lambda M)=\lambda^2
  \left(F^{\rm 4d}(\tau,a,M)+|G|\frac{M^2}{2}\log \lambda\right)\ .
\end{equation}
Applying this to $F^{\rm 4d}(\tau_D,a_D,m)$, one obtains
\begin{equation}
  F^{\rm 4d}(\tau_D,a_D/\tau, M/\tau)=\tau^{-2}F^{\rm 4d}(\tau_D,a_D,M)
  -\frac{M^2}{2\tau^2}\log\tau\ .
\end{equation}
So the left hand side of (\ref{4d-S-dual}) can be written as
\begin{equation}\label{F-S-dual-scaling}
  F_{\textrm{S-dual}}^{\rm 4d}(\tau_D,a_D,M)=
  \tau^2 F^{\rm 4d}_{\textrm{S-dual}}(\tau_D,a_D/\tau,M/\tau)
  +\frac{|G|M^2}{2}\log\tau+\tau^2 F_{\rm anom}^{\rm 4d}(\tau_D,M/\tau)
  -F^{\rm 4d}_{\rm anom}(\tau_D,M)\ .
\end{equation}
Let us consider the structure of $F^{\rm 4d}_{\rm anom}$.
Since the prepotential has mass dimension $2$, one may think that 
its $M$ dependence is simply $M^2$. However, the perturbative part 
(\ref{4d-Fpert}) shows that there is a term $\frac{rM^2}{2}\log M$ in 
$F^{\rm 4d}$ which scales in an odd manner. In the computational framework of 
\cite{Billo':2015ria}, which we shall explain below in our 6d version, 
$F_{\textrm{S-dual}}^{\rm 4d}$ is by construction taken to be a series expansion 
in $M^2$. This means that the odd term $\frac{rM^2}{2}\log M$ should have been 
put in $F^{\rm 4d}_{\rm anom}$. Therefore, had one been doing the calculation 
of \cite{Billo':2015ria} using (\ref{4d-Fpert}) as the perturbative part, 
one would have found that $F^{\rm 4d}_{\rm anom}=\frac{rM^2}{2}\log M + M^2(\cdots)$, 
where $(\cdots)$ only depends on $\tau$. Using this structure, 
(\ref{F-S-dual-scaling}) can be rewritten as
\begin{equation}
  F_{\textrm{S-dual}}^{\rm 4d}(\tau_D,a_D,M)=
  \tau^2 F^{\rm 4d}_{\textrm{S-dual}}(\tau_D,a_D/\tau,M/\tau)
  +(|G|-r)\frac{M^2}{2}\log\tau\ .
\end{equation}
So defining 
\begin{equation}\label{f-S-dual-redefine}
  \tilde{F}^{\rm 4d}_{\textrm{S-dual}}(\tau,a,M)
  =F^{\rm 4d}_{\rm S-dual}(\tau,a,M)-\frac{|G|-r}{2}M^2\log M\ ,
\end{equation}
one finds that $\tilde{F}^{\rm 4d}_{\textrm{S-dual}}$ satisfies
\begin{equation}\label{4d-S-dual-2}
  \tau^2 \tilde{F}^{\rm 4d}_{\textrm{S-dual}}(\tau_D,a_D/\tau,M/\tau)=
  \tilde{F}^{\rm 4d}_{\textrm{S-dual}}(\tau,a,M)
  -a\frac{\partial \tilde{F}^{\rm 4d}_{\textrm{S-dual}}}{\partial a}(\tau,a,M)\ ,
\end{equation}
instead of (\ref{4d-S-dual}). To summarize, by trivially redefining 
$F_{\textrm{S-dual}}^{\rm 4d}$ and $F_{\rm anom}^{\rm 4d}$ by the last term 
of (\ref{f-S-dual-redefine}), one can reformulate the standard S-duality 
(\ref{4d-S-dual}) as (\ref{4d-S-dual-2}). Only (\ref{4d-S-dual-2}) will naturally
generalize to the S-duality on $\mathbb{R}^4\times T^2$.

Now we seek for the S-duality of the 6d prepotential.
Note that in 4d, (\ref{4d-S-dual}) and (\ref{4d-S-dual-2}) are equivalent
by making a minor redefinition of $F_{\rm anom}^{\rm 4d}$, using (\ref{homogeneous}).
In 6d, a property like
(\ref{homogeneous}) does not hold. Before making a quantitative study of the 6d 
S-duality, we first explain that (\ref{4d-S-dual-2}) 
is more natural in 6d. To discuss the 6d prepotential,
it is convenient to work with the dimensionless
parameters $v,m,\epsilon_{1,2}$.

Firstly, in the 6d theory compactified on $T^2$, the complex mass parameter
$m$ is simply the holonomy of the background gauge field for $SU(2)_L$ global
symmetry, along the two sides of $T^2$. Then after making an S-duality of the
torus, exchanging two sides of $T^2$, one naturally expects $m_D=\frac{m}{\tau}$.
Let us briefly review this by taking a rectangular torus, for simplicity.
In this case, the complex structure $\tau$ of the torus is
purely imaginary. $\tau$ is related to the two radii of $T^2$ by
\begin{equation}
  \tau=i\frac{R}{R^\prime}\ ,
\end{equation}
where $R^\prime$ is the radius of the circle which compactifies the 6d theory to
5d SYM, and $R$ is the radius of another circle which compactifies
the 5d theory to 4d. The S-duality transformation exchanges 
$R\leftrightarrow R^\prime$. So the dual complex structure is $\tau_D=i\frac{R^\prime}{R}=-\frac{1}{\tau}$.
More precisely, S-duality rotates the torus by $90$ degrees on a plane.
It also transforms the two $SU(2)_L$ holonomies along the two circles.
Let ${\rm Re}(M)$ be the holonomy on the circle with radius
$R$, and ${\rm Im}(M)$ that on the circle with radius $R^\prime$.
Under S-duality, one finds
${\rm Re}(M_D)={\rm Im}(M)$, ${\rm Im}(M_D)=-{\rm Re}(M)$. So one finds
$M_D=-iM$. In $F$, $M$ appears in the dimensionless combination
$m\equiv RM$, which transforms as
\begin{equation}
  m_D=R^\prime M_D=-iR^\prime M=-i\frac{R^\prime}{R}m=\frac{m}{\tau}\ .
\end{equation}
The final result holds for complex $\tau$.
Similar property holds for $\epsilon_{1,2}\equiv R\varepsilon_{1,2}$,
i.e. $\epsilon_{1,2}^D=\frac{\epsilon_{1,2}}{\tau}$. This makes 
the appearance of $\frac{M}{\tau}$ to be more natural on the 
left hand side of (\ref{4d-S-dual-2}).

Secondly, let us discuss how $a$ should transform. In 4d, we already stated that
\begin{equation}\label{aD-conventional}
  a_D=\tau a+\frac{1}{2\pi i}\frac{\partial f}{\partial a}
\end{equation}
naturally appears on the left hand side of (\ref{4d-S-dual}).
For simplicity, let us discuss these variables in the
limit of large Coulomb VEV, $v\equiv Ra \gg 1$, $a\gg m$. 
The second term can
be ignored in this limit, yielding the semi-classical result $a_D=\tau a$. 
In this limit, we shall discuss what is the natural S-dual variable using the
Abelian 6d $(2,0)$ theory. In 4d, $a_D=\tau a$ is a natural
aspect of S-duality being electromagnetic duality. Also, it makes sense
to multiply $a$ by a complex number $\tau$, since $a$ is a complex variable
living on a plane. However, in 6d
CFT on $T^2$, $a$ lives on a cylinder. The real part of $a$ is the 
VEV of the real scalar in the 6d self-dual
tensor multiplet, which is noncompact. On the other hand, the imaginary
part of $a$ comes from the holonomy of the 2-form tensor field $B$ on $T^2$,
implying that it is a periodic variable. So it does not good make sense to rotate
$a$ living on a cylinder by complex $\tau$.
More precisely, the 6d scalar
$\phi$ and the 5d scalar $a$ are related by $a\sim R^\prime\phi$. So one finds
\begin{equation}
  a\sim R^\prime(\phi+iB_{12})\ ,
\end{equation}
where $1$ and $2$ denote two directions of $T^2$.
Thus, $v=Ra\sim RR^\prime(\phi+iB_{12})$ is invariant
under $R\leftrightarrow R^\prime$, meaning that 
it makes more sense to set $v_D\approx v$ in the limit $v\gg 1$. 
Using the dimensionful variables,
This requires one to use $\frac{a_D}{\tau}\approx a$ as the dual variable, 
instead of $a_D\approx \tau a$. This does not rotate the
variable $a$ by a complex number, so makes better sense in 6d.
Incidently, we have already found the alternative (but equivalent) 
statement (\ref{4d-S-dual-2}) of S-duality which uses $\frac{a_D}{\tau}$ 
as the dual variable, instead of $a_D$. Note that the usage 
of $\frac{a_D}{\tau}=a+\frac{1}{2\pi i\tau}\frac{\partial f^{\rm 4d}}{\partial a}$
is valid even beyond the semi-classical limit $a\gg M$.
Thus, in the 6d uplift, it is natural and consistent to regard 
$v_D\equiv \frac{Ra_D}{\tau}
=v+\frac{1}{2\pi i\tau}\frac{\partial f}{\partial v}$.
as the dual variable.\footnote{Here, one may
wonder that $f$ appearing on the right hand side should have been 
$R^2 f$. However, we shall define the prepotential as the
coefficient of the dimensionless $\frac{1}{\epsilon_1\epsilon_2}$,
$-\log Z\sim\frac{f}{\epsilon_1\epsilon_2}$, rather than
$\frac{f^{\rm 4d}}{\varepsilon_1\varepsilon_2}$ that is conventional in 
the Seiberg-Witten theory, making $f$ dimensionless. Namely, 
$f_{\rm ours}$ in 6d is related to the conventionally normalized 
prepotential by $f_{\rm ours}=R^2f_{\rm conventional}$.}

So it appears natural to seek for a 6d generalization of (\ref{4d-S-dual-2}) 
rather than (\ref{4d-S-dual}). This is what we shall
establish in the rest of this section. Namely, 
we shall find that the 6d prepotential is divided into two,
\begin{equation}
  F=F_{\textrm{S-dual}}(\tau,v,m)+F_{\rm anom}(\tau,m)
\end{equation}
where $v=Ra$, $m=RM$, and $F_{\rm anom}$ is independent of the Coulomb VEV.
$F_{\textrm{S-dual}}$ satisfies
\begin{equation}
  \tau^2F_{\textrm{S-dual}}\left(\tau_D=-\frac{1}{\tau},v_D=v+\frac{1}{2\pi i\tau}
  \frac{\partial f}{\partial v},\frac{m}{\tau}\right)
  =F_{\textrm{S-dual}}(\tau,v,m)-v\frac{\partial F_{\textrm{S-dual}}}{\partial v}
  (\tau,v,m)\ .
\end{equation}
We have some freedom 
to choose $F_{\rm anom}$, by adding/subtracting $v$ independent S-dual 
expressions to $F_{\rm anom}$, $F_{\textrm{S-dual}}$.
We shall explain that one can choose $F_{\rm anom}$ as 
\begin{equation}\label{S-dual-anomaly}
  F_{\rm anom}=Nf_{U(1)}(\tau,m)+\frac{N^3-N}{288}m^4E_2(\tau)
\end{equation}
where $f_{U(1)}$ is the $U(1)$ prepotential (\ref{Abelian-f}).
The first term $Nf_{U(1)}$ comes from the $N$ 6d Abelian tensor 
multiplets in $U(1)^N$, which has their own S-duality anomaly. The second term 
of $F_{\rm anom}$ is one of the key findings of this paper, which comes from 
the charged part of the partition function.
After replacing $m=MR$, and multiplying $\frac{1}{R^2}$ to the above
$F_{\rm anom}$ to get to the conventionally normalized prepotential, 
one can take the 4d limit of $F_{\rm anom}$. The second term proportional to 
$N^3-N$ vanishes in the 4d limit $R\rightarrow 0$, as it is proportional 
to $M^4R^2$.

With the motivations and results given, we now properly set up 
the calculation and show the claims made above. As in 4d, we
decompose the 6d prepotential as
\begin{equation}
  F(\tau,v,m)=F_{\rm cl}+F_{\rm pert}+F_{\rm inst}\equiv F_{\rm cl}+f\ ,
\end{equation}
where $F_{\rm cl}\equiv \pi i\tau v^2$. The prepotential is S-dual
if it satisfies
\begin{equation}\label{S-dual-5d}
  \tau^2 F\left(\tau_D=-\frac{1}{\tau},v_D=v+\frac{1}{2\pi i\tau}
  \frac{\partial f}{\partial v},m_D=\frac{m}{\tau}\right)=
  F(\tau,v,m)-v\frac{\partial F}{\partial v}(\tau,v,m)\ .
\end{equation}
We first study the structures of this equation, before showing 
that it is satisfied by our $F_{\textrm{S-dual}}$.
Firstly, replacing $F$ by $F_{\rm cl}$, one can check
that S-duality trivially holds at the classical level:
\begin{equation}\label{S-dual-classical}
  \tau^2F_{\rm cl}(\tau_D,v_D)=\tau^2\left[-\frac{\pi i}{\tau}v^2\right]
  =-\pi i\tau v^2
  =F_{\rm cl}(\tau,v)-v\frac{\partial F_{\rm cl}}{\partial v}(\tau,v)\ ,
\end{equation}
where $v_D$ is replaced by its classical value
$v_D=v$ (formally at $f=0$).
Now we subtract (\ref{S-dual-5d}) by (\ref{S-dual-classical})
to find the following condition for the quantum prepotential $f$:
\begin{equation}\label{S-dual-requirement}
  \tau^2f\left(-\frac{1}{\tau},\ v+\frac{1}{2\pi i\tau}\frac{\partial f}{\partial v},
  \ \frac{m}{\tau}\right)=f(\tau,v,m)+\frac{1}{4\pi i\tau}\left(\frac{\partial f}
  {\partial v}(\tau,v,m)\right)^2\ .
\end{equation}
We are going to study the last equation.
Note again that the effective action in the Coulomb branch only contains
$v$ derivatives of $F$, or $f$.
Thus, in Seiberg-Witten theory, $f$ is ambiguous by addition of $v$
independent functions, possibly depending on $\tau$ and $m$. However, the
S-duality requirement (\ref{S-dual-requirement}) is sensitive to the value of $f$,
including the $v$ independent part. So when one tries to establish the S-duality
of the Coulomb branch effective action, one should have in mind that one may have
to add suitable Coulomb VEV independent terms to $f$ computed microscopically 
from $Z$.

Following \cite{Billo':2015ria}, we shall establish the
S-duality (\ref{S-dual-requirement}) and its anomaly (\ref{S-dual-anomaly})
by expanding $f$ in the mass $m$ when it is small enough. We shall still
get an exact statement (\ref{S-dual-anomaly}), which we check for
certain orders in $m$. One should however have in mind that the exact statement
(\ref{S-dual-anomaly}) may be valid only within a finite region of $m,v$
in the complex planes. In section 2.2, we shed more lights on
the exactness of (\ref{S-dual-anomaly}), by making an M-string expansion \cite{Haghighat:2013gba}.

As studied in the 4d limit \cite{Billo':2015ria}, there is a natural way of
achieving the S-duality requirement (\ref{S-dual-requirement}). This is to require
that $f$ is expanded in quasi-modular forms of suitable weights. To precisely explain
its meaning, we first expand $f$ in $m$ as
\begin{equation}\label{m-series}
  f(\tau,v,m)=\sum_{n=1}^\infty m^{2n}f_n(\tau,v)\ .
\end{equation}
This series makes sense as follows. Firstly,
the $m\rightarrow 0$ limit exhibits enhanced maximal supersymmetry. 
So at $m=0$, the classical prepotential $F_{\rm cl}=\pi i\tau v^2$ acquires no 
quantum corrections, meaning that $f$ vanishes at $m=0$.
Also, the prepotential is an even function of $m$,
which restrict the expansion as above.\footnote{Strictly speaking, there is 
a term $\frac{rm^2}{2}\log m$ in the perturbative part, which is easiest
to see from the 4d limit (\ref{4d-Fpert}). However, we shall expand 
$f_{\textrm{S-dual}}$ as (\ref{m-series}), while the term $\frac{rm^2}{2}\log m$
is moved to $F_{\rm anom}$.}
Then, following \cite{Billo':2015ria}, we require that $f_n$ is a quasi-modular form
of weight $2n-2$, which means the following. Quasi-modular forms are polynomials of
the first three Eisenstein series $E_2$, $E_4$, $E_6$, where each series has
weight $2,4,6$ respectively under S-duality in the following sense:
\begin{equation}
  E_2(-1/\tau)=\tau^2\left(E_2+\frac{6}{\pi i\tau}\right)\ ,\ \
  E_4(-1/\tau)=\tau^4E_4(\tau)\ ,\ \ E_6(-1/\tau)=\tau^6E_6(\tau)\ .
\end{equation}
More concretely, they are given by
\begin{equation}
  E_2=1-24\sum_{n=1}^\infty\frac{nq^n}{1-q^n}\ ,\ \
  E_4(\tau)=1+240\sum_{n=1}^\infty\frac{n^3q^n}{1-q^n}\ ,\ \
  E_6=1-504\sum_{n=1}^\infty\frac{n^5q^n}{1-q^n}\ .
\end{equation}
Higher Eisenstein series $E_{2n}$ are polynomials of $E_4,E_6$
with weight $2n$. To study the quasi-modular property,
it is helpful to decompose their dependence on $\tau$ into the dependence
through $E_2$ and the dependence through $E_4,E_6$. We thus write $f_n(\tau,v,E_2(\tau))$, where the $\tau$
dependence through $E_2$ is explicitly shown. A weight $2n-2$ quasi-modular 
form $f_n$ satisfies
\begin{equation}
  f_n(-1/\tau,v,E_2(-1/\tau))=\tau^{2n-2}f_n(\tau,v,E_2(\tau)+\delta)\ ,
\end{equation}
where $\delta\equiv\frac{6}{\pi i\tau}$. In terms of $f$,
this is equivalent to
\begin{equation}\label{quasi-modular}
  \tau^2f\left(-\frac{1}{\tau},v,\frac{m}{\tau},E_2(-1/\tau)\right)
  =f(\tau,v,m,E_2(\tau)+\delta)\ .
\end{equation}
We now investigate how quasi-modularity is related to the S-duality
(\ref{S-dual-requirement}). One can make (\ref{quasi-modular}) to be
equivalent to (\ref{S-dual-requirement}) by specifying the $E_2$ dependence of $f$,
which we now turn to.

Let us first try to find the desired $E_2$ dependence, by requiring both
(\ref{S-dual-requirement}) and (\ref{quasi-modular}).
By applying (\ref{quasi-modular}) to
$f(-\frac{1}{\tau},v_D,\frac{m}{\tau},E_2(-1/\tau))$, one obtains
\begin{equation}\label{modular}
  \tau^2f\left(-\frac{1}{\tau},v+\frac{\delta}{12}\frac{\partial f}{\partial v},
  \frac{m}{\tau},E_2(-1/\tau)\right)=
  f\left(\tau,v+\frac{\delta}{12}\frac{\partial f}{\partial v},m,E_2(\tau)+\delta\right)\ ,
\end{equation}
where again recall that $\delta\equiv\frac{6}{\pi i\tau}$.
Combining this with (\ref{S-dual-requirement}), one obtains
\begin{equation}\label{S-dual-v2}
  f\left(\tau,v+\frac{\delta}{12}\frac{\partial f}{\partial v},m,E_2(\tau)+\delta\right)
  =f(\tau,v,m,E_2(\tau))+\frac{\delta}{24}
  \left(\frac{\partial f}{\partial v}(\tau,v,m,E_2(\tau))\right)^2\ .
\end{equation}
We want to make this equation to hold, by specifying a particular
$E_2$ dependence of $f$. \cite{Billo':2015ria} showed that 
the desired $E_2$ dependence is
\begin{equation}\label{modular-anomaly}
  \frac{\partial f}{\partial E_2}
  =-\frac{1}{24}\left(\frac{\partial f}{\partial v}\right)^2\ .
\end{equation}
For the sake of completeness, we repeat the logics
presented in \cite{Billo':2015ria} and expand it to make a proof.
In fact, we shall make a stronger claim than needed. Namely,
we need to find the $E_2$ dependence of $f$ which guarantees (\ref{S-dual-v2})
only at $\delta=\frac{6}{\pi i\tau}$. However, we shall show that
(\ref{modular-anomaly}) guarantees (\ref{S-dual-v2})
for arbitrary independent parameter $\delta$, and then set
$\delta=\frac{6}{\pi i\tau}$ later.

As a warm-up, we follow \cite{Billo':2015ria} to make a series expansion of 
the left hand side of (\ref{S-dual-v2}) in small $\delta$, to see how 
(\ref{modular-anomaly}) guarantees (\ref{S-dual-v2}) at low orders. One 
finds that
\begin{equation}\label{zeroth-first}
  ({\rm LHS})=f+\frac{\delta}{12}\left(\frac{\partial f}{\partial v}\right)^2
  +\delta\frac{\partial f}{\partial E_2}+\mathcal{O}(\delta^2)\ .
\end{equation}
So at $\delta^0$ and $\delta^1$ orders, one finds
that it agrees with the right hand side if
(\ref{modular-anomaly}) is met.

Now assuming (\ref{modular-anomaly}), we consider whether (\ref{S-dual-v2})
is satisfied in full generality. To this end, we take $\delta$ derivative
of both sides of (\ref{S-dual-v2}), at fixed $\tau,v,E_2$,
\begin{equation}\label{derivative-S-dual}
  \frac{\partial}{\partial\delta}({\rm LHS})=
  \frac{1}{12}\frac{\partial f}{\partial v}\frac{\partial\tilde{f}}{\partial\tilde{v}}
  +\frac{\partial\tilde{f}}{\partial\tilde{E}_2}=
  \frac{1}{12}\frac{\partial f}{\partial v}\frac{\partial\tilde{f}}{\partial\tilde{v}}
  -\frac{1}{24}\left(\frac{\partial\tilde{f}}{\partial\tilde{v}}\right)^2\ ,\ \ \
  \frac{\partial}{\partial\delta}({\rm RHS})
  =\frac{1}{24}\left(\frac{\partial f}{\partial v}\right)^2
\end{equation}
where for simplicity, we defined
\begin{equation}
  \tilde{v}=v+\frac{\delta}{12}\frac{\partial f}{\partial v}\ ,\ \
  \tilde{E}_2=E_2+\delta\ ,\ \ \tilde{f}=f(\tau,\tilde{v},\tilde{E}_2)\ .
\end{equation}
Note that at the second step of the first equation in (\ref{derivative-S-dual}),
we used $\frac{\partial\tilde{f}}{\partial\tilde{E}_2}
=-\frac{1}{24}\left(\frac{\partial\tilde{f}}{\partial\tilde{v}}\right)^2$
which follows from (\ref{modular-anomaly}). If (\ref{S-dual-v2}) holds
for general $\delta$, its first derivative would yield 
\begin{equation}\label{derivative-S-dual-v2}
  \frac{\partial\tilde{f}}{\partial\tilde{v}}=\frac{\partial f}{\partial v}\ ,
\end{equation}
which one can show by using (\ref{derivative-S-dual}).
On the other hand, (\ref{derivative-S-dual-v2}) together with the
$\mathcal{O}(\delta^0)$ component of (\ref{S-dual-v2}) is equivalent to
(\ref{S-dual-v2}), since the $\mathcal{O}(\delta^0)$ component is the only
information lost by taking $\delta$ derivative. However, we have already
shown around (\ref{zeroth-first}) that the $\mathcal{O}(\delta^0)$
component of (\ref{S-dual-v2})
is satisfied. Therefore, showing (\ref{derivative-S-dual-v2}) will be
equivalent to showing (\ref{S-dual-v2}).
So will show (\ref{derivative-S-dual-v2}) by assuming (\ref{modular-anomaly}).
We take $\delta$ derivative of
$\frac{\partial\tilde{f}}{\partial\tilde{v}}-\frac{\partial f}{\partial v}$
at fixed $v,E_2$. Again using (\ref{modular-anomaly}), one obtains
\begin{equation}\label{2nd-derivative}
  \frac{\partial}{\partial\delta}\left(
  \frac{\partial\tilde{f}}{\partial\tilde{v}}-\frac{\partial f}{\partial v}\right)
  =-\frac{1}{12}\frac{\partial^2\tilde{f}}{\partial\tilde{v}^2}\left(
  \frac{\partial\tilde{f}}{\partial\tilde{v}}-\frac{\partial f}{\partial v}\right)\ .
\end{equation}
So if $\frac{\partial\tilde{f}}{\partial\tilde{v}}-\frac{\partial f}{\partial v}$ is
zero at a particular value of $\delta$, (\ref{2nd-derivative}) guarantees
that it is zero at different values of $\delta$. Since we already checked around
(\ref{zeroth-first}) that (\ref{S-dual-v2}) is true up to $\mathcal{O}(\delta^1)$,
we have shown that (\ref{derivative-S-dual-v2}) holds at $\mathcal{O}(\delta^0)$,
or that $\frac{\partial\tilde{f}}{\partial\tilde{v}}-\frac{\partial f}{\partial v}=0$
at $\delta=0$. This establishes that (\ref{modular-anomaly}) implies (\ref{derivative-S-dual-v2}), and in turn that (\ref{modular-anomaly}) implies
(\ref{S-dual-v2}). Finally, we insert $\delta=\frac{6}{\pi i\tau}$.

To summarize till here, (\ref{S-dual-v2}) holds
if $f$ satisfies (\ref{modular-anomaly}).
But (\ref{S-dual-v2}) and (\ref{quasi-modular}) implies the S-duality
relation (\ref{S-dual-requirement}). Therefore, S-duality requirement
(\ref{S-dual-requirement}) is satisfied if $f$ satisfies
the quasi-modular property (\ref{quasi-modular}) and
the modular anomaly equation (\ref{modular-anomaly}). In the rest of this
subsection, we shall discuss the last two equations.

Following and extending \cite{Billo':2015ria}, we show that the
prepotential $f$ obeys the two
properties (\ref{quasi-modular}), (\ref{modular-anomaly}), up to an anomalous
part which is independent of the Coulomb VEV $v$. Again following 
\cite{Billo':2015ria}, our strategy is to first find a prepotential 
$f_{\textrm{S-dual}}$
in a series of $m^2$ which satisfies both (\ref{quasi-modular}) and
(\ref{modular-anomaly}). Then we show that $f-f_{\textrm{S-dual}}$ is 
independent of $v$.

We expand $f_{\textrm{S-dual}}$ like (\ref{m-series}),
$f_{\textrm{S-dual}}=\sum_{n=1}^\infty m^{2n}f_n(\tau,v)$.
(\ref{modular-anomaly}) is given in terms of $f_n$ by
\begin{equation}\label{modular-anomaly-rec}
  \frac{\partial f_n}{\partial E_2}=-\frac{1}{24}\sum_{m=1}^{n-1}
  \frac{\partial f_m}{\partial v}\frac{\partial f_{n-m}}{\partial v}
\end{equation}
for $n\geq 2$. This equation can be used to recursively compute
$f_n$. Namely, once we know $f_{m}$ for $m=1,\cdots,n-1$, one can integrate
the right hand side of (\ref{modular-anomaly-rec}) with $E_2$ to get $f_n$,
up to an integration constant independent
of $E_2$. The integration constant is a polynomial of $E_4$ and $E_6$
with modular weight $2n-2$, whose coefficients depend only
on $v$. These integration constants depending on $v$
can be fixed once we know a few low order coefficients of $f$ in $q$ expansion.
Also, to start the recursive construction, the first coefficient $f_1$ at
$m^2$ should be known. It will turn out that this can be also fixed by the 
known perturbative part $f_{\rm pert}$ \cite{Billo':2015ria}. This way, one 
can recursively generate the coefficients of $f_{\textrm{S-dual}}$ from 
(\ref{quasi-modular}), (\ref{modular-anomaly}) and the knowledge of the few 
low order coefficients of $f$ in $q$ expansion.
We emphasize here that our purpose of making a recursive construction of
$f_{\textrm{S-dual}}$ is to show that the Coulomb VEV dependent part of $f$
is S-dual. Therefore, while fixing the integration constants and $f_1$ in
$f_{\textrm{S-dual}}$ by using the low order $q$ expansion coefficients of $f$,
it suffices to use $f$ up to the addition of any convenient expression independent
of $v$. So for technical reasons, we shall fit these integration constants
and $f_1$ by comparing $f_{\textrm{S-dual}}$ with
\begin{equation}
  f(\tau,v,m)-Nf_{U(1)}(\tau,m)
\end{equation}
rather than $f$ itself. Note that  $Nf_{U(1)}$ is the
prepotential contribution from $U(1)^N$ Cartan part, coming from 
D0-branes bound to D4-branes but unbound to W-bosons which
see $v$. One reason for comparing with $f-Nf_{U(1)}$ is that
$f_{U(1)}$ does not admit a power series expansion
in $m^2$ like (\ref{m-series}). The S-duality anomaly of $Nf_{U(1)}$ 
can be calculated separately from (\ref{anom-series}).

With these understood, we start the recursive contruction by determining
$f_1$. This can be fixed solely from the perturbative part 
of $(f-Nf_{U(1)})_{\rm pert}$ \cite{Billo':2015ria}. Namely, when instantons are
bound to W-bosons, there are fermion zero modes which provide
at least a factor of $m^4$ in $f$. This means that $m^2$ term $f_1$ should 
come from the perturbative part only.
This fact can also be straightforwardly
checked from the microscopic calculus. So one finds
\begin{equation}
  f_1=\left.(f-Nf_{U(1)})_{\rm pert}\frac{}{}\!\right|_{m^2}
  =-\frac{1}{2}\sum_{\alpha\in\Delta}{\rm Li}_1(e^{\alpha(v)})
  =\frac{1}{2}\sum_{\alpha\in\Delta}\log(1-e^{\alpha(v)})\ ,
\end{equation}
where $\Delta$ is the set of roots of $U(N)$. 

One can then compute $f_2$ using (\ref{modular-anomaly-rec}) at $n=2$,
\begin{equation}
  \frac{\partial f_2}{\partial E_2}
  =-\frac{1}{24}\left(\frac{\partial f_1}{\partial v}\right)^2
  =-\frac{1}{96}\sum_{\alpha,\beta\in\Delta}
  \frac{\alpha\cdot\beta}{(1-e^{\alpha(v)})(1-e^{\beta(v)})}\ .
\end{equation}
One can integrate it with $E_2$, to obtain
\begin{equation}
  f_2=-\frac{E_2(\tau)}{96}\sum_{\alpha,\beta\in\Delta}
  \frac{\alpha\cdot\beta}{(1-e^{\alpha(v)})(1-e^{\beta(v)})}\ .
\end{equation}
There is no integration constant at weight $2$.
To proceed, we study the properties of the $U(N)$ roots.
$\Delta$ consists of vectors of the form $e_i-e_j$, $i\neq j$, $i,j=1,\cdots,N$,
where $e_i$ are $N$ orthogonal unit vectors.
$\alpha\cdot\beta$ takes following possible values,
\begin{eqnarray}
  \alpha\cdot\beta=\pm 2&&{\textrm{if  }}\pm\beta=\alpha\nonumber\\
  \alpha\cdot\beta=\pm 1&&{\textrm{if }}\pm\beta\in\Psi(\alpha)\nonumber\\
  \alpha\cdot\beta=0&&\textrm{otherwise}\ ,
\end{eqnarray}
where $\Psi(\alpha)$ is given for $\alpha=e_i-e_j$ by
\begin{equation}
  \Psi(e_i-e_j)=\{k\neq i,j:e_i-e_k,e_k-e_j\}\ .
\end{equation}
For a given $\alpha$, there are $2(N-2)$ elements of $\Psi(\alpha)$.
Using this, one finds
\begin{eqnarray}
  \hspace*{-1cm}f_2&=&-\frac{E_2(\tau)}{96}\left[
  \sum_{\alpha\in\Delta}\frac{2}{1-e^{\alpha(v)}}
  \left(\frac{1}{1-e^{\alpha(v)}}-\frac{1}{1-e^{-\alpha(v)}}\right)
  +\sum_{\alpha\in\Delta}\sum_{\beta\in\Psi(\alpha)}
  \frac{1}{1-e^{\alpha(v)}}
  \left(\frac{1}{1-e^{\beta(v)}}-\frac{1}{1-e^{-\beta(v)}}\right)\right]\nonumber\\
  \hspace*{-1cm}&=&-\frac{E_2(\tau)}{96}\left[
  \sum_{\alpha\in\Delta}\frac{4}{(1-e^{\alpha(v)})^2}
  -\sum_{\alpha\in\Delta}\frac{2(N-1)}{1-e^{\alpha(v)}}
  +\sum_{\alpha\in\Delta}\sum_{\beta\in\Psi(\alpha)}
  \frac{2}{(1-e^{\alpha(v)})(1-e^{\beta(v)})}\right]\nonumber
\end{eqnarray}
where we used
$\frac{1}{1-e^{-\alpha(v)}}-\frac{1}{1-e^{-\alpha(v)}}
=\frac{2}{1-e^{\alpha(v)}}-1$. Using $\frac{1}{(1-e^{\alpha(v)})^2}=\frac{1}{1-e^{\alpha(v)}}
+\frac{e^{\alpha(v)}}{(1-e^{\alpha(v)})^2}=\frac{1}{1-e^{\alpha(v)}}
+{\rm Li}_{-1}(e^{-\alpha(v)})$, the first term can be rewritten so that
\begin{equation}
  f_2=-\frac{E_2(\tau)}{96}\left[
  4\sum_{\alpha\in\Delta}{\rm Li}_{-1}(e^{\alpha(v)})
  -\sum_{\alpha\in\Delta}\frac{2(N-3)}{1-e^{\alpha(v)}}
  +\sum_{\alpha\in\Delta}\sum_{\beta\in\Psi(\alpha)}
  \frac{2}{(1-e^{\alpha(v)})(1-e^{\beta(v)})}\right]\ .
\end{equation}
Here, one can simplify the second term by using
\begin{equation}\label{root-sum}
  \sum_{\alpha\in\Delta}\frac{1}{1-e^{\alpha(v)}}=
  \frac{1}{2}\sum_{\alpha\in\Delta}\left(\frac{1}{1-e^{\alpha(v)}}
  +\frac{1}{1-e^{-\alpha(v)}}\right)=\frac{1}{2}\sum_{\alpha\in\Delta}1=
  \frac{N(N-1)}{2}\ .\nonumber
\end{equation}
Also, using
$\frac{1}{(1-e^x)(1-e^y)}+\frac{1}{(1-e^{-x})(1-e^{y-x})}
+\frac{1}{(1-e^{x-y})(1-e^{-y})}=1$, one finds
\begin{eqnarray}
  &&\sum_{\alpha\in\Delta}\sum_{\beta\in\Psi(\alpha)}
  \frac{1}{(1-e^{\alpha(v)})(1-e^{\beta(v)})}
  =\sum_{i\neq j}\sum_{k\neq i,j}\left[\frac{1}{(1-e^{v_i-v_j})(1-e^{v_i-v_k})}
  +\frac{1}{(1-e^{v_i-v_j})(1-e^{v_k-v_j})}\right]\nonumber\\
  &&\hspace{1cm}=\frac{1}{3}\sum_{i\neq j\neq k \neq i}\left[(i,j,k)+(j,k,i)+(k,i,j)\right]
  =\sum_{i\neq j\neq k\neq i}\frac{2}{3}=\frac{2}{3}N(N-1)(N-2)\ ,
\end{eqnarray}
where at the second step we symmetrized the summand by making a cyclic 
permutation of $i,j,k$. This simplifies the third term. One thus finds
\begin{equation}
  f_2=-\frac{E_2(\tau)}{24}\left[
  \sum_{\alpha\in\Delta}{\rm Li}_{-1}(e^{\alpha(v)})
  +\frac{N^3-N}{12}\right]
\end{equation}
at $\mathcal{O}(m^4)$ order.

Before proceeding to higher order coefficients $f_n$ with $n\geq 3$,
let us first discuss $f_2$ that we computed by requiring S-duality of
$f_{\textrm{S-dual}}$. Note that at $m^4$, we have obtained an all order
result in the instanton expansion, coming from
$E_2(\tau)=1-24q-72q^2-96q^3-168q^4\cdots$. So from the microscopic
instanton calculus, one can expand $f(\tau,v,m)$ in small $m$, and we can
compare $f$ and $f_{\textrm{S-dual}}$ at $m^4$ order. We find that
\begin{equation}
  \left.(f-Nf_{U(1)})-f_{\textrm{S-dual}}\frac{}{}\!\right|_{m^4}
  =\frac{N^3-N}{288}m^4E_2(\tau)\ ,
\end{equation}
which we checked till $q^2$ order for general $N$, and
till $q^3$ for $N=2,3$. Therefore, we find that
the microscopic prepotential is compatible with S-duality at $m^4$ order,
up to the addition of an `anomalous'
term on the right hand side independent of the Coulomb VEV.

One can make further recursive calculations of $f_n$ for $n\geq 3$,
using (\ref{modular-anomaly-rec}), and test the consistency of
$f_{\textrm{S-dual}}$ with our microscopic $f$. The next recursion relation 
of (\ref{modular-anomaly-rec}) is
\begin{equation}
  \frac{\partial f_3}{\partial E_2}=-\frac{1}{12}
  \frac{\partial f_1}{\partial v}\frac{\partial f_2}{\partial v}\ .\label{f3-rec}
\end{equation}
Knowing $f_1,f_2$, one can integrate (\ref{f3-rec}) to obtain
\begin{equation}\label{f3-result}
  f_3=-\frac{E_2(\tau)^2}{1152}\left[\sum_{\alpha\in\Delta}
  (2{\rm Li}_{-3}(e^{\alpha(v)})-4{\rm Li}_{-1}(e^{\alpha(v)})^2)
  +2\sum_{\alpha\in\Delta}\sum_{\beta\in\Psi(\alpha)}
  {\rm Li}_0(e^{\alpha(v)}){\rm Li}_{-2}(e^{\beta(v)})\right]+c_3(v)E_4(\tau)\ .
\end{equation}
The integration constant $c_3(v)$ can be determined by expanding
$f_3$ in $q$, and comparing the $q^0$ order with the 
perturbative contribution $(f-Nf_{U(1)})_{\rm pert}$ at $m^6$ order. One obtains
\begin{equation}
  c_3(v)=\frac{1}{2880}\sum_{\alpha\in\Delta}{\rm Li}_{-3}(e^{\alpha(v)})
  -\frac{1}{288}\sum_{\alpha\in\Delta}{\rm Li}_{-2}(e^{\alpha(v)})^2
  +\frac{1}{576}\sum_{\alpha\in\Delta}\sum_{\beta\in\Psi(\alpha)}
  {\rm Li}_0(e^{\alpha(v)}){\rm Li}_{-2}(e^{\beta(v)})\ .
\end{equation}
Inserting this $c_3(v)$ in (\ref{f3-result}), one can further study
the higher order coefficients of $f_3$ in $q$ expansion, against the
microscopic result $f$. We find that
\begin{equation}
  \left.(f-Nf_{U(1)})-f_{\textrm{S-dual}}\frac{}{}\!\right|_{m^6}=0\ ,
\end{equation}
which we checked till $q^2$ order for general $N$, 
and till $q^3$ order for $N=2,3$.

Integrating (\ref{modular-anomaly-rec}) to get
higher $f_n$'s, the integration constants take the following form,
\begin{equation}
  f_n(\tau,v)\leftarrow\sum_{4a+6b=2n-2,~a\geq 0,~b\geq 0}
  c_{a,b}(v)E_4(\tau)^aE_6(\tau)^b\ .
\end{equation}
More concretely, one would get
\begin{equation}
  f_4\leftarrow c_{0,1}E_6\ ,\ f_5\leftarrow c_{2,0}E_4^2
  \ ,\ f_6\leftarrow c_{1,1}E_4E_6\ ,\
  f_7\leftarrow c_{3,0}E_4^3+c_{0,2}E_6^2\ ,\
  f_8\leftarrow c_{2,1}E_4^2E_6\ ,\cdots
\end{equation}
and so on. To fix the coefficients $c_{p,q}(v)$, one should use
some low order data of $f-Nf_{U(1)}$. If there are $k+1$ independent $c_{p,q}$'s, 
one should use up to $k$ instanton coefficients of $f-Nf_{U(1)}$ to fix them. 
Then from $k+1$ or higher instantons, one can
test $f_n$ by comparing with $f-Nf_{U(1)}$. For general $N$, we tested
$f_{\textrm{S-dual}}$ against $f-Nf_{U(1)}$ till $m^6$ and $q^2$ orders, as 
already reported above.
For $U(2)$, we tested it till $m^{14}$ (i.e. $f_7$) and $q^3$ orders.
For $U(3)$, we tested it till $m^8$ (i.e. $f_4$) and $q^3$ orders.
The result is that there is no further difference between $f_{\textrm{S-dual}}$ 
and $f-Nf_{U(1)}$ from $m^6$ and higher orders. Namely, we find that
\begin{equation}\label{S-dual-summary}
  f(\tau,v,m)=f_{\textrm{S-dual}}(\tau,v,m)+Nf_{U(1)}(\tau,m)
  +\frac{N^3-N}{288}m^4E_2(\tau)\ .
\end{equation}
The S-duality transformation of $f_{U(1)}$ 
can be separately derived from (\ref{anom-series}), which is
\begin{eqnarray}\label{Abelian-S-dual}
  \tau^{2}f_{U(1)}(-1/\tau,m/\tau)-f_{U(1)}(\tau,m)&=&
  \frac{m^2}{2}\log\frac{\phi(-1/\tau)^2}{\tau\phi(\tau)^2}
  +\frac{m^4}{288}(\tau^2 E_2(-1/\tau)-E_2(\tau))\nonumber\\
  &=&\frac{m^2}{2}\left[\frac{\pi i\tau}{6}+\frac{\pi i}{6\tau}
  +\log(-i)\right]+\frac{m^4}{48\pi i\tau}\ ,
\end{eqnarray}
where $\eta(-1/\tau)=\sqrt{-i\tau}\eta(\tau)$.
Although we checked (\ref{S-dual-summary}) in a serious
expansion in small $m$, we believe it is an exact
property, valid for finite $m$. In section 2.2, we shall provide another
test of (\ref{S-dual-summary}) by keeping $m$ finite, but expanding $f$ as
a series of $e^{-v}$ at positive Coulomb VEV $v$. Also, in section 2.3, 
we extend (\ref{S-dual-summary}) to all ADE theories.

\subsection{Derivation from M-strings}

In this subsection, we provide another derivation of the S-duality
of prepotential. The analysis here will provide 
more evidence for our S-duality anomaly at finite $m$. It may also 
provide strong hints towards the S-duality of the partition function
$Z(\tau,v,m,\epsilon_{1,2})$ at finite $\epsilon_{1,2}$, but we postpone 
more extensive discussions on $Z$ to a separate project.

We use an alternative partition
function $\check{Z}(\tau,v,m,\epsilon_{1,2})=\check{Z}_{\rm pert}Z_{\rm inst}$.
This differs from $Z$ by a different perturbative partition 
function $\check{Z}_{\rm pert}$. For the purpose of studying the prepotential 
in the limit $\epsilon_{1,2}\rightarrow 0$, we already saw that 
they differ only by a trivial term, as shown in (\ref{pert-difference})
in a suitable regime on the complex $v$ plane. The last term of 
(\ref{pert-difference}) is separately invariant under the S-duality transformation
$(\tau,m,\epsilon_{1,2})\rightarrow(-\frac{1}{\tau},\frac{m}{\tau},
\frac{\epsilon_{1,2}}{\tau})$, and is independent of $v$, so it is completely 
fine to drop this constant factor when discussing the S-duality of prepotential.
$\check{Z}$ can be expanded in $e^{-\alpha_i(v)}$ when $\alpha_i(v)$ are positive 
and sufficiently large, where $\alpha_i$ ($i=1,\cdots,N-1$) are the simple roots 
of $SU(N)$. The expansion takes the form of \cite{Haghighat:2013gba}
\begin{equation}
  Z(\tau,v,m,\epsilon_{1,2})=e^{-\varepsilon_0}
  Z_{U(1)}(\tau,m,\epsilon_{1,2})^N
  \sum_{n_1,\cdots,n_{N-1}=0}^\infty e^{-\sum_{i=1}^{N-1}n_i\alpha_i(v)}
  Z_{(n_i)}(\tau,m,\epsilon_{1,2})\equiv
  e^{-\varepsilon_0}Z_{U(1)}^N\hat{Z}\ .
\end{equation}
$Z_{U(1)}^N$ is for the D0-branes unbound to the W-bosons, just bound to
one of the $N$ D4-branes, which is this independent of $e^{-\alpha_i(v)}$.
$\varepsilon_0$ is defined at the beginning of section 2 by
\begin{equation}
  \varepsilon_0=\frac{m^2-\epsilon_+^2}{2\epsilon_1\epsilon_2}
  \sum_{\alpha>0}(\alpha(v)+\pi i)\ .
\end{equation}
This factor has to be multiplied to guarantee (\ref{pert-difference}). 
This expansion is reliable in a particular Weyl chamber of the 
Coulomb branch, with sufficiently large $\alpha_i(v)$ for all simple roots.
The coefficient $Z_{(n_i)}(\tau,m,\epsilon_{1,2})$ at given 
self-dual string numbers $n_i$
is computed from `M-strings,' which are M2-branes suspended
between separated M5-branes. $Z_{(n_i)}$ is the elliptic genus of the
2d quiver gauge theory with $U(n_1)\times\cdots U(n_{N-1})$ gauge group.
See \cite{Haghighat:2013tka} for the details.
$Z_{(n_i)}(\tau,m,\epsilon_{1,2})$ is given by 
\cite{Haghighat:2013gba} (we follow the notations of \cite{Kim:2015gha})
\begin{equation}\label{M-string-elliptic}
  Z_{(n_i)}=\sum_{Y_1,\cdots,Y_{N-1};|Y_i|=n_i}\prod_{i=1}^N
  \prod_{s\in Y_i}\frac{\theta_1(\tau|\frac{E_{i,i+1}(s)-m+\epsilon_-}{2\pi i})
  \theta_1(\tau|\frac{E_{i,i-1}(s)+m+\epsilon_-}{2\pi i})}
  {\theta_1(\tau|\frac{E_{i,i}(s)+\epsilon_1}{2\pi i})
  \theta_1(\tau|\frac{E_{i,i}(s)-\epsilon_2}{2\pi i})}\ ,
\end{equation}
where $s=(a,b)$ denotes the position of each box in the Young diagram, and
\begin{equation}
  E_{i,j}(s=(a,b))=(Y_{i,a}-b)\epsilon_1-(Y^T_{j,b}-a)\epsilon_2\ .
\end{equation}
$Y_{i,a}$ is the length of the $a$'th row of $Y_i$, and $Y^T_{j,b}$ 
is the length of the $b$'th column of $Y_j$. $Y_0$ and $Y_N$ are empty 
by definition.

We study the S-duality of $\check{Z}$ in this setting. By using 
\begin{equation}\label{theta-S-dual}
  \frac{\theta_1(-\frac{1}{\tau},\frac{z}{\tau})}{\eta(-\frac{1}{\tau})}
  =e^{\frac{\pi iz^2}{\tau}}\frac{\theta_1(\tau,z)}{\eta(\tau)}\ ,
\end{equation}
one can show that $Z_{(n_i)}$ transforms as 
\begin{equation}\label{M-string-S-dual}
  Z_{(n_i)}\left(-\frac{1}{\tau},\frac{m}{\tau},\frac{\epsilon_{1,2}}{\tau}\right)
  =\exp\left[\frac{1}{4\pi i\tau}\left(
  \epsilon_1\epsilon_2\sum_{i,j=1}^{N-1}A_{ij}n_in_j
  +2(m^2-\epsilon_+^2)\sum_{i=1}^{N-1}n_i\right)\right]
  Z_{(n_i)}(\tau,m,\epsilon_{1,2})\ .
\end{equation}
$A_{ij}$ is the 
Cartan matrix for $SU(N)$, given by $A_{ii}=2$, $A_{i,i+1}=A_{i,i-1}=-1$
and $A_{ij}=0$ otherwise. This transformation can be expressed as 
a modular anomaly equation, as follows. First note that Jacobi's
theta functions can be expressed as 
\begin{equation}
  \theta_1(\tau|z)=2\pi iz\ \eta(\tau)^3\exp\left[\sum_{k=1}^\infty
  \frac{B_{2k}}{(2k)(2k)!}E_{2k}(\tau)(2\pi iz)^{2k}\right]\ .
\end{equation}
$\tau$ dependence of $Z_{(n_i)}$ can be understood as its dependence 
through $E_{2n}(\tau)$, since $\eta(\tau)^3$ factors cancel out in 
(\ref{M-string-elliptic}). As we did in section 2.1 for the prepotential, 
the $\tau$ dependence of $Z_{(n_i)}$ can again be decomposed into its 
dependence through $E_2(\tau)$, and the rest. To emphasize this, 
we write $Z_{(n_i)}(\tau,m,\epsilon_{1,2},E_2)$.
In this setting, the modular anomaly $e^{\frac{\pi iz^2}{\tau}}$
of (\ref{theta-S-dual}) appears due to the dependence of $\theta_1$ on $E_2(\tau)$, which is not modular. Therefore, we can rephrase (\ref{M-string-S-dual}) as
\begin{equation}
  \frac{\partial}{\partial E_2}Z_{(n_i)}(\tau,m,\epsilon_{1,2};E_2)
  =\frac{1}{24}\left(\epsilon_1\epsilon_2\sum_{i,j=1}^{N-1}
  A_{ij}n_in_j+2(m^2-\epsilon_+^2)\sum_{i=1}^{N-1}n_i\right)Z_{(n_i)}\ .
\end{equation}
Let us consider $\hat{Z}(\tau,\lambda,m,\epsilon_{1,2};E_2)=
\sum_{n_i=0}^\infty e^{-n_i\lambda_i}Z_{(n_i)}(\tau,m,\epsilon_{1,2};E_2)$, 
where we defined
$\lambda_i\equiv v_i-v_{i+1}>0$ with the choice of simple roots
$\alpha_i=e_i-e_{i+1}$. Here we view $\hat{Z}$ as a function of $\lambda$'s,
since it depends only on the differences of $v$'s.
Now with the replacement $n_i\rightarrow-\frac{\partial}{\partial\lambda_i}$
in the Laplace transformation, one obtains
\begin{equation}\label{modular-anomaly-M-1}
  \frac{\partial}{\partial E_2}\hat{Z}(\tau,\lambda,m,\epsilon_{1,2};E_2)
  =\frac{1}{24}\left(\epsilon_1\epsilon_2\sum_{i,j=1}^{N-1}
  A_{ij}\frac{\partial^2}{\partial\lambda_i\partial\lambda_j}
  -2(m^2-\epsilon_+^2)\sum_{i=1}^{N-1}\frac{\partial}{\partial\lambda_i}
  \right)\hat{Z}\ .
\end{equation}
This is essentially the `holomorphic anomaly equation,' eqn.(3.54) of 
\cite{Haghighat:2013gba}. In our context, we continue to study it as the 
modular anomaly equation.

We shall get better intuitions on the modular anomaly equation. 
Consider $\tilde{Z}\equiv e^{-\varepsilon_0}\hat{Z}$ with
$\varepsilon_0=\frac{m^2-\epsilon_+^2}{2\epsilon_1\epsilon_2}
\sum_{\alpha>0}(\alpha(v)+\pi i)$, 
still without $(Z_{U(1)})^N$ included. Since the $v$ independent part 
of $\varepsilon_0$ is independent of $E_2,\lambda$, 
it does not affect (\ref{modular-anomaly-M-1}).
The $v$ dependent part can be written as
\begin{eqnarray}
  \sum_{\alpha>0}\alpha(v)&=&\sum_{i<j}(v_i-v_j)=
  \sum_{i=1}^N[(N-i)-(i-1)]v_i=\sum_{i=1}^N(N+1-2i)v_i\ .
\end{eqnarray}
Since $\frac{\partial\lambda_i(v)}{\partial v_j}=\delta_{i,j}-\delta_{i+1,j}$,
$\frac{\partial}{\partial v_j}$ acting on a function of $\lambda_i$'s can be 
written as $\frac{\partial}{\partial v_j}=\frac{\partial\lambda_i(v)}{\partial v_j}
\frac{\partial}{\partial\lambda_i}$,
\begin{equation}
  \frac{\partial}{\partial v_1}=\frac{\partial}{\partial\lambda_1}\ ,\ \ 
  \frac{\partial}{\partial v_j}=\frac{\partial}{\partial\lambda_j}
  -\frac{\partial}{\partial\lambda_{j-1}}\ \ \ (j=2,\cdots,N-1)\ ,\ 
  \frac{\partial}{\partial v_N}=-\frac{\partial}{\partial\lambda_{N-1}}.
\end{equation}
The Laplacian of $v$ acting on a function of $\lambda$
is given by
\begin{equation}
  \nabla^2\equiv\sum_{i=1}^N\frac{\partial^2}{\partial v_i^2}=
  A_{ij}\frac{\partial^2}{\partial\lambda_i\partial\lambda_j}\ .
\end{equation}
One also finds that
\begin{equation}
  \sum_{i=1}^{N-1}\frac{\partial}{\partial\lambda_i}=
  \sum_{i=1}^N(a-i)\partial_{v_i}
\end{equation}
for any number $a$. Acting $\nabla^2$ on $\tilde{Z}=e^{-\varepsilon_0}\hat{Z}$, 
one obtains
\begin{eqnarray}
  \nabla^2\tilde{Z}&=&e^{-\varepsilon_0}\nabla^2\hat{Z}
  -2\frac{m^2-\epsilon_+^2}{2\epsilon_1\epsilon_2}
  e^{-\varepsilon_0}\sum_{i=1}^N(N+1-2i)\partial_{v_i}\hat{Z}+
  \left(\frac{m^2-\epsilon_+^2}{2\epsilon_1\epsilon_2}\right)^2
  e^{-\varepsilon_0}\sum_{i=1}^N(2i-N-1)^2\hat{Z}\nonumber\\
  &=&e^{-\varepsilon_0}\left[\nabla^2\hat{Z}
  -2\frac{m^2-\epsilon_+^2}{\epsilon_1\epsilon_2}\sum_{i=1}^{N-1}
  \frac{\partial}{\partial\lambda_i}\hat{Z}
  +\frac{N^3-N}{3}\left(\frac{m^2-\epsilon_+^2}{2\epsilon_1\epsilon_2}\right)^2
  \hat{Z}\right]
\end{eqnarray}
where we used $\frac{\partial\varepsilon_0}{\partial v_i}
=\frac{m^2-\epsilon_+^2}{2\epsilon_1\epsilon_2}(N+1-2i)$.
Using (\ref{modular-anomaly-M-1}), this can be rewritten as
\begin{equation}
  \nabla^2\tilde{Z}=e^{-\varepsilon_0}
  \left[\frac{24}{\epsilon_1\epsilon_2}\frac{\partial}{\partial E_2}\hat{Z}
  +\frac{N^3-N}{3}\left(\frac{m^2-\epsilon_+^2}{2\epsilon_1\epsilon_2}\right)^2
  \hat{Z}\right]\ .
\end{equation}
Thus, one finds that the partition function 
\begin{equation}\label{Z-S-dual}
  Z_{\textrm{S-dual}}\equiv\exp\left[\frac{\epsilon_1\epsilon_2}{24}\frac{N^3-N}{12}
  \left(\frac{m^2-\epsilon_+^2}{\epsilon_1\epsilon_2}\right)^2E_2\right]\tilde{Z}
  =\exp\left[\frac{N^3-N}{288}
  \frac{(m^2-\epsilon_+^2)^2}{\epsilon_1\epsilon_2}E_2
  -\varepsilon_0\right]\hat{Z}
\end{equation}
satisfies the modular anomaly equation
\begin{equation}\label{Z-modular-anomaly}
  \frac{\partial}{\partial E_2}Z_{\textrm{S-dual}}(\tau,v,m,\epsilon_{1,2};E_2)
  =\frac{\epsilon_1\epsilon_2}{24}\nabla^2Z_{\textrm{S-dual}}\ .
\end{equation}
Also, from the M-string expansion form of $\hat{Z}$, and the the 
form of the prefactors we multiplied to define $Z_{\textrm{S-dual}}$, 
$Z_{\textrm{S-dual}}$ satisfies the following quasi-modularity condition,
\begin{equation}\label{Z-quasi-modular}
  Z_{\textrm{S-dual}}\left(-\frac{1}{\tau},v,\frac{m}{\tau},
  \frac{\epsilon_{1,2}}{\tau};E_2(-\frac{1}{\tau})\right)
  =Z_{\textrm{S-dual}}(\tau,v,m,\epsilon_{1,2},E_2(\tau)+\delta)
\end{equation}
where $\delta=\frac{6}{\pi i\tau}$. (\ref{Z-modular-anomaly}) and 
(\ref{Z-quasi-modular}) are the two main properties of $Z_{\textrm{S-dual}}$.

Using (\ref{Z-modular-anomaly}) and (\ref{Z-quasi-modular}), 
we would like to study the relation between  
$Z_{\textrm{S-dual}}(\tau,v,m,\epsilon_{1,2},E_2)$ and 
$Z_{\textrm{S-dual}}(-\frac{1}{\tau},v,\frac{m}{\tau},
\frac{\epsilon_{1,2}}{\tau},E_2(-\frac{1}{\tau}))$. Since 
$\tilde{Z}$ is related to $Z_{\text{S-dual}}$ in a simple 
manner, answering this question will tell us how $\tilde{Z}$ transforms 
under S-duality. Then, since we already understand how the $U(1)^N$ part 
$Z_{U(1)}^N$ transforms under S-duality \cite{Kim:2013nva}, we 
shall in turn know the S-duality transformation of $\check{Z}$.
Using (\ref{Z-quasi-modular}), we should understand how 
$Z_{\textrm{S-dual}}(\tau,v,m,\epsilon_{1,2};E_2(\tau)+\delta)$ 
and $Z_{\textrm{S-dual}}(\tau,v,m,\epsilon_{1,2};E_2)$ are related to 
each other, at same values of $\tau,v,m,\epsilon_{1,2}$ but with a 
shift of $E_2$ by $\delta=\frac{6}{\pi i\tau}$. Let us formally 
regard $E_2$ as time variable, and $Z_{\textrm{S-dual}}$ as a 
wavefunction on the space formed by $v_i$. Then
(\ref{Z-modular-anomaly}) takes the form of heat equation, or Euclidean 
Schr\"{o}dinger equation for a free particle. More precisely, taking $\tau$ to be 
purely imaginary for simplicity, $\delta$ is real and negative. So $-E_2$ 
plays the role of time, and we would like to understand the time evolution 
$Z_{\textrm{S-dual}}(E_2+\delta)$ by $-\delta$ following the heat equation.
The heat equation comes with proper sign when $\epsilon_1\epsilon_2<0$.
In this case, the time evolution is described by evolving 
$Z_{\textrm{S-dual}}(E_2)$ by the Gaussian heat kernel as
\begin{equation}\label{Z-S-dual-Gaussian}
  Z_{\textrm{S-dual}}(\tau,v,m,\epsilon_{1,2};E_2(\tau)+\delta)
  =\int_{-\infty}^\infty\prod_{i=1}^{N}dv^\prime_i\ K(v,v^\prime)
  Z_{\textrm{S-dual}}(\tau,v^\prime,m,\epsilon_{1,2};E_2(\tau))\ ,
\end{equation}
where
\begin{equation}
  K(v,v^\prime)=\left(\frac{i\tau}{\epsilon_1\epsilon_2}\right)^{\frac{N}{2}}
  \exp\left[-\frac{\pi i\tau}{\epsilon_1\epsilon_2}(v-v^\prime)^2\right]
\end{equation}
is the heat kernel which approaches $K(v,v^\prime)\rightarrow\delta^{(N)}(v-v^\prime)$ 
when $\frac{i\tau}{\epsilon_1\epsilon_2}\rightarrow 0$.
When $\epsilon_1\epsilon_2>0$, the `time evolution' from $Z_{\textrm{S-dual}}(E_2)$ 
to $Z_{\textrm{S-dual}}(E_2+\delta)$ is described by the time reversal of the heat 
equation. Therefore, one finds
\begin{equation}\label{Z-S-dual-inverse}
  Z_{\textrm{S-dual}}(\tau,v,m,\epsilon_{1,2};E_2(\tau))=
  \int_{-\infty}^\infty\prod_{i=1}^{N}dv^\prime_i\ K_-(v,v^\prime)
  Z_{\textrm{S-dual}}(\tau,v^\prime,m,\epsilon_{1,2};E_2(\tau)+\delta)
\end{equation}
with $K_-(v,v^\prime)=\left(-\frac{i\tau}{\epsilon_1\epsilon_2}\right)^{\frac{N}{2}}
\exp\left[\frac{\pi i\tau}{\epsilon_1\epsilon_2}(v-v^\prime)^2\right]$
for $\epsilon_1\epsilon_2>0$.

The S-duality of $f_{\textrm{S-dual}}$ can be studied 
from (\ref{Z-S-dual-Gaussian}) or (\ref{Z-S-dual-inverse}) 
by a saddle point approximation of the $v^\prime$ integral at 
$\epsilon_{1,2}\rightarrow 0$. Using both equations yield idential results.
One finds 
\begin{equation} 
  Z_{\textrm{S-dual}}=e^{-\frac{f_{\textrm{S-dual}}}{\epsilon_1\epsilon_2}}
  \sim\exp\left[-\frac{\hat{f}}{\epsilon_1\epsilon_2}
  +\frac{N^3-N}{288}\frac{m^4}{\epsilon_1\epsilon_2}E_2(\tau)
  -\frac{m^2}{2\epsilon_1\epsilon_2}\sum_{\alpha>0}(\alpha(v)+\pi i)\right]
\end{equation}
in the limit $\epsilon_{1,2}\rightarrow 0$, so that 
\begin{equation}\label{f-S-dual-M}
  f_{\textrm{S-dual}}=\hat{f}-\frac{N^3-N}{288}m^4E_2(\tau)
  +\frac{m^2}{2}\sum_{\alpha>0}(\alpha(v)+\pi i)\ .
\end{equation}
We shall show that $f_{\textrm{S-dual}}$ defined this way is the 
same $f_{\textrm{S-dual}}$ defined and computed in section 2.1.
Firstly, note that
\begin{equation}
  \hat{f}=\check{f}-Nf_{U(1)}-\frac{m^2}{2}\sum_{\alpha<0}(\alpha(v)+\pi i)
  =f-Nf_{U(1)}-\frac{m^2}{2}\sum_{\alpha<0}(\alpha(v)+\pi i)
  -\frac{\pi im^2|\Delta_+|}{2}
\end{equation}
from the relations $\check{Z}=e^{-\varepsilon_0}[Z_{U(1)}]^N\hat{Z}$ 
and $\check{Z}\sim Ze^{\frac{\pi im^2|\Delta_+|}{2\epsilon_1\epsilon_2}}$. 
Inserting this in (\ref{f-S-dual-M}), one obtains
\begin{equation}
  f_{\textrm{S-dual}}=f-Nf_{U(1)}-\frac{N^3-N}{288}m^4E_2(\tau)  
  -\frac{\pi im^2|\Delta_+|}{2}\ .
\end{equation}
This is completely the same as the relation between $f$ and $f_{\textrm{S-dual}}$ 
that we found in section 2.1, except the last term on the right hand side. 
However, we know that the last term comes from using slightly different perturbative 
partition function in $\check{Z}$, (\ref{pert-difference}).
Also this term can be completely 
ignored for studying S-duality since it is separately invariant under 
S-duality. Therefore, after discarding this last term, 
we find that $f_{\textrm{S-dual}}$ is the same as $f_{\textrm{S-dual}}$ defined 
in section 2.1. Now one can independently check that $f_{\textrm{S-dual}}$ is 
S-dual. Firstly, Since $Z_{\textrm{S-dual}}$ is quasi-modular, so is 
$f_{\textrm{S-dual}}$, i.e.
\begin{equation}
  \tau^2f_{\textrm{S-dual}}
  \left(-\frac{1}{\tau},v,\frac{m}{\tau},E_2(-1/\tau)\right)
  =f_{\textrm{S-dual}}(\tau,v,m,E_2(\tau)+\delta)
\end{equation}
Secondly, inserting 
$Z_{\textrm{S-dual}}\sim
\exp\left[-\frac{f_{\textrm{S-dual}}}{\epsilon_1\epsilon_2}\right]$ into 
(\ref{Z-modular-anomaly})
and keeping the leading terms in the limit $\epsilon_{1,2}\rightarrow 0$, 
one finds $\frac{\partial f_{\textrm{S-dual}}}{\partial E_2}=-\frac{1}{24}
\left(\frac{\partial f_{\textrm{S-dual}}}{\partial v}\right)^2$, the 
same modular anomaly equation that we studied in section 2.1 \cite{Billo:2013jba}.
The last two equations guarantee the S-duality of $f_{\textrm{S-dual}}$, 
completing an alternative proof of S-duality based on M-strings.

While making an alternative derivation of the S-duality and its anomaly, 
we did not assume the smallness of $m$.
On the other hand, around (\ref{pert-difference}), we required 
${\rm Re}(\alpha(v))>\pm{\rm Re}(m)$ for the positive roots $\alpha$, 
and that ${\rm Im}(\alpha(v)\pm m)$ for positive roots be in the range 
$(0,2\pi]$, to justify the uses of $\check{F}_{\rm pert}$ and 
$\check{Z}_{\rm pert}$. So at least in this range, 
$F_{\rm anom}=Nf_{U(1)}+\frac{N^3-N}{288}m^4E_2(\tau)$ is exact at finite $m$.
One can scan the whole complex planes of $v$ and $m$, considering the multiple 
values of ${\rm Li}_3$ function, to find the most general form of the S-duality 
anomaly when (\ref{pert-difference}) is violated. We shall not do this exercise here. 

Before closing this subsection, 
we comment on the nature of the S-duality transformation of $Z_{\textrm{S-dual}}$ 
or $\check{Z}$, and what it implies to the S-duality of $Z$. This 
issue is also related to the S-duality of the 4d limit of $Z$, which 
was studied in the context of AGT correspondence \cite{Alday:2009aq}. 
Taking the 4d limit $R\rightarrow 0$ with fixed $a,M,\varepsilon_{1,2},m,\tau$,  (\ref{Z-S-dual-Gaussian}) asserts that 
$Z^{\rm 4d}(\tau,a,m,\varepsilon_{1,2})$ is related to its S-dual 
by a Gaussian S-duality kernel. This is because the $R$ dependence
$K$ is given by
\begin{equation}
  K(a,a^\prime)=R^{-N}
  \left(\frac{i\tau}{\varepsilon_1\varepsilon_2}\right)^{\frac{N}{2}}
  \exp\left[-\frac{\pi i\tau}{\varepsilon_1\varepsilon_2}(a-a^\prime)^2\right]\ ,
\end{equation}
and the overall $R$ dependence is absorbed into the $v^\prime$ integration to 
be $\prod_ida^\prime_i$. Expanding the exponent of the kernel, and inserting
$a=\frac{a_D}{\tau}$, one obtains
\begin{equation}
  \exp\left[-\frac{F_{\rm cl}(a^\prime,\tau)-F_{\rm cl}(a_D,-\frac{1}{\tau})}
  {\varepsilon_1\varepsilon_2}\right]
  \exp\left[\frac{\pi i a_Da^\prime}{\varepsilon_1\varepsilon_2}\right]\ ,
\end{equation}
where $F_{\rm cl}(a,\tau)=\pi i\tau a^2$. The two $F_{\rm cl}$'s can be absorbed into
two $Z^{\rm 4d}$'s on the left and right hand sides of (\ref{Z-S-dual-Gaussian}). 
Then, the 4d limit of (\ref{Z-S-dual-Gaussian}) states that S-dualization is
Fourier transformation. Our studies imply that the same result holds 
for $Z_{\textrm{S-dual}}$ in 6d.

In fact, the proper S-duality transformation of the 4d partition function 
is known \textit{not} to be the Fourier transformation. Instead, based on the 
AGT correspondence, the S-duality kernel is asserted to be a nontrivial function 
given by the partition function of the 3d $T[SU(N)]$ theory
on $S^3$ \cite{Drukker:2010jp,Hosomichi:2010vh}. On the other hand, 
it has been found that the S-duality of the Omega deformed partition function
is the Fourier transformation at all perturbative orders in
$\varepsilon_1,\varepsilon_2$ 
\cite{Billo:2013jba,Galakhov:2013jma,Bullimore:2014awa,Nemkov:2015zha}.
An explanation of this was given in \cite{Galakhov:2013jma}, which 
finds that the choice of proper `normalization factor' independent of $\tau$ 
dressing the instanton partition funciton 
yields such a nonperturbative correction in $\varepsilon_{1,2}$. 
This is the `choice' of $Z_{\rm pert}^{\rm 4d}$, which 
was called $N(a)$ and $N_s(a)$ in \cite{Galakhov:2013jma}. Incidently, what we 
find in 6d is analogous to the findings of \cite{Galakhov:2013jma}. Namely, $Z$ 
was defined in section 2 with $Z_{\rm pert}$ which is manifestly invariant under
Weyl symmetry. However, $\check{Z}$ was defined with $\check{Z}_{\rm pert}$
which is not invariant under Weyl symmetry. In fact, we checked that the 
ratio $\frac{\check{Z}_{\rm pert}}{Z_{\rm pert}}$ at small $\epsilon_{1,2}$ is 
nonperturbative in $\epsilon_{1,2}$, which is qualitatively consistent with 
\cite{Galakhov:2013jma}. So along this line, it will be 
interesting to pursue the 6d extensions of \cite{Galakhov:2013jma}.
We stress again that, all our findings in this subsection concerns 
the prepotential in the $\epsilon_{1,2}\rightarrow 0$ limit, for which 
the distinction of $Z_{\rm pert}$ or $\check{Z}_{\rm pert}$ is 
irrelevant.

\subsection{6d $(2,0)$ theories of $D_N$ and $E_N$ types}

We generalize some studies we made for $A_{N-1}$
type $(2,0)$ theories to $D_N$ and $E_N$ type theories. For $D_N$ type
theories, $f_{\textrm{S-dual}}$ can be compared with microscopic instanton
calculus for the 5d $SO(2N)$ $\mathcal{N}=1^\ast$ theory, or
the D0-D4-O4 matrix quantum mechanics \cite{Hwang:2016gfw}. For $E_N$ types, 
we make a prediction of the S-duality and our knowledge of 5d perturbative 
prepotential. All ADE results will be partly tested in section 3.2 from 
6d chiral anomalies.

In the setting of section 2.1, the leading coefficient $f_1$ of $f_{\textrm{S-dual}}=\sum_{n=1}^\infty m^{2n}f_n$
is obtained from $(f-rf_{U(1)})_{\rm pert}$, where $r$ is the rank
of the gauge group. The result is
\begin{equation}
  f_1=\frac{1}{2}\sum_{\alpha\in\Delta}\log(1-e^{\alpha(v)})\ .
\end{equation}
Then using (\ref{modular-anomaly-rec}) at $n=2$, one finds
\begin{equation}
  f_2=-\frac{E_2(\tau)}{96}\sum_{\alpha,\beta\in\Delta}
  \frac{\alpha\cdot\beta}{(1-e^{\alpha(v)})(1-e^{\beta(v)})}\ .
\end{equation}
To proceed, we classify the roots $\beta$ depending on
their norm with $\alpha$. The possibilities are
\begin{eqnarray}
  (1)&:&\alpha\cdot\beta=\pm 2\textrm{ if }\pm\beta=\alpha\nonumber\\
  (2)&:&\alpha\cdot\beta=\pm 1\textrm{ if }\pm\beta\in\Psi(\alpha)\nonumber\\
  (3)&:&\alpha\cdot\beta=0\textrm{ otherwise }\ .
\end{eqnarray}
It is again important to understand the set $\Psi(\alpha)$ for ADE, 
which we explain now.

For $D_N=SO(2N)$, the $2N^2-2N$ roots in $\Delta$ 
are given by $\pm e_i\pm e_j$, where $i,j=1,\cdots,N$ and $i<j$.
Elements of $\Psi(\alpha)$ are given for various $\alpha$ by
\begin{eqnarray}
  \alpha=e_i-e_j&:&\Psi(\alpha)=\{k\neq i,j:e_i\pm e_k, \pm e_k-e_j\}\ ,\ \
  4(N-2)\textrm{ elements}\nonumber\\
  \alpha=e_i+e_j&:&\Psi(\alpha)=\{k\neq i,j:e_i\pm e_k,e_j\pm e_k\}\ ,\ \
  4(N-2)\textrm{ elements}\ .
\end{eqnarray}
For $E_6$, the number of roots is $|\Delta|=72$. $40$ roots take the form 
of $\pm e_i\pm e_j$ where $i\neq j$ and $i,j=1,\cdots,5$, from the $SO(10)$
subalgebra. Additional $32$ roots take the form of
$\pm\frac{1}{2}(\pm e_1\pm\cdots\pm e_5-e_6-e_7+e_8)$, where
the total number of $-$ signs is even. The structure
of $\Psi(\alpha)$ is given for various $\alpha$ as follows. Firstly, 
when $\alpha=e_i-e_j$, then 
\begin{equation}
  \Psi(\alpha)=\{k\neq i,j:e_i\pm e_k, \pm e_k-e_j\}
  \cup\{\frac{1}{2}(e_i-e_j+\cdots)\}
\end{equation}
where $\cdots$ means that all possible signs are allowed in the $32$ 
spinorial elements. Thus, one finds $12+8=20$ elements of $\Psi(\alpha)$ in this case.
Similarly, for $\alpha=e_i+e_j$, one finds
\begin{equation}
  \Psi(\alpha)=\{k\neq i,j:e_i\pm e_k, e_j\pm e_k\}
  \cup\{\frac{1}{2}(e_i+e_j+\cdots)\}
\end{equation}
where $\cdots$ means the same. So again, one finds $|\Psi(\alpha)|=12+8=20$.
For $\alpha=-e_i-e_j$, one can do a similar analysis.
Finally, $\alpha$ can be one of the $32$ spinorial elements, 
$\alpha=\frac{s_0}{2}(s_1e_1+\cdots+s_5e_5-e_6-e_7+e_8)$ with 
$s_0,\cdots,s_5=\pm 1$ and $s_1\cdots s_5=1$. Then, 
\begin{equation}
  \Psi(\alpha)=\{s_0(s_ie_i+s_je_j)\}\cup
  \{\alpha-s_0(s_ie_i+s_je_j)\}\ ,
\end{equation}
so $|\Psi(\alpha)|={}_{5}C_{2}+{}_{5}C_2=20$. 
For $E_7$, $|\Delta|=126$. $60$ roots take the form of 
$\pm e_i\pm e_j$, $i,j=1,\cdots,6$, from $SO(12)$ subalgebra. Additional 
$64$ roots take the form of $\pm\frac{1}{2}(\pm e_1\pm\cdots\pm e_6-e_7+e_8)$,
with total number of $-$ signs being even. Finally, 
$2$ more roots are given by $\pm(e_7-e_8)$. When 
$\alpha=\pm e_i\pm e_j$, $\Psi(\alpha)$ takes the same structure as 
that shown for $E_6$. For instance, for $\alpha=e_i+e_j$, one finds 
$\Psi(\alpha)=\{k\neq i,j|e_i\pm e_k,e_j\pm e_k\}\cup\{\frac{1}{2}(e_i+e_j\cdots)\}$
with $|\Psi(\alpha)|=16+16=32$. When 
$\alpha=\frac{s_0}{2}(s_1e_1+\cdots+s_6e_6-e_7+e_8)$, 
with $s_1\cdots s_6=1$, one finds
\begin{equation}
  \Psi(\alpha)=\{s_0(s_ie_i+s_je_j)\}\cup
  \{\alpha-s_0(s_ie_i+s_je_j)\}\cup\{s_0(e_8-e_7),\alpha+s_0(e_7-e_8)\}
\end{equation}
with $|\Psi(\alpha)|={}_{6}C_{2}+{}_{6}C_2+2=32$. Finally, when 
$\alpha=e_7-e_8$, one finds
\begin{equation}
  \Psi(\alpha)=\{s_1\cdots s_6=-1\left.\frac{}{}\!\!\right|
  \frac{1}{2}(s_1e_1+\cdots s_6e_6+e_7-e_8)\}\ ,
\end{equation}
with $|\Psi(\alpha)|=32$. The case with $\alpha=e_8-e_7$ is similar. 
For $E_8$, $|\Delta|=240$. $112$ roots take the form of $\pm e_i\pm e_j$, 
$i,j=1,\cdots,8$, from $SO(16)$ subalgebra. Additional $128$ roots take 
the form of $\frac{1}{2}(\pm e_1\pm \cdots\pm e_8)$ with number of $-$ signs 
being even, forming the $SO(16)$ spinor representation. For 
$\alpha=e_i+e_j$, one finds $\Psi(\alpha)=\{k\neq i,j|e_i\pm e_k,e_j\pm e_k\}
\cup\{\frac{1}{2}(e_i+e_j\cdots)\}$, with $|\Psi(\alpha)|=24+32=56$. 
Other cases with roots of the form $\alpha=\pm e_i\pm e_j$ can be studied 
similarly. For $\alpha=\frac{1}{2}(s_1e_1+\cdots s_8e_8)$ 
with $s_1\cdots s_8=1$, one finds
\begin{equation}
  \Psi(\alpha)=\{s_ie_i+s_je_j\}\cup\{\alpha-(s_ie_i+s_je_j)\}
\end{equation}
with $|\Psi(\alpha)|={}_{8}C_2+{}_{8}C_2=56$.
Including the $SU(N)$ case studied in section 2.1, one finds
$|\Psi(\alpha)|=2c_2-4$, where $c_2$ is the dual Coxeter number. 
See Table \ref{ADE}.
Another useful fact that can be checked with all 
$\Psi(\alpha)$ we listed above is that,
if $\beta\in\Psi(\alpha)$, then $\alpha-\beta$ is also a root. 
One also finds $\alpha-\beta\in\Psi(\alpha)$, since $\alpha\cdot(\alpha-\beta)=1$.
So at given $\alpha$, one finds
\begin{equation}\label{Psi-identity}
  \sum_{\beta\in\Psi(\alpha)}f_{\alpha,\beta}=
  \sum_{\beta\in\Psi(\alpha)}f_{\alpha,\alpha-\beta}
\end{equation}
for any expression $f_{\alpha,\beta}$.
\begin{table}[t]
\begin{center}
  \begin{tabular}{  c || c | c | c | c | c  }
    \hline
    $G$ & $A_{N-1}$ & $D_N$ & $E_6$ & $E_7$ & $E_8$ \\
    \hline
    $r$ & $N-1$ & $N$ & $6$ & $7$ & $8$ \\
    \hline
    $|G|$ & $N^2-1$ & $2N^2-N$ & $78$ & $133$ & $248$ \\
    \hline
    $c_2$ & $N$ & $2N-2$ & $12$ & $18$ & $30$ \\
    \hline
 \end{tabular}
\end{center}
\caption{Data on simply laced Lie algebras}\label{ADE}
\end{table}

By following the analysis for the $U(N)$ case, till (\ref{root-sum}),
one finds
\begin{equation}\label{f2-general-int}
  f_2=-\frac{E_2}{96}\left[4\sum_{\alpha\in\Delta}{\rm Li}_{-1}(e^{\alpha(v)})
  -(c_2-3)(|G|-r)
  +\sum_{\alpha\in\Delta}\sum_{\beta\in\Psi(\alpha)}
  \frac{2}{(1-e^{\alpha(v)})(1-e^{\beta(v)})}\right]\ .
\end{equation}
Now we use the identity (\ref{Psi-identity}) to rewrite the last term
in the parenthesis as
\begin{equation}\label{perm-ADE}
  \frac{2}{3}\sum_{\alpha\in\Delta}\sum_{\beta\in\Psi(\alpha)}
  \left[\frac{1}{(1-e^{\alpha(v)})(1-e^{\beta(v)})}+
  \frac{1}{(1-e^{-\alpha(v)})(1-e^{\beta(v)-\alpha(v)})}+
  \frac{1}{(1-e^{\alpha(v)-\beta(v)})(1-e^{-\beta(v)})}\right]\ .
\end{equation}
On the second term, we relabeled $\alpha$ into $-\alpha$ in the first
sum, and then took $\beta-\alpha$ with $\beta\in\Psi(\alpha)$ as
labeling the elements of $\Psi(-\alpha)$. The third term is simply the 
second term with renaming $\alpha\leftrightarrow\beta$. 
Using the identity
$\frac{1}{(1-e^x)(1-e^y)}+\frac{1}{(1-e^{-x})(1-e^{y-x})}
+\frac{1}{(1-e^{x-y})(1-e^{-y})}=1$, (\ref{perm-ADE}) becomes
$\frac{2}{3}\sum_{\alpha\in\Delta}\sum_{\beta\in\Psi(\alpha)}1=
\frac{4}{3}(|G|-r)(c_2-2)$. Thus, one obtains
\begin{equation}\label{ADE-m4}
  f_2=-\frac{E_2(\tau)}{24}\left[\sum_{\alpha\in\Delta}
  {\rm Li}_{-1}(e^{\alpha(v)})+\frac{1}{12}(c_2+1)(|G|-r)\right]=
  -\frac{E_2(\tau)}{24}\left[\sum_{\alpha\in\Delta}
  {\rm Li}_{-1}(e^{\alpha(v)})+\frac{c_2|G|}{12}\right]\ ,
\end{equation}
where at the last step we used the identity $|G|=r(c_2+1)$
for simply-laced Lie algebra.

$f_2$ contains $E_2(\tau)=1-24q+\cdots$, 
so makes a prediction on the instanton corrections.
For $G=SO(2N)$, one can compare this against microscopic instanton calculus 
for the 5d $\mathcal{N}=1^\ast$ theory \cite{Hwang:2016gfw}. We compared
the two results at 1 instanton level for $SO(8)$. Namely, 
(\ref{ADE-m4}) implies
\begin{equation}
  \left.f_2\right|_{q^1}=\left.f_{\textrm{S-dual}}\right|_{m^4q^1}=
  \sum_{\alpha\in\Delta}{\rm Li}_{-1}(e^{\alpha(v)})+\frac{c_2|G|}{12}\ ,
\end{equation}
where ${\rm Li}_{-1}(x)=\frac{x}{(1-x)^2}$. On the other hand, the single 
instanton partition function $Z_1$ for the $SO(2N)$ theory can be obtained 
by starting from the Witten index for the quantum mechanics describing 
an O4$^-$ plane, $2$ D0-branes and $2N$ D4-branes (in the covering space).
The index is a complicated residue sum. One should further 
subtract the contributions from D0-branes unbound to 
D4-O4, which was explained in \cite{Hwang:2016gfw}. Following this procedure, 
we checked that
\begin{equation}
  \left.\frac{}{}\!\!f_{\textrm{S-dual}}-(f_{SO(8)}-4f_{U(1)})\right|_{m^4q^1}
  =\frac{c_2|G|}{12}\ .
\end{equation}
One can continue to generate higher order $f_n$'s, and 
also the microscopic instanton calculus for general $D_N$ at higher order in 
$q$, and compare them. Here we simply conjecture 
\begin{equation}\label{S-dual-ADE}
  f(\tau,v,m)=f_{\textrm{S-dual}}(\tau,v,m)+rf_{U(1)}(\tau,m)
  +\frac{c_2|G|}{288}m^4E_2(\tau)
\end{equation}
for all $G=SU(N),SO(2N),E_N$, where $r$ is the rank of $G$. 
For $G=SU(N)$, we have tested it
extensively in section 2.1, after adding one free tensor multiplet to make 
it $U(N)$. For $G=SO(2N)$, we tested it till $m^4$, $q^1$ order only at $N=4$, 
but in principle one can do all the calculus of section 2.1, following 
the methods of \cite{Hwang:2016gfw}. For $E_N$, this is just a prediction 
by assuming S-duality and 5d perturbative results. The last term proportional 
to $c_2|G|$ will be further tested in section 3, from the 6d chiral anomaly 
of $SO(5)$ R-symmetry.

\section{High temperature limit of the index}

In this section, we compute the asymptotic form of the prepotential at strong 
coupling, or high `temperature' $\tau\rightarrow i0$. This is the limit in which 
the compactification radius $R^\prime$ of the sixth circle becomes large, 
or equivalently in which D0-branes become light.
The key technique of computation
will be the anomalous S-duality that we developed in section 2.

Our convention is that the strong coupling theory of our interest is
the `S-dualized' theory. So we take $\tau_D\rightarrow i0^+$, and
$\tau=-\frac{1}{\tau_D}\rightarrow i\infty$.
Recall $f_{\textrm{S-dual}}$ satisfies
\begin{equation}\label{S-dual-relation}
  \tau^2f_{\textrm{S-dual}}(\tau_D,v_D,m_D)=f_{\textrm{S-dual}}(\tau,v,m)
  +\frac{1}{4\pi i\tau}\left(\frac{\partial f}{\partial v}\right)^2
\end{equation}
where $\tau_D=-\frac{1}{\tau}$,
$v_D=v+\frac{1}{2\pi i\tau}\frac{\partial f}{\partial v}$,
$m_D=\frac{m}{\tau}$. We replaced $f_{\textrm{S-dual}}$
by $f$ when it appears with $v$ derivatives, since $f_{\rm anom}$ is
independent of $v$. Inserting $f_{\textrm{S-dual}}=f-f_{\rm anom}$,
one finds that
\begin{equation}\label{micro-dual}
  \tau^2f(\tau_D,v_D,m_D)=f(\tau,v,m)
  +\frac{1}{4\pi i\tau}\left(\frac{\partial f}{\partial v}\right)^2
  +\tau^2f_{\rm anom}(\tau_D,m_D)-f_{\rm anom}(\tau,m)\ .
\end{equation}
Using (\ref{Abelian-S-dual}) and $E_2(-1/\tau)=\tau^2
\left(E_2(\tau)+\frac{6}{\pi i\tau}\right)$, one obtains
\begin{equation}
  \tau^2f_{\rm anom}(\tau_D,m_D)-f_{\rm anom}(\tau,m)
  =\frac{Nm^2}{2}\left(\log(-i)-\frac{\pi i\tau_D}{6}
  +\frac{\pi i\tau}{6}\right)+\frac{N^3m^4}{48\pi i\tau}\ .
\end{equation}
Inserting this in (\ref{micro-dual}), one obtains
\begin{equation}\label{micro-dual-3}
  f(\tau_D,v_D,m_D)=\tau^{-2}f(\tau,v,\tau m_D)
  +\frac{1}{4\pi i\tau^3}\left(\frac{\partial f}{\partial v}\right)^2
  +\frac{Nm_D^2}{2}\left(\log(-i)+\frac{\pi i}{6\tau}
  +\frac{\pi i\tau}{6}\right)+\frac{N^3m_D^4}{48\pi i}\tau\ .
\end{equation}
We shall study it in the limit $\tau\rightarrow i\infty$ with $m_D$ and 
$v_D$ fixed.

The limit $\tau\rightarrow i\infty$ on the right hand side 
has to be understood with care, since $m=\tau m_D$
scales with $\tau$. Also, we should study how $v$ scales
with $\tau\rightarrow i\infty$, at fixed $v_D$.
Had $v,m$ not scaled with $\tau$, one would have naively
expected that the instanton corrections in $f$ would have been suppressed 
at $q\ll 1$, so that we could replace $f$ on the right hand side by $f_{\rm pert}$.
Let us check when this is correct. 
This expectation is correct if $F_{k}(v,m)$ does not scale to be larger
than $q^k$. From (\ref{Zk-Young}),
$F_k$ scales like $F_k\sim e^{\pm kNm}$ at $\pm{\rm Re}(m)\gg 1$. For
this factor to be smaller than $q^k$, one should require
$\left|{\rm Re}(\tau m_D)\right|<-\frac{2\pi i\tau}{N}$. Let us take 
$\tau$ to be purely imaginary for convenience (although most of our final results 
are valid for complex $\tau$). Then, $F_k$ can be ignored if
\begin{equation}\label{mass-requirement}
  \left|{\rm Im}(m_D)\right|<\frac{2\pi}{N}\ .
\end{equation}
When ${\rm Im}(m_D)$ reaches $\pm\frac{2\pi }{N}$,
we encounter a phase transition, beyond which one should make a new
$q$ expansion on the right hand side. The correct nature of this
phase transition will be commented on later. To make the
simplest calculus at $\tau_D\rightarrow i0^+$,
we take $m_D$ to satisfy (\ref{mass-requirement}). 

Let us also discuss how $v$ should scale at fixed $v_D$.
We shall first assume that $v$ is finite at finite $v_D$, and then show that 
it is consistent with ignoring $f_{\rm inst}$. If $f_{\rm inst}$ can be ignored, 
then the relation between $v$ and $v_D$ can be simplified as
\begin{equation}\label{v-relation}
  v=v_D-\frac{1}{2\pi i\tau}\frac{\partial f_{\rm pert}}{\partial v}
  \left(v,\tau m_D\right)\ .
\end{equation}
$\frac{\partial f_{\rm pert}}{\partial v}$ is given by
\begin{equation}\label{pert-root-sum}
  \frac{\partial f_{\rm pert}}{\partial v}=\sum_{\alpha\in\Delta}
  \alpha\left[{\rm Li}_2(e^{\alpha\cdot v})
  -\frac{1}{2}{\rm Li}_2(e^{\alpha\cdot v\pm m})\right]\ .
\end{equation}
Since we assume that $v$ is finite, the first term not containing $m$ 
yields a subleading contribution, from the $\frac{1}{\tau}\rightarrow 0$ factor 
in (\ref{v-relation}). To be definite, we take ${\rm Im}(m_D)<0$ so that 
${\rm Re}(m)\gg 1$. Then, 
\begin{eqnarray}
  {\rm Li}_2(e^{\alpha(v)+m})&=&-{\rm Li}_2(e^{-\alpha(v)-m})-\frac{\pi^2}{6}
  -\frac{1}{2}\left(\log(-1)+\alpha(v)+m\right)^2
\end{eqnarray}
where we used ${\rm Li}_2(e^x)+{\rm Li}_2(e^{-x})=-\frac{\pi^2}{6}
-\frac{1}{2}\left(\log(-e^x)\right)^2$ with the branch cut at 
$e^x\in (1,\infty)$. So one can approximate
\begin{equation}
  \frac{\partial f_{\rm pert}}{\partial v}\sim
  -\frac{1}{2}\sum_{\alpha\in\Delta}\alpha
  \left[-\frac{1}{2}(m+\alpha(v)+\log(-1))^2\right]\ ,
\end{equation}
where $-{\rm Li}_2(e^{-\alpha(v)-m})$ can be ignored at ${\rm Re}(m)\gg 1$.
We ignored all the terms that vanish after summing over $\alpha$, or are 
subleading in the $\frac{1}{\tau}\rightarrow 0$ limit. Expanding the square on the 
right hand side, the term proportional to $m^2=m_D^2\tau^2$ will vanish 
upon summing over $\alpha$. The next term proportional to $m\alpha(v)$ 
will be the nonzero leading term. One obtains
\begin{equation}
  v^i\approx v^i_D-\frac{m_D}{4\pi i}\sum_{\alpha\in\Delta}\alpha^i
  \alpha\cdot v=v^i_D-\frac{Nm_D}{2\pi i}(Pv)^i\ ,
\end{equation}
where we used
\begin{equation}
  \sum_{\alpha\in\Delta}\alpha\otimes\alpha=
  \sum_{i\neq j}(e^i-e^j)\otimes (e^i-e^j)=
  2(N-1){\bf 1}_{N\times N}-2\sum_{i\neq j}e^i\otimes e^j\equiv 2NP\ .
\end{equation}
Here, $P$ is the $N\times N$ projection 
to $SU(N)$. Decomposing $v=v_{U(1)}+v_{SU(N)}$, one finds
that
\begin{equation}
  v_{U(1)}=(v_{U(1)})_D\ ,\ \
  v_{SU(N)}\approx\frac{1}{1+\frac{Nm_D}{2\pi i}}(v_{SU(N)})_D
\end{equation}
at $\tau\rightarrow i\infty$.\footnote{At $m_D=-\frac{2\pi i}{N}$, 
one finds that $v_{SU(N)}$ diverges. In this case, one has to approximate 
(\ref{v-relation}) by assuming that $v$ can scale with $\tau$. At 
$m_D=-\frac{2\pi i}{N}$, we checked for $N=2,3$ that $v_{SU(N)}$ scales like 
$\sqrt{\tau}$, which grows large but is much smaller than $m=\tau m_D$. 
Due to this fact, $v$ does not affect the asymptotic free energy, and our final
result for $f_{\rm asymp}$ below will be reliable even at $m_D=-\frac{2\pi i}{N}$.}
Inserting this back to
$\frac{\partial f_{\rm pert}}{\partial v}$, one obtains
\begin{equation}
  \frac{\partial f_{\rm pert}}{\partial v^i}\approx
  -\frac{\tau m_D}{2}\sum_{\alpha\in\Delta}\alpha^i\alpha\cdot v
  =-N\tau m_D (v_{SU(N)})^i=-\frac{Nm_D}{1+\frac{Nm_D}{2\pi i}}\tau
  (v_{SU(N)})^i_D\ .
\end{equation}
Also $f_{\rm pert}$ itself is given by
\begin{equation}
  f_{\rm pert}=\sum_{\alpha\in{\bf adj}}
  \left({\rm Li}_3(e^{\alpha(v)})-\frac{1}{2}{\rm Li}_3(e^{\alpha(v)\pm m})\right)
  \approx\frac{1}{12}\sum_{\alpha\in{\bf adj}}m^3=\frac{N^2\tau^3m_D^3}{12}
\end{equation}
where we used ${\rm Li}_3(e^x)\approx-\frac{x^3}{6}-\frac{\pi ix^2}{2}
+\frac{\pi^2x}{3}$ if the real part of
$x$ is positive and large.

Therefore, the asymptotic prepotential is given by
\begin{eqnarray}
  f(\tau_D,v_D,m_D)&\rightarrow&\tau^{-2}f(\tau,v,\tau m_D)
  +\frac{1}{4\pi i\tau^3}\left(\frac{\partial f}{\partial v}\right)^2
  +\frac{\pi i Nm_D^2\tau}{12}+\frac{N^3m_D^4\tau}{48\pi i}\nonumber\\
  &\approx&\frac{N^2m_D^3\tau}{12}+\frac{\pi i Nm_D^2\tau}{12}+\frac{N^3m_D^4\tau}{48\pi i}\ .
\end{eqnarray}
In particular, one finds that the Coulomb VEV $v_D$ does not appear in
the asymptotic limit. This is natural since the Coulomb VEV is a dimensionful
parameter, which should not be visible in the large momentum limit.
This is a result for $-\frac{2\pi}{N}<{\rm Im}(m_D)<0$.
When $0<{\rm Im}(m_D)<\frac{2\pi}{N}$, all the analysis above is same except
the step of approximating $\frac{1}{\tau^2}f_{\rm pert}(v,\tau m_D)$. In this case,
$\frac{N^2m_D^3}{12}\tau$ is replaced
by $-\frac{N^2m_D^3}{12}\tau$. Combining the two cases, one obtains
\begin{equation}
  f_{\rm asymp}^{(\pm)}=-\frac{\pi^3i\tau}{3N}\left[\left(\frac{iNm_D}{2\pi}\right)^2
  \pm 2\left(\frac{iNm_D}{2\pi}\right)^3+\left(\frac{iNm_D}{2\pi}\right)^4\right]\ ,
\end{equation}
where $\pm$ signs are for $0<\pm{\rm Im}(m_D)<\frac{2\pi}{N}$, respectively. 
Finally, when ${\rm Im}(m_D)=0$, $\frac{1}{\tau^2}f_{\rm pert}(v,\tau m_D)$ 
provides subleading contribution in $\tau$ so that one finds
\begin{equation}
  f_{\rm asymp}^{(0)}=-\frac{\pi^3i\tau}{3N}\left[\left(\frac{iNm_D}{2\pi}\right)^2
  +\left(\frac{iNm_D}{2\pi}\right)^4\right]\ ,
\end{equation}
where the superscript $(0)$ means vanishing imaginary part of $m_D$. 
At this stage, we note that $f_{\rm asymp}$ at ${\rm Im}(m_D)\neq 0$ 
can be written as the following holomorphic function with a branch cut,
\begin{equation}\label{high-T-Li4}
  f_{\rm asymp}=-\frac{i\tau}{2\pi N}
  \left(2{\rm Li}_4(1)-{\rm Li}_4(e^{Nm_D})-{\rm Li}_4(e^{-Nm_D})\right)\ .
\end{equation}
This expression will be helpful later.

We first investigate $f_{\rm asymp}$ for purely imaginary
$m_D\equiv ix$, at $-\frac{2\pi}{N}<x<\frac{2\pi}{N}$. One finds
\begin{equation}\label{high-T-imaginary}
  f_{\rm asymp}=-\frac{\pi^3i\tau}{3N}
  \left[\left(\frac{Nx}{2\pi}\right)^2-2\left|\frac{Nx}{2\pi}\right|^3
  +\left(\frac{Nx}{2\pi}\right)^4\right]\ .
\end{equation}
The partition function undergoes a
phase transition at $x=0$, from certain perturbative 
particles being massless at $m_D\sim x=0$. One may wonder how $f_{\rm asymp}$ 
behaves beyond $x=\pm\frac{2\pi}{N}$. At $x=\pm\frac{2\pi}{N}$, one finds 
from the S-dual picture that $f_{\rm inst}$ cannot be ignored, 
since $F_k q^k\sim (e^{\pm Nm}e^{2\pi i\tau})^k\sim\mathcal{O}(1)$ at 
$m=\tau m_D\rightarrow\mp 2\pi i\tau$.
This means that particles with nonzero instanton number become
light at these points. One can get some insights on these nonperturbative 
massless particles.

\begin{figure}[t!]
  \begin{center}
    \includegraphics[width=13cm]{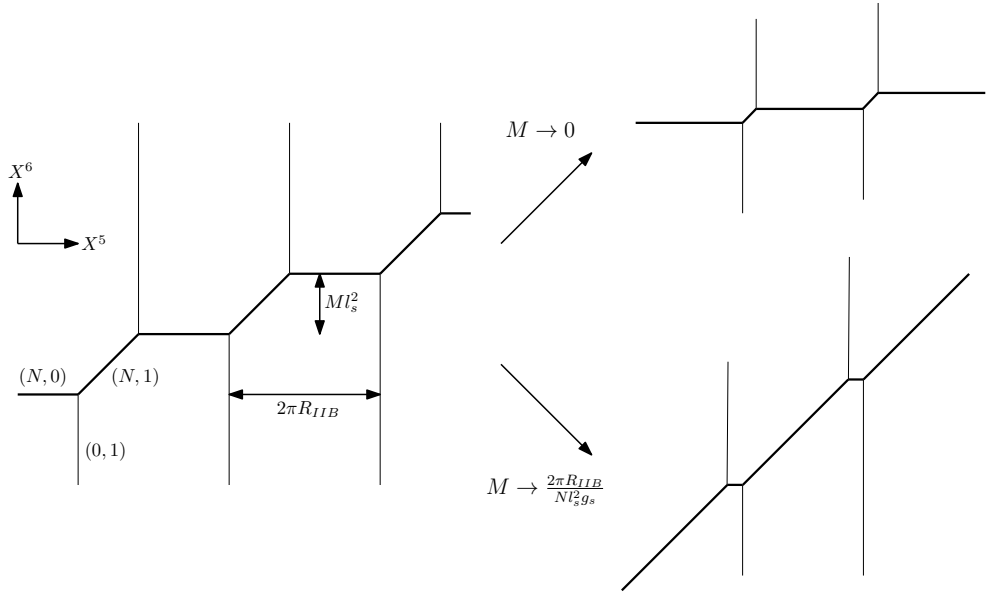}
\caption{Type IIB 5-brane web for the 5d 
$\mathcal{N}=1^\ast$ system}\label{web}
  \end{center}
\end{figure}
To see this, it is helpful to recall the type IIB 5-brane web realization of 
the 5d $\mathcal{N}=1^\ast$ system. More precisely, we realize the `S-dualized' 
setting at $\tau\rightarrow i\infty$, using weakly coupled type IIB string 
theory. The brane web first consists of $N$ D5-branes and $1$ NS5-brane, 
all extended on $01234$ directions, transverse to $789$, and forming a web on 
the $x^5$-$x^6$ plane. One makes a twisted compactification
$(x^5,x^6)\sim (x^5+2\pi R_{\rm IIB}, x^6+M\ell_s^2)$. The D5-branes 
wrap $x^5$ direction, and form a web with the NS5-brane extended along 
$x^6$, as shown in Fig.\ref{web}. The twisted compactification guarantees 
that the open strings with tension $\tau_{\rm F1}=\frac{1}{2\pi\alpha^\prime}$ 
(where $\alpha^\prime=\ell_s^2$) suspended between D5-branes across the web 
have mass $\frac{M}{2\pi}$. D1-branes wrapping $x^5$ ending on NS5-brane are 
identified as Yang-Mills instantons. Unit instanton's mass is given by
$\tau_{\rm D1}\cdot 2\pi R_{\rm IIB}=\frac{2\pi R_{\rm IIB}}{2\pi\alpha^\prime g_s}$, 
which should be identified with $\frac{1}{R^\prime}$ in our M5-brane setting.
So one obtains $2\pi R_{\rm IIB}=\frac{2\pi\alpha^\prime g_s}{R^\prime}$.
On the other hand, $\tau$ is given at zero axion by 
$\tau=\frac{i}{g_s}$, which should be identified in our M5-brane setting 
as $i\frac{R}{R^\prime}$. So one finds $g_s=\frac{R^\prime}{R}$. These 
provide the relations between the parameters $R_{\rm IIB}$, $g_s$ and 
the M5-brane parameters $R,R^\prime$. 
The slope of the $(N,1)$ 5-brane is 
$\frac{\Delta x^6}{\Delta x^5}=\frac{1}{g_s N}$, where $\Delta x^5$ and 
$\Delta x^6$ are the distances between the two ends of the segment 
on Fig.\ref{web}. We stated above that $\Delta x^6=M\alpha^\prime$, so
one finds $\Delta x^5=NMg_s\alpha^\prime$.

In this setting, the segment of $(N,1)$ 5-brane shrinks at $M=0$.
Here, the perturbative hypermultiplet particle becomes massless, corresponding 
to the fundamental strings connecting D5-branes across the NS5-brane.
This causes the so-called flop phase transition. The singular term 
proportional to $|x|^3$ in (\ref{high-T-imaginary}) is caused by $f_{\rm pert}$
in the S-dual setting, from the particles becoming massless at $M=0$.
So we conclude that the cusp $\propto|x|^3$ is due to the flop transition.

As one increases positive $M$, the next transition happens when
the $(N,1)$ brane segment goes around the circle in $x^5$ direction, as shown 
on the bottom-right side of Fig.\ref{web}. This happens at 
$\Delta x^5=NMg_s\alpha^\prime=2\pi R_{\rm IIB}$,
\begin{equation}
  NMg_s\alpha^\prime=2\pi R_{\rm IIB}=\frac{2\pi\alpha^\prime g_s}{R^\prime}
  =\frac{2\pi\alpha^\prime}{R}\ .
\end{equation}
So one finds that the transition happens at
\begin{equation}
  m_D=\frac{m}{\tau}=\frac{MR}{i/g_s}=-\frac{2\pi i}{N}\ ,
\end{equation}
precisely when $F_{\rm inst}$ cannot be ignored. 
Across $x=\pm\frac{2\pi}{N}$, i.e. $m=\mp\frac{2\pi i\tau}{N}$, 
the $N$ D5-brane segment shrinks. So across this value, another transition 
happens, with the D1-brane segment extended along the shrinking segment 
being massless. 

As one continues to change $M$, transitions due to non-perturbative massless 
particles will happen at $x=\frac{2\pi n}{N}$ with $n$ being integers. 
At $n=1,2,\cdots,N-1$, the nature of this transition is hard to study. 
This is because the massless particles are nontrivial bound states of 
D1-branes. Also, studying the $\tau\rightarrow i\infty$ approximations 
around $x=0$, not all massless particles were responsible for the cusp at 
$x=0$. So it will be important to know which
types of massless particles contribute to the cusp of $f_{\rm asymp}$ 
at $x=\frac{2\pi n}{N}$. However, if $n$ is a multiple of $N$, one finds 
from the 5-brane web diagram that the transition is an $SL(2,\mathbb{Z})$ 
transformation of the transition at $x=0$, so that the same type of cusp 
will happen. Indeed this has to be the case, since $x\sim x+2\pi$ 
(or $m_D\sim m_D+2\pi i$) is the periodicity of the instanton partition function.

\begin{figure}[t!]
  \begin{center}
    \includegraphics[width=12cm]{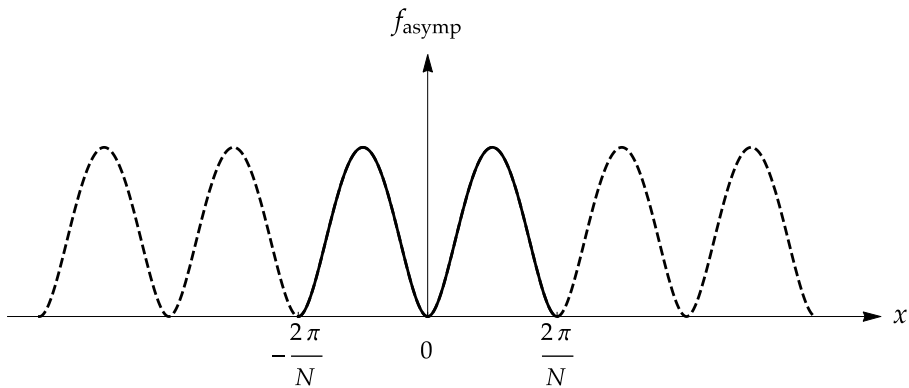}
\caption{Continuation of the asymptotic free energy
across flop transitions}\label{continuation}
  \end{center}
\end{figure}
Interestingly, if one takes the holomorphic extension (\ref{high-T-Li4}) within 
$|{\rm Im}(m_D)|<\frac{2\pi}{N}$
to the whole region of $m_D$, one gets a definite prediction on $f_{\rm asymp}$ 
as a function of real $x$, and also on the nature of phase transitions at 
all $n$. Plotting (\ref{high-T-Li4}) for the entire real $x$, one finds 
Fig.\ref{continuation}. $f_{\rm asymp}(x)$ for 
$\frac{2\pi n}{N}<x<\frac{2\pi(n+1)}{N}$ is given by simply translating 
the function in the range $0<x<\frac{2\pi}{N}$ by $\frac{2\pi n}{N}$. 
This means that all the cusp structures are completely the same at all 
$n$, at least in $f_{\rm asymp}(x)$. It will be interesting to understand 
how the non-perturbative massless particles cause the same cusp in 
(\ref{high-T-Li4}). Also, in (\ref{high-T-Li4}) or in Fig.\ref{continuation},
$f_{\rm asymp}$ has a shorter period $x\sim x+\frac{2\pi}{N}$. 
It will be interesting to see if the reduced period has to do with 
multiple-wrapping of M5-branes on $S^1$, analogous to the mutiple-winding 
fundamental strings \cite{Dijkgraaf:1996xw}.

Now we study $f_{\rm asymp}$ for purely real $m_D$.
The asymptotic free energy is given by
\begin{equation}
  -\log Z\sim\frac{f^{(0)}_{\rm asymp}}{\epsilon_1\epsilon_2}=
  \frac{i}{3\epsilon_1\epsilon_2\tau_D}
  \left[\frac{N^3m_D^4}{16\pi}-\frac{\pi Nm_D^2}{4}\right]\ .
\end{equation}
Holding real $m_D$ fixed, and further taking the large $N$ limit, 
one finds that the free energy is proportional to $N^3$. Namely, 
one finds that the single particle index $f(\tau,\epsilon_1,\epsilon_2,m,v)$ 
in the limit $\epsilon_{1,2}\rightarrow 0$, $\tau\rightarrow i0^+$ is 
given by
\begin{equation}
  \sum_{n=1}^\infty\frac{1}{n}f(n\tau,\epsilon_{1,2}=0,nm,nv)
  \rightarrow-\frac{i}{3\epsilon_1\epsilon_2\tau}
  \left[\frac{N^3m^4}{16\pi}-\frac{\pi Nm^2}{4}\right]\ ,
\end{equation}
where we dropped the $D$ subscripts. This shows that the microscopic 
entropy (with minus sign for fermions) of light D0-branes bound to $N$ 
D4-branes exhibit large number of bound states proportional to $N^3$.
The second term proportional to $N$ clearly comes from $N$ free tensor multiplets, 
as this comes from the S-dualization of $Nf_{U(1)}$. One can understand that 
the first term proportional to $N^3m^4$ is a remnant of the
cancelation between bosonic and fermionic states in the index, since 
this term vanishes at $m=0$. It will be interesting to guess what kind 
of index $f(\tau,\epsilon_{1,2}=0,m,v)$ would exhibit the above behavior 
in the high temperature limit. In particular, having the analytic expression 
(\ref{high-T-Li4}) given in terms of ${\rm Li}_4$ functions, with chemical 
potentials multiplied by $N$, it will be interesting to seek for an 
interpretation using multiple-wrapping of M5-branes, or instanton partons 
\cite{Collie:2009iz}.

Finally, we comment that one can obtain the asymptotic free energy 
at $\tau_D\rightarrow i0^+$ for all ADE theories, starting from (\ref{S-dual-ADE}) 
and following the analysis of this section. To make a similar calculation, 
one also needs to know 
the perturbative partition function, and the range of ${\rm Im}(m_D)$ in
which the instanton correction $f_{\rm inst}$ can be ignored on the right 
hand side. The perturbative prepotential is straightforward for all ADE. 
As for the instanton part, we should know when $F_k q^k$ is much smaller 
than $1$ at $q\rightarrow 0$ for large real part of $m=\tau m_D$.
The leading behavior of $F_k$ for large real $m$ can be easily inferred, 
by knowing the correct parameter scalings between the 5d $\mathcal{N}=1^\ast$ 
theory and the pure $\mathcal{N}=1$ theory. Namely, one finds
\begin{equation}
  F_kq^k\sim e^{kc_2 m}q^k\ ,
\end{equation}
where $c_2$ is the dual Coxeter number of the gauge group $G$. This 
is because the pure 5d $\mathcal{N}=1$ theory is obtained by taking 
the limit $m\rightarrow \infty$, $q\rightarrow 0$, with 
$\Lambda\sim e^{c_2 m}q$ held fixed. This means that 
one can ignore the instanton part in the region 
$-\frac{2\pi }{c_2}<{\rm Im}(m_D)<\frac{2\pi}{c_2}$.
By following the analysis for the $U(N)$ case,
the asymptotic free energies of ADE theories are given by
\begin{equation}\label{ADE-asymp}
  f_{\rm asymp}^{(\pm)}=-\frac{\pi^3i\tau}{3}
  \left[r\left(\frac{im_D}{2\pi}\right)^2
  \pm 2|G|\left(\frac{im_D}{2\pi}\right)^3
  +(c_2|G|+r)\left(\frac{im_D}{2\pi}\right)^4\right]\ ,
\end{equation}
where $\pm$ signs are for $0<\pm {\rm Im}(m_D)<\frac{2\pi}{c_2}$, and
\begin{equation}
  f_{\rm asymp}^{(0)}=-\frac{\pi^3i\tau}{3}
  \left[r\left(\frac{im_D}{2\pi}\right)^2
  +(c_2|G|+r)\left(\frac{im_D}{2\pi}\right)^4\right]
\end{equation}
for ${\rm Im}(m_D)=0$.

\subsection{Tests with $U(1)$ partition function}

We provide a small consistency check of $f_{\rm asymp}$ for the $U(1)$ case. 
By this exercise, one can also get better intuitions on the true nature of the 
approximations and phase transitions, which perhaps may be a bit obscure in 
our S-duality based approach.

In the previous S-duality based approach, we first took $\epsilon_1,\epsilon_2\rightarrow 0$
limit of the partition function, to focus on the prepotential only. Then using the
S-duality, we extracted out the $\beta\rightarrow 0$ asymptotics of the prepotential,
where $q=e^{2\pi i\tau}=e^{-\beta}$, at finite $m$ and $N$. We reconsider the same
limits directly with the $U(1)$ instanton partition function. The instanton 
partition function is given by 
\begin{equation}\label{U1-prepotential}
  Z_{\rm inst}=
  \exp\left[\sum_{n=1}^\infty\frac{1}{n}\frac{\sinh\frac{n(m\pm\epsilon_-)}{2}}
  {\sinh\frac{n\epsilon_1}{2}\sinh\frac{n\epsilon_2}{2}}\frac{e^{-n\beta}}
  {1-e^{-n\beta}}\right]\sim\exp\left[\frac{4}{\epsilon_1\epsilon_2}
  \sum_{n=1}^\infty\frac{\sinh^2\frac{nm}{2}}{n^3}\frac{e^{-n\beta}}{1-e^{-n\beta}}\right]
\end{equation}
in the $\epsilon_1,\epsilon_2\rightarrow 0$ limit. Now we take
the $\beta\rightarrow 0$ limit at fixed $m$. This is somewhat tricky at real $m$,
which we also take to be positive. This is because the above formula is valid
for $m<\beta$ when $m$ is real. Physically,
this is because the partition function $Z$ has poles at
$m=n\beta$ for all positive integers $n$. So with fixed real $m$, one would hit
many poles as one takes the $\beta\rightarrow 0$ limit.
To deal with this situation more easily, we first continue $m$ to be 
purely imaginary, $m=ix$, and continue back later to complex $m$.

Inserting $m=ix$ and taking $\beta\rightarrow 0$ limit, one obtains
\begin{eqnarray}\label{U1-Li4}
  Z&\sim&\exp\left[-\frac{4}{\epsilon_1\epsilon_2\beta}
  \sum_{n=1}^\infty\frac{\sin^2\frac{nx}{2}}{n^4}\right]=
  \exp\left[\frac{1}{\epsilon_1\epsilon_2\beta}
  \sum_{n=1}^\infty\frac{1}{n^4}\left(e^{inx}+e^{-inx}-2\right)\right]\nonumber\\
  &=&\exp\left[\frac{1}{\epsilon_1\epsilon_2\beta}
  \left({\rm Li}_4(e^{ix})+{\rm Li}_4(e^{-ix})-2{\rm Li}_4(1)\right)\right]\ .
\end{eqnarray}
The final expression can be 
continued to complex $x$. Here, we use the property
\begin{equation}
  {\rm Li}_n(e^{2\pi ix})+(-1)^n{\rm Li}_n(e^{-2\pi ix})=
  -\frac{(2\pi i)^n}{n!}B_n(x)\ ,
\end{equation}
where $0\leq {\rm Re}(x)<1$ for ${\rm Im}(x)\geq 0$. $B_n(x)$ are the Bernoulli
polynomials, given by
\begin{equation}
  \frac{te^{xt}}{e^t-1}=\sum_{n=0}^\infty B_n(x)\frac{t^n}{n!}\ .
\end{equation}
In particular, one finds $B_4(x)=\frac{1}{30}\left(-1+30x^2-60x^3+30x^4\right)$, so
that
\begin{equation}
  {\rm Li}_4(e^{ix})+{\rm Li}_4(e^{-ix})=
  -\frac{(2\pi)^4}{24}B_4(x/2\pi)=\frac{2\pi^4}{90}
  -\frac{2\pi^4}{3}\left[\left(\frac{x}{2\pi}\right)^2
  -2\left(\frac{x}{2\pi}\right)^3+\left(\frac{x}{2\pi}\right)^4\right]
\end{equation}
for $0\leq x<2\pi$. This leads to the asymptotic formula
\begin{equation}
  -\log Z\sim\frac{2\pi^4}{3\epsilon_1\epsilon_2\beta}
  \left[\left(\frac{x}{2\pi}\right)^2
  -2\left(\frac{x}{2\pi}\right)^3+\left(\frac{x}{2\pi}\right)^4\right]
\end{equation}
for $0\leq x< 2\pi$, which is in complete agreement with the S-duality-based result,
(\ref{high-T-imaginary}), upon inserting $-i\tau\rightarrow\frac{2\pi}{\beta}$
and $N=1$. When $-2\pi <x\leq 0$, we use a different identity of ${\rm Li}_4$ 
function to find a similar expression, with the sign of the 
$\mathcal{O}(x^3)$ term flipped. This also shows that
the continuation (\ref{high-T-Li4}) beyond $-2\pi<x<2\pi$ by the 
${\rm Li}_4$ functions to complex $x$ is indeed correct.

\subsection{6d chiral anomalies on Omega-deformed $\mathbb{R}^4\times T^2$}

In this subsection, we shall discuss the connection between the S-duality 
anomaly and the 6d chiral anomalies of global symmetries. In particular, 
we shall independently compute some part of our asymptotic free energy 
$f_{\rm asymp}$ based on chiral anomaly only. However, let us start by 
giving a general comment, on why one should naturally expect S-duality 
anomaly of the partition function if the system has chiral anomaly.

Consider a partition function of even dimensional chiral theories on $T^2$, like
2d theories on $T^2$ or our system on $\mathbb{R}^4\times T^2$. For a global 
symmetry, one
turns on a background gauge field $A$. In particular, let us turn on the flat 
connection of $A$ on $T^2$. We shall only be interested in Abelian flat connections, 
characterized by the commuting holonomies along the two circles of $T^2$. 
Large gauge transformations would have made both holonomies to be periodic, 
had there been no chiral anomalies. For simplicity, let us take a rectangular 
torus with two radii $r_1$, $r_2$, respectively. Then the large gauge 
transformations would
have given the periods $A_1\sim A_1+\frac{1}{r_1}$ and $A_2\sim A_2+\frac{1}{r_2}$.
With matter fields having integral charge $q$ of this global symmetry, the modes 
of these fields would have frequencies 
$(\omega_1,\omega_2)=(\frac{n_1}{r_1}+qA_1,\frac{n_2}{r_2}+qA_2)$ on $T^2$, 
with integral $n_1,n_2$, which is invariant under the periodic shifts of 
$A_1$, $A_2$. This is a consequence of these gauge symmetries.
However, in quantum observables like the partition function on $T^2$,
these large gauge transformations may fail to be symmetries for 
theories with chiral anomalies. This is because one has to regularize 
the path integral over these modes, by regarding one of the two 
directions as temporal circle \cite{Assel:2015nca}. By this procedure, 
one of the two holonomies $A_1,A_2$ fail to be periodic in the partition 
functions. This is precisely what happen for the 2d elliptic
genera \cite{Benini:2013xpa}. We expect that similar things will happen 
to 6d chiral theories on $\mathbb{R}^4\times T^2$, but we cannot make 
this expectation more precise here. We shall simply assume the failure of 
double periodicity of background holonomies due to chiral anomalies, 
and then explain that it forces the partition function to have S-duality 
anomaly, as we found in section 2 by nonzero $F_{\rm anom}$.

Let us write the background holonomies as a complex number $m$. 
Had a free energy $F(\tau,m)$ on $T^2$ been exactly S-dual, 
then its exact S-duality $F(-1/\tau,m/\tau)=F(\tau,m)$ means that 
$m$ has double period. This is because if the right hand side has period 
in one direction, say $F(\tau,m)=F(\tau,m+1)$, the left 
hand side forces $F(-\frac{1}{\tau},\frac{m}{\tau})
=F(-\frac{1}{\tau},\frac{m}{\tau}+\frac{1}{\tau})$, and thus
$F(\tau,m)=F(\tau,m-\tau)$, contradicting the obstruction of 
double periodicity from chiral anomaly. This comment 
applies to our 6d partition functions. So we naturally expect  
S-duality anomaly. 

With these motivations in mind, rather than trying to elaborate on 
it, we shall make a concretely calculation which shows that a 
particular term in our asymptotic high temperature free energy dictated 
by $F_{\rm anom}$ can be computed using 6d chiral anomaly only.

Let us first explain the anomalies of the 6d $(2,0)$ theory of 
$A_{N-1}$ type. More precisely, we shall
consider the anomaly of the interacting $A_{N-1}$ type theory times a decoupled
free self-dual tensor multiplet theory. This corresponds to the system of $N$
M5-branes including the decoupled center-of-mass multiplet. 
The anomaly polynomial $8$-form is given by
\begin{equation}\label{anomaly-U(N)}
  I_8=NI_8(1)+N(N^2-1)\frac{p_2(N)}{24}
\end{equation}
where $I_8(1)$ is the anomaly of the single M5-brane theory, or one
free $(2,0)$ tensor multiplet,
\begin{equation}
  I_8(1)=\frac{1}{48}\left[p_2(N)-p_2(T)+\frac{1}{4}\left(p_1(T)-p_1(N)\right)^2\right]\ .
\end{equation}
The Pontryagin classes are defined by
\begin{eqnarray}
  p_1
  =-\frac{1}{2(2\pi)^2}{\rm tr}R^2\ \ ,\ \ \
  p_2
  =\frac{1}{(2\pi)^4}\left[-\frac{1}{4}{\rm tr}R^4+\frac{1}{8}({\rm tr}R^2)^2\right]\ .
\end{eqnarray}
Here, traces are acting on either $6\times 6$ matrices for 
$SO(5,1)$ tangent bundle $T$, or $5\times 5$ matrices for $SO(5)$ 
normal bundle $N$.
Taking their curvatures to be $R$ and $F$, respectively, one finds
\begin{eqnarray}
  (2\pi)^4I_8&=&\frac{N}{48}\left[-\frac{1}{4}{\rm tr}F^4+\frac{1}{8}({\rm tr}F^2)^2
  +\frac{1}{4}{\rm tr}R^4-\frac{1}{8}({\rm tr}R^2)^2
  +\frac{1}{16}({\rm tr}R^2-{\rm tr}F^2)^2\right]\nonumber\\
  &&
  +\frac{N^3-N}{24}\left(-\frac{1}{4}{\rm tr}F^4+\frac{1}{8}({\rm tr}F^2)^2\right)\ .
\end{eqnarray}
We shall restrict $F$ to a Cartan part. In particular, since we shall be taking the Omega backgrounds to be small,
the Cartan for $SU(2)_R$ will have much smaller background field than
$SU(2)_L$, from $\epsilon_+\ll m$. So we shall only turn on the background
field for the Cartan in $SU(2)_L\subset SO(5)$,
corresponding to our $\mathcal{N}=1^\ast$ mass $m$. $F$ 
is a $5\times 5$ matrix-valued 2-form, whose components are $F^{ab}=-F^{ba}$
with $a,b=1,\cdots,5$. The component corresponding to the Cartan of $SU(2)_L$
is obtained by keeping $F^{12}=-F^{21}=-F^{34}=F^{43}\equiv F$ only.
With this restriction, one finds ${\rm tr}(F^2)\rightarrow -4F^2$,
${\rm tr}(F^4)\rightarrow 4F^4$.
Inserting these, the $SO(5,1)$ and $U(1)\subset SU(2)_L$
anomalies are given by
\begin{equation}\label{anomaly-pol-reduced}
  (2\pi)^4I_8\rightarrow \frac{N^3}{24}F^4
  +\frac{N}{48}\left[\frac{1}{2}F^2 {\rm tr}R^2+\frac{1}{4}{\rm tr}R^4
  -\frac{1}{8}({\rm tr}R^2)^2\right]\ .
\end{equation}
Only the first term $\frac{N^3}{24}F^4$ will be relevant for the 
computations below.

Our goal is to compute some part of the asymptotic
free energy at high temperature $\tau_D\rightarrow 0$, using 6d chiral 
anomalies. Recall that we found 
\begin{equation}\label{Seff}
  S_{\rm eff}=-\log Z\rightarrow\frac{f_{\rm asympt}}{\epsilon_1\epsilon_2}
  =\frac{i}{2^4\cdot 3\pi\epsilon_1\epsilon_2\tau_D}
  \left[N^3m^4-4\pi^2Nm^2+\cdots\right]
\end{equation}
where $\cdots$ stands for the $m^3$ term which exists when $m$ has 
imaginary component. The $m^3$ term will not be of our interest in this 
subsection. We obtained this expression at $\epsilon_{1,2}\ll 1$ and 
$\tau_D\rightarrow 0$,
where $\tau_D\equiv\frac{\beta}{4\pi}(\mu+i)$ is the same $\tau_D$ used
before. Often, we used purely imaginary $\tau_D$ with
$\mu=i$, but we keep real $\mu$ in this subsection to see a clear
relation to chiral anomalies. For a reason to be explained below, 
we would like to study the asymptotic free energy when all the 
chemical potentials $\epsilon_{1,2},m$ are purely imaginary.
So inserting $i\epsilon_{1,2}$, $im$ (with real $\epsilon_{1,2},m$) 
in the places of $\epsilon_{1,2},m$ in (\ref{Seff}), one obtains 
$S_{\rm eff}=-\frac{i}{2^4\cdot 3\pi\epsilon_1\epsilon_2\tau_D}
\left[N^3m^4+4\pi^2Nm^2+\mathcal{O}(m^3)\right]$. In this setting, 
we focus on the imaginary part of the effective action,
\begin{equation}\label{imaginary}
  {\rm Im}(S_{\rm eff})=
  -\frac{\mu}{12\epsilon_1\epsilon_2\beta(1+\mu^2)}
  \left[N^3m^4+4\pi^2Nm^2+\mathcal{O}(m^3)\right]\ ,
\end{equation}
and compute it from 6d chiral anomalies. Especially, we shall compute
part of ${\rm Im}(S_{\rm eff})$ from the 5d effective action approach
for the 6d theory on small temporal circle. 6d chiral anomaly determines
a special class of terms in the 5d effective action.
It turns out that, knowing the terms determined by anomaly, one can only
compute the term proportional to $m^4$. So we shall pay attention
to the first term
\begin{equation}\label{imaginary-normal}
  \left.\frac{}{}\!\!{\rm Im}(S_{\rm eff})\right|_{m^4}=
  -\frac{\mu N^3m^4}{12\epsilon_1\epsilon_2\beta(1+\mu^2)}\ .
\end{equation}
We shall argue below that this term is completely dictated by
6d chiral anomaly, and then we re-compute this term
using chiral anomaly only. This will provide another strong test of
our findings from the D0-D4 calculus. Then, since one naturally expects 
that supersymmetrization of (\ref{imaginary-normal}) is holomorphic in 
$\tau_D$, one can reconstruct the term
$-\frac{iN^3m^4}{2^4\cdot 3\pi\epsilon_1\epsilon_2\tau_D}$ in
(\ref{Seff}).

We shall consider the 6d anomaly from the viewpoint of 5d effective action,
obtained by compactification on a small circle of circumference 
$\beta\ll 1$, and discuss our asymptotic free energy $f_{\rm asympt}$ on 
$\mathbb{R}^4_{\epsilon_{1,2}}\times T^2$ in this setting.
On $T^2$, regarding one circle as the temporal circle, the partition function
is an index of the form
\begin{equation}
  Z(\tau,v,m,\epsilon_{1,2})={\rm Tr}\left[(-1)^F
  e^{-\frac{\beta}{2}(H-i\mu P)}e^{\sum_{a=1}^2
  \epsilon_a(J_a+J_R)}e^{2mJ_L}e^{-v_iq_i}\right]\ .
\end{equation}
Real $\epsilon_{1,2},m$ is consistent with the conventions for the 
partition function presented at the beginning of section 2. In this setting, 
the chemical potentials
$\epsilon_{1,2},m$ will twist the translation on the temporal circle in 
a way that the twisted time evolution is not unitary (simply because
the factors in the trace are not unitary transformations). 
This would cause a complex deformation of the Euclidean action by twisting 
with chemical potentials.\footnote{Strictly speaking,
Lagrangian formulation is not known in 6d. So when we refer to a Lagrangian
description, we mean a 5d Lagrangian after reducing
on a small circle. See also comments in \cite{DiPietro:2014bca} 
concerning the conversion between twistings and background gauge fields in 
the presence of anomalies.}
For a technical reason, it will be convenient to keep these twistings to 
preserve the reality of the action. So we replace 
\begin{equation}
  e^{\sum_{a=1}^2
  \epsilon_a(J_a+J_R)}e^{2mJ_L}e^{-v_iq_i}\rightarrow 
  e^{i\sum_{a=1}^2\epsilon_a(J_a+J_R)}e^{2imJ_L}e^{-iv_iq_i}\ ,
\end{equation}
which will make real twists of the Euclidean action. This is equivalent to 
the insertions of $i\epsilon_{1,2},im$ around (\ref{imaginary-normal}).
The factor $e^{-\frac{\beta}{2}H}$ demands us to consider
a 6d Euclidean theory whose temporal coordinate $y$ satisfies
periodicity $y\sim y+\frac{\beta}{2}$. This forms a circle of the $T^2$. 
Another circle factor is labeled by $x$, which we take to
have periodicity $x\sim x+2\pi$. 
Defining $\tau_D=\frac{\beta}{4\pi}(\mu+i)$, one obtains
\begin{equation}
  e^{-\frac{\beta}{2}(H-i\mu P)}\equiv e^{2\pi i\tau_D\frac{H+P}{2}}
  e^{-2\pi i\bar\tau_D\frac{H-P}{2}}
  =e^{-2\pi {\rm Im}\tau_D H+2\pi i{\rm Re}\tau_D P}\ .
\end{equation}
So $\tau_D$ is the complex structure of $T^2$. This torus is
endowed with the metric
\begin{equation}
  ds^2(T^2)=(dx-\mu dy)^2+dy^2
\end{equation}
and periods $(x,y)\sim (x+2\pi,y)\sim
(x+\frac{\beta\mu}{2},y+\frac{\beta}{2})$. Including the
chemical potential $\epsilon_a$, the metric of
$\mathbb{R}^4\times T^2$ is given by
\begin{equation}\label{6d-metric}
  ds^2(\mathbb{R}^4\times T^2)=\sum_{a=1,2}
  \left|dz_a-\frac{2i\epsilon_a}{\beta}z_ady\right|^2
  +(dx-\mu dy)^2+dy^2\ ,
\end{equation}
where $z_a$ are complex coordinates of $\mathbb{C}^2\sim\mathbb{R}^4$
with charges $J_a[z_b]=\delta_{ab}$.
Finally, the chemical potential $m$ is realized as the background
gauge field $A=\frac{2m}{\beta}dy$ for $U(1)\subset SU(2)_L$.
Also, $\frac{H-P}{2}\sim\{Q,\bar{Q}\}$, where $Q$ is a supercharge 
preserved by the index. So $Z$ is independent of $\bar\tau$.

Following \cite{DiPietro:2014bca} (see also \cite{Banerjee:2012iz}),
we shall make a KK reduction on the small circle along $y$,
for small inverse-temperature $\beta\ll 1$. To this end,
one rewrites the background in the form of
\begin{equation}
  ds^2=e^{2\phi}(dy+a)^2+h_{ij}dx^idx^j\ ,
\end{equation}
where $h_{ij}$ with $i,j=1,\cdots 5$ is the 5d metric,  $e^{2\phi}=1+\mu^2+\frac{4}{\beta^2}\sum_a\epsilon_a^2|z_a|^2$ is the
dilaton, and
\begin{equation}
  a=\frac{1}{1+\mu^2+\frac{4\epsilon_a^2|z_a|^2}{\beta^2}}
  \left(-\mu dx-\frac{2\epsilon_a|z_a|^2}{\beta}d\phi_a\right)
\end{equation}
is the gravi-photon field, where $z_a=|z_a|e^{i\phi_a}$.
The 6d background gauge field $A$ for $U(1)\subset SU(2)_L$
is also rewritten in the form $A=A_6(dy+a)+\mathcal{A}$, where $\mathcal{A}$ is
the 5d background gauge field and $A_6$ is the 5d scalar. So one finds
$A_6=\frac{2m}{\beta}$ and $\mathcal{A}=-A_6 a$.

If the 6d theory compactified on a small circle has no 5d massless modes, 
one can express the thermal partition function in terms
of a 5d local effective field theory of background fields, where the
5d derivative expansion corresponds to a  $\beta$ series expansion. 
As noted in \cite{DiPietro:2014bca}, with massless
modes in 5d, there could be nonlocal part of  
the effective action which is smooth in the $\beta\rightarrow 0$ limit.
In our case, the non-local part comes from the 5d perturbative maximal 
SYM. There is additional difficulty in using 
the derivative expansion in our setting, since some of our background
fields are proportional to $\beta^{-1}$, which may spoil the orderings 
provided by the derivative expansion. So it appears tricky to
directly employ the formalism of \cite{DiPietro:2014bca,Banerjee:2012iz}.

However, one can study the imaginary part (\ref{imaginary}) of our 
asymptotic free energy using the 5d approach. The imaginary part can be 
computed completely by knowing the 5d Chern-Simons like 
terms. To explain this, note first that we have been careful to set 
all our background fields to be real, e.g. by setting our chemical 
potentials to be imaginary. With real background fields turned on, suppose 
that we first reduce the 6d theory on a small circle to a general 5d
\textit{Lorentzian} spacetime. Then the 5d effective action is real, 
since Hermiticity is not broken in the Lorentzian
theory. Now we Wick-rotate the `time' direction in this 5d setting. Since 
all background fields are real, the only possible 
step which may cause complex effective action is the Wick rotation
to Euclidean 5d space. Here, note that we are seeking for an effective action 
of the vectors $a,\mathcal{A},\omega$ (spin connection), tensor $h_{ij}$, and 
scalars $A_6$, $\phi$.
To compute the imaginary part, one can focus on the local terms. 
This is because the nonlocal terms come from the determinant of 5d maximal SYM whose
fields are covariantized by real background fields, which is real.
Among the local terms obtained from scalar Lagrangian density, 
we should seek for terms containing the tensor $\epsilon^{ijklm}$ to 
obtain imaginary contribution after Wick rotation. It should be contracted 
with antisymmetric tensors formed
by the background fields. There are many possibilities, arranged in 
derivative expansion. For instance, there could complicated terms like 
$\sim da\wedge d\mathcal{A}\wedge d\phi f(\phi,A_6)$, and so on.

Although there are many terms, let us comment that there 
can be gauge invariant terms and gauge non-invariant terms in the imaginary 
action. The latter class should exist because the 5d effective action should 
realize 6d chiral anomalies. The coefficients of the terms in the latter class 
are thus completely determined by known 6d anomalies 
\cite{DiPietro:2014bca,Banerjee:2012iz}. Among the gauge
invariant terms, there can be action coming from gauge invariant Lagrangian 
density, like the term that we illustrated in the last paragraph. Finally, 
there may be Chern-Simons terms in which Lagrangian densities are not gauge 
invariant but their integrals are.
So the imaginary action takes the following structure,
\begin{eqnarray}\label{5d-derivative}
  S_{\rm CS}&=&S_{\rm CS}^{(1)}+S_{\rm CS}^{(2)}+S_{\rm GI}\\
  S_{\rm CS}^{(1)}&=&\frac{i\kappa_1}{\beta^3}\int a\wedge da\wedge da+
  \frac{i\kappa_2}{\beta}\int \mathcal{A}\wedge d\mathcal{A}\wedge da
  +\frac{i\kappa_3}{\beta}\int a\wedge R\wedge R+i\kappa_4\int
  \mathcal{A}\wedge d\mathcal{A}\wedge d\mathcal{A}+\cdots\nonumber\\
  S_{\rm CS}^{(2)}&=&-\frac{iDr_1}{96\pi^2}\int \left(A_6^4a\wedge da\wedge da
  +4A_6^3\mathcal{A}\wedge da\wedge da
  +6A_6^2\mathcal{A}\wedge d\mathcal{A}\wedge da+4A_6\mathcal{A}\wedge d\mathcal{A}\wedge\mathcal{A}\right)+\cdots\ ,\nonumber
\end{eqnarray}
where $r_1=\frac{\beta}{4\pi}$ is the radius of the small sixth circle
with circumference $\frac{\beta}{2}$. $S_{\rm CS}^{(1)}$ consists of the
gauge invariant Chern-Simons terms. $S_{\rm CS}^{(2)}$ is
part of the gauge non-invariant Chern-Simons terms that comes from
$U(1)\subset SU(2)_L\subset SO(5)_R$
normal bundle anomaly in 6d, namely the first term $\sim\frac{N^3}{24}F^4$ of
(\ref{anomaly-pol-reduced}). Anomaly matching fixes $D=N^3$, as well as
the relative coefficients as shown on the second line.\footnote{Following
\cite{DiPietro:2014bca}, we show the form of the action with constant
value of $A_6$, taking into account the covariant anomaly rather than
the consistent anomaly. This is sufficient for our calculus of the free energy.}
The omitted terms $\cdots$ in $S^{(1)}_{\rm CS}$ are other Chern-Simons terms 
containing $\omega$, which we do not need here. The omitted terms
in $S_{\rm CS}^{(2)}$ can all be computed from mixed anomalies and gravitational
anomalies of (\ref{anomaly-pol-reduced}), which we do not work out here as
we shall not need them. Finally, $S_{\rm GI}$ is the action containing 
$\epsilon^{ijklm}$ associated with gauge invariant Lagrangian density, e.g.
$da\wedge da\wedge d\phi f(\phi,A_6)$, $d\mathcal{A}\wedge d\mathcal{A}\wedge 
d[(da)^{ij}(d\mathcal{A})_{ij}]g(\phi,A_6)$, and so on. One point we emphasize is 
that $S_{\rm GI}$ can come in infinite series of derivative expansion, while 
$S^{(1)}_{\rm CS}$ and $S^{(2)}_{\rm CS}$ consist of finite number of terms 
and can be completely classified.

The imaginary terms have rich possibilities.
Here we consider the terms which are nonzero 
with our background, and also the leading terms
in small $\epsilon_{1,2}$, proportional to 
$\frac{1}{\epsilon_1\epsilon_2}$. $A_6=\frac{2m}{\beta}$ 
is constant in our background. Also, $\mathcal{A}=-A_6a$ is constant 
times the graviphoton. Plugging in these values, one obtains
\begin{equation}\label{imaginary-reduced}
  (A_6)^n\epsilon^{ijklm}(\textrm{rank 5 antisymmetric tensor of }
  a,\phi,\omega,h)\ .
\end{equation}
The parenthesis consists of the fields reduced from 6d metric (\ref{6d-metric}). 
Note that, after plugging in constant $A_6$ and $\mathcal{A}=-A_6a$, all terms 
should be formally gauge invariant in the remaining fields. This is because 
the only possible gauge non-invariant terms $S^{(2)}_{\rm CS}$, completely dictated 
by anomaly, also become gauge invariant like 
$A_6^4a\wedge da\wedge da$ with constant $A_6$.

Now we note the fact that, in the 6d metric, all $z_a$ coordinates of 
$\mathbb{R}^4$ are multiplied by $\epsilon_a$. So in the parenthesis of 
(\ref{imaginary-reduced}), the only $z^a$'s not associated with $\epsilon_a$ 
are derivatives. So one makes a formal derivative expansion of this term, 
assigning the `mass dimensions' $[a]=0$, $[\phi]=0$, $[h]=0$, $[\omega]=1$.
The lowest order term comes in two derivatives, and is proportional to 
$a\wedge da\wedge da$. There can be no other gauge-invariant terms 
at this order. This term indeed yields the desired $\frac{1}{\epsilon_1\epsilon_2}$ 
scaling. Firstly, the integral $dx d^2z_1d^2z_1$ can be scaled into 
$(\frac{\beta^2}{\epsilon_1\epsilon_2})^2$ times a measure depending on 
$\frac{\epsilon_az_a}{\beta}$. Also, two derivatives in $a\wedge da\wedge da$ 
can also be scaled with $\frac{\epsilon_a}{\beta}$, yielding another overall 
factor $\frac{\epsilon_1\epsilon_2}{\beta^2}$. $z_a$ in the remaining
integral appear in the combination $\frac{\epsilon_az_a}{\beta}$, including 
the integration variable, so is independent of $\epsilon_a$. So this 
term yields the right scaling $\sim\frac{1}{\epsilon_1\epsilon_2}$. Therefore, 
to compute (\ref{imaginary-normal}), we only need to consider those terms 
that reduce to 
\begin{equation}\label{CS-leading}
  (A_6)^n a\wedge da\wedge da
\end{equation} 
upon plugging in our background.
This implies that one does not have to consider $S_{\rm GI}$ of 
(\ref{5d-derivative}), since they are associated with local Lagrangian 
density and cannot provide terms like (\ref{CS-leading}).
 
So we only consider $S^{(1)}_{\rm CS}$ and $S^{(2)}_{\rm CS}$ of (\ref{5d-derivative}).
Unlike the coefficients of $S^{(2)}_{\rm CS}$, 
coefficients of $S^{(1)}_{\rm CS}$ cannot be determined with
our limited knowledge of the 6d theory. 
So even after restricting our interest to
the imaginary part (\ref{imaginary}) of the effective action, we cannot 
compute them all due to our ignorance on these coefficients. Since 
the second term of $S_{\rm CS}^{(1)}$ is quadratic in $\mathcal{A}$, 
we cannot compute the $\mathcal{O}(m^2)$ term of (\ref{imaginary}). 
This is why we shall not need the mixed anomaly contributions 
in $S_{\rm CS}^{(2)}$ coming from the term 
$\sim F^2{\rm tr}R^2$ in (\ref{anomaly-pol-reduced}), which will 
also yield a contribution at $\mathcal{O}(m^2)$, since knowing them is incomplete 
to compute the whole $\mathcal{O}(m^2)$ contributions. 
Also, the $\mathcal{O}(m^3)$ term cannot be computed since we do not know
$\kappa_4$. However, the Chern-Simons terms that are quartic in $\mathcal{A}$ 
and $A_6$ are completely dictated by 6d anomalies, as shown on the second line of 
(\ref{5d-derivative}). Note that quartic Chern-Simons term is allowed 
precisely because we allow gauge non-invariant Chern-Simons term, to 
match 6d anomalies which are fourth 
order in the fields. Thus, we can compute (\ref{imaginary-normal}) from 
$S_{\rm CS}^{(2)}$ of (\ref{5d-derivative}).
Note also that, for imaginary chemical potentials, we have found 
earlier in this section that $f_{\rm asymp}$ undergoes phase transitions 
due to massless particles. This only changes $\mathcal{O}(m^3)$ or lower 
order terms, so that the $m^4$ order that we are going to compute 
is unaffected.

We also note in passing that, we can turn the logic around 
and use our D0-D4 results to constrain the 5d effective action. Namely, we know 
from our D0-D4 calculus the $\mathcal{O}(m^2)$ and $\mathcal{O}(m^3)$ 
coefficients of ${\rm Im}(f_{\rm asymp})$, and also the vanishing of the 
$\mathcal{O}(m^0)$ coefficient. This knowledge can be used to constrain 
$\kappa_1,\kappa_2,\kappa_3,\cdots$ of (\ref{5d-derivative}).
This information may be useful for studying other high temperature partition 
functions of the 6d $(2,0)$ theories.

Coming back to the computation of (\ref{imaginary-normal}), we plug $\mathcal{A}=-A_6da$ and $A_6=\frac{2m}{\beta}$ into
$S_{\rm CS}^{(2)}$ of (\ref{5d-derivative}) to obtain
\begin{equation}
  \frac{iN^3(A_6)^4r_1}{96\pi^2}\int a\wedge da\wedge da\ .
\end{equation}
To compute this, one should evaluate the gravi-photon Chern-Simons term,
\begin{equation}\label{adada}
  \int a\wedge da\wedge da=
  \int\left(1+\mu^2+\frac{4\epsilon_a^2|z_a|^2}{\beta^2}\right)^{-3}
  \left(-\mu dx\right)\wedge 2\frac{4\epsilon_1\epsilon_2}{\beta^2}
  4dx_1\wedge dy_1\wedge dx_2\wedge dy_2
\end{equation}
where $z_a\equiv x_a+iy_a$, with $x_1,y_1,x_2,y_2$ being the Cartesian coordinates 
of $\mathbb{R}^4$. Since $\int dx=2\pi$, $\int dx_ady_a=\pi\int d(r_a^2)$,
(\ref{adada}) becomes
\begin{equation}
  -\frac{64\pi^3\mu\epsilon_1\epsilon_2}{\beta^2}
  \int_0^\infty \frac{d(r_1^2)d(r_2^2)}
  {\left(1+\mu^2+\frac{4\epsilon_a^2r_a^2}{\beta^2}\right)^3}
  =-\frac{4\pi^3\mu\beta^2}{\epsilon_1\epsilon_2}\int_0^\infty \frac{dXdY}{(1+\mu^2+X+Y)^3}=-\frac{2\pi^3\mu\beta^2}{(1+\mu^2)
  \epsilon_1\epsilon_2}\ ,
\end{equation}
where $X=\frac{4\epsilon_1^2r_1^2}{\beta^2}$, $Y=\frac{4\epsilon_2^2r_2^2}{\beta^2}$.
So one obtains
\begin{equation}
  \frac{iN^3(A_6)^4r_1}{96\pi^2}\int a\wedge da\wedge da
  =-i\frac{N^3\beta}{3\cdot 2^7\pi^3}\cdot\frac{16m^4}{\beta^4}\cdot
  \frac{2\pi^3\mu\beta^2}{(1+\mu^2)\epsilon_1\epsilon_2}=
  -i\frac{N^3m^4\mu}{12\epsilon_1\epsilon_2\beta(1+\mu^2)}\ ,
\end{equation}
where we plugged in $r_1=\frac{\beta}{4\pi}$.
This precisely agrees with (\ref{imaginary-normal}), based on D0-D4 calculus. 

Finally, let us comment that the same calculation can be done to test 
some part of (\ref{ADE-asymp}) for all ADE theories. For ADE, 
(\ref{ADE-asymp}) yields the imaginary part 
\begin{equation}\label{Seff-ADE}
  \left.\frac{}{}\!\!{\rm Im}(S_{\rm eff})\right|_{m^4}=
  -\frac{\mu(c_2|G|+r)m^4}{12\epsilon_1\epsilon_2\beta(1+\mu^2)}\ ,
\end{equation}
simply by changing the coefficient $N^3\rightarrow c_2|G|+r$ from (\ref{ADE-asymp}).
On the other hand, the anomaly polynomial (\ref{anomaly-U(N)}) is replaced by 
the following polynomial
\begin{equation}
  I_8=rI_8(1)+c_2|G|\frac{p_2(N)}{24}
\end{equation}
for ADE. Again after restricting $SO(5)_R$ to $U(1)\subset SU(2)_L$, 
the term $\frac{N^3}{24}F^4$ of (\ref{anomaly-pol-reduced}) is replaced by 
$\frac{c_2|G|+r}{24}F^4$. So the calculations of this subsection can be 
done by replacing all $N^3$ by $c_2|G|+r$, completely reproducing 
(\ref{Seff-ADE}).

\section{Conclusions and remarks}

In this paper, we explored S-duality of the prepotential of the 
6d $(2,0)$ theories compactified on $T^2$, on the Coulomb branch. 
We found evidences of S-duality and its anomaly. Using this result, we computed 
the asymptotic free energy of this system compactified on $S^1$ (in the index version), when the Omega background parameters $\epsilon_{1,2}$ and the chemical potential 
$\beta$ conjugate to the KK momentum are small. The asymptotic free 
energy is proportional to $N^3$ in a suitable large $N$ limit, showing 
that the light KK fields exhibit the $N^3$ degrees of freedom.
After suitably complexifying the chemical potentials, we showed that 
the imaginary part of the free energy proportional to $N^3$ is completely 
reproduced from the 6d chiral anomaly of the $SO(5)$ R-symmetry. 
Most results are generalized to the ADE class of $(2,0)$ theories.

In the literatures, the $N^3$ scalings of various observables of 
6d $(2,0)$ theory have been found, using various approaches. Thermal entropy of black
M5-branes \cite{Klebanov:1996un} or various other quantities
are computed from the gravity dual. Chiral anomalies are computed from the 
anomaly inflow mechanism \cite{Harvey:1998bx}. The supersymmetric Casimir 
energy on $S^5$ was computed from the superconformal index
\cite{Kim:2012ava,Lockhart:2012vp,Kim:2012qf,Kim:2013nva,Kallen:2012zn,Bobev:2015kza}. Perhaps among these, the
mysteries of 6d CFT may be most directly addressed from the thermal 
partition function calculus of \cite{Klebanov:1996un}. So it would be 
desirable to have a microscopic view of this phenomenon by directly counting 
states of the 6d CFTs. As far as we are aware of, such a direct account for 
$N^3$ scaling of states has not been available from a mircroscopic quantum 
calculus yet. Our studies show the $N^3$ scalings of the microscopically counted 
degrees of freedom. More precisely, we compactified the
6d SCFT on $S^1$, so $N^3$ degrees of freedom are absent at low energy. However,
at high temperature compared to the inverse-radius
of the circle, we expect the 6d CFT physics to be visible, hopefully in our $F$.
One subtlety is that fermionic states are counted 
with minus sign in the index, so there may be cancelation between 
bosons and fermions. Even after this possible cancelation, we find that
the uncanceled free energy still exhibits $N^3$ scaling, which proves
that the 6d CFT has $N^3$ degrees of freedom. We have provided an alternative study
of the asymptotic free energy based on 6d chiral anomalies, which completely
agrees with our D0-D4 calculus.

Our studies based on D0-D4 system also shows that the light D0-brane
particles are responsible for the UV enhancement of degrees of freedom.
Since D0-branes are the key objects which construct M-theory at strong coupling
limit of the type IIA strings, it is natural to see that they are also responsible
for the $N^3$ degrees of freedom of the 6d $(2,0)$ theory. It will be interesting 
to better understand the the single particle
index $f(\tau,\epsilon_{1,2},m,v)$ which yields this behavior. In particular, 
conjectures on instanton partons \cite{Collie:2009iz} may be addressed in 
more detail.

The Coulomb branch partition function on $\mathbb{R}^4\times T^2$ was used 
as building blocks of interesting CFT indices in the symmetric phase. 
We comment that our asymptotic free energy proportional to $N^3$ 
does not appear in these symmetric phase indices. Let us explain this 
with the 6d superconformal index, and the DLCQ index.

Firstly, it has been proposed that the D0-D4 partition function, or more precisely
this partition function multiplied by the 5d perturbative part, is a building
block for the 6d superconformal indices
\cite{Kim:2012ava,Lockhart:2012vp,Kim:2012qf,Kim:2013nva} on $S^5\times S^1$. 
So one might wonder
whether our finding $\log Z\propto\frac{N^3m^4}{\epsilon_1\epsilon_2\beta}$ 
(with $\tau_D=\frac{i\beta}{2\pi}$)
at high temperature has implications to the supercofonrmal
index. One can immedidately see that the answer is negative. For this discussion,
the relevant formula is presented in \cite{Kim:2013nva}, which uses the product 
of $3$ copies of Coulomb branch partition functions on $\mathbb{R}^4\times T^2$ as 
the integrand. The angular momentum chemical potentials of 
$U(1)^2\subset SO(6)$ on $S^5$ are labeled by three numbers $a_1,a_2,a_3$ 
satisfying $a_1+a_2+a_3=0$. In this setting, the $3$ sets of Omega deformation 
parameters are given by $(\epsilon_1,\epsilon_2)=(a_2-a_1,a_3-a_1)$, 
$(a_3-a_2,a_1-a_2)$, $(a_1-a_3,a_2-a_3)$ respectively. Since the asymptotic 
formula for $Z$ is obtained in the limit of small
$\epsilon_1,\epsilon_2$, one can study the superconformal index in the limit 
of small $a_1,a_2,a_3$. In this limit, the most divergent part in 
$\epsilon_{1,2}$ is given by
\begin{equation}\label{cancel-SC}
  \log Z_{S^5\times S^1}\sim
  \frac{N^3m^4}{\beta}\left[\frac{1}{(a_2-a_1)(a_3-a_1)}
  +\frac{1}{(a_3-a_2)(a_1-a_2)}+\frac{1}{(a_1-a_3)(a_2-a_3)}\right]\ .
\end{equation}
It is an identity that the sum in the square bracket vanishes, 
so that the leading asymptotic part proportional to $N^3$ vanishes on 
$S^5\times S^1$. So our $f_{\rm asymp}$ has no implication to the 
superconformal index. However, study of the the subleading part
$\mathcal{O}(\epsilon_{1,2})^0$ will be interesting, along the lines of 
our section 2.2. We hope to come back to this problem in the near future.

Secondly, the M5-brane theory compactified on a lightlike circle 
can be studied using the D0-D4 quantum mechanics \cite{Aharony:1997th,Aharony:1997an}.
Its index at DLCQ momentum $k$ can be computed by integrating 
the D0-D4 index in the Coulomb branch suitably with the Coulomb VEV $v$, as
explained in \cite{Kim:2011mv}. So one finds (again with 
$\tau_D=\frac{i\beta}{2\pi}\rightarrow 0$) 
\begin{equation}
  Z_{\rm DLCQ}\sim\exp\left[-\frac{N^3m^4}{24\epsilon_1\epsilon_2\beta}\right]\ .
\end{equation}
Here, unlike the partition function on $\mathbb{R}^4\times T^2$, where 
we have notion of multi-particles so that $\log Z$ itself is meaningful 
as the singe particle index, the DLCQ index is defined with a confining 
harmonic potential on $\mathbb{R}^4$ \cite{Kim:2011mv}. Thus, the exponent 
cannot be physically meaningful separately. Also, the definition of 
$Z_{\rm DLCQ}$ is such that $\epsilon_+=\frac{\epsilon_1+\epsilon_2}{2}$ 
has to be real and bigger than other fugacities, as $e^{-\epsilon_+}<1$ 
plays the role of main convergence parameter. 
So one has to set $\epsilon_1\epsilon_2>0$. This implies that $Z_{\rm DLCQ}$ does not 
exhibit exponential growth, but is rather highly suppressed at small $\beta$, 
presumably due to boson/fermion cancelation.

From these observations on the superconformal index and the DLCQ index, 
one realizes that $Z_{\mathbb{R}^4\times T^2}$ contains interesting
dynamical information which may be wiped out in other observables.

Omega deformed partition functions can also be used to study 6d
$(1,0)$ superconformal field theories. In fact, for many 6d $(1,0)$ systems, 
the index on $\mathbb{R}^4\times T^2$ is known in the `self-dual string expansion,' 
similar to the M-string expansion explained in our section 2.2.
The coefficients like $Z_{(n_i)}$ of section 2.2 are elliptic genera of 2d CFTs 
for the 6d self-dual strings in the tensor branch. Those elliptic genera 
have been studied for various 6d $(1,0)$ theories 
\cite{Kim:2014dza,Haghighat:2014vxa,Gadde:2015tra,Kim:2015fxa,Kim:2016foj,DelZotto:2016pvm}.
The S-duality anomaly and the high temperature asymptotic free energies could 
be studied using the approaches explored in this paper. This may be an 
interesting approach to explore the rich physics of 6d CFTs and their 
compactifications to 5d/4d.

It would also be interesting to further study the S-duality of the 
full index of the $(2,0)$ theory, based on some ideas sketched 
in our section 2.2. Following \cite{Galakhov:2013jma}, we find it interesting 
to study the Wilson/'t Hooft line defects uplifed to 6d surface operators. 
S-dualities of other defect operators should also be interesting.

Finally, one may ask if a suitable M2-brane partition function on
$\mathbb{R}^2_\epsilon\times S^1$ can exhibit $N^{\frac{3}{2}}$ scaling,
where $\epsilon$ is the Omega deformation parameter.
Although this scaling has been microscopically computed from the $S^3$ partition
function, or the entanglement entropy, perhaps better physical intuitions can be
obtained by directly accounting for where such degrees of freedom come from,
like we did for 6d SCFTs on $S^1$ from D0-branes (instanton solitons).

\vskip 0.5cm

\hspace*{-0.8cm} {\bf\large Acknowledgements}
\vskip 0.2cm

\hspace*{-0.75cm} We thank Prarit Agarwal, Joonho Kim, Kimyeong Lee, Jaemo Park, 
Jaewon Song, Shuichi Yokoyama for helpful discussions, and especially Hee-Cheol Kim 
for many inspiring discussions and comments. We also thank Joonho Kim for helping us 
with the $SO(8)$ instanton calculus. This work is supported in part by NRF Grant 
2015R1A2A2A01003124 (SK, JN), and Hyundai Motor Chung Mong-Koo Foundation (JN).

\end{document}